\documentclass[twocolumn, tighten]{aastex631} 

\usepackage{float}
\usepackage{amssymb, amsmath}
\usepackage[flushleft]{threeparttable}
\usepackage[english]{babel} 
\usepackage{multirow}
\usepackage{xcolor}
\usepackage{enumitem} 
\usepackage{rotating}

\newcommand{\Ms}{$\textrm{M}_{\odot}$}
\newcommand{\Msperyr}{\textrm{M}$_{\odot}$~\textrm{yr}$^{-1}$}
\newcommand{\kms}{$\textrm{km~s$^{-1}$}$}
\newcommand{\oi}{\,{\sc i}}
\newcommand{\ii}{\,{\sc ii}}
\newcommand{\iii}{\,{\sc iii}}
\newcommand{\Ha}{H$\alpha$}
\newcommand{\Hb}{H$\beta$}
\newcommand{\Hg}{H$\gamma$}

\newcommand{\nb}{\textsc{NBursts}}

\definecolor{MyOrange}{RGB}{255,165,0} 
\definecolor{MyPurple}{RGB}{128,0,128} 
\definecolor{MyMagenta}{RGB}{255,0,255} 
\graphicspath{{./}{}}

\shorttitle{Stellar counter-rotation in MaNGA}
\shortauthors{Gasymov et al.}

\begin{document}

\title{Stellar CoRGI in MaNGA: Stellar Counterrotating Galaxies Identified in the MaNGA Survey}

\correspondingauthor{Damir Gasymov}
\email{damir.gasymov@stud.uni-heidelberg.de}

\author[0000-0002-1750-2096]{Damir Gasymov}
\affiliation{Astronomisches Rechen-Institut, Zentrum für Astronomie der Universität Heidelberg, Mönchhofstr. 12-14, 69120 Heidelberg, Germany}
\affiliation{Sternberg Astronomical Institute, Lomonosov Moscow State University, Universitetskij pr., 13,  Moscow, 119234, Russia}

\author[0000-0002-6425-6879]{Ivan Yu. Katkov}
\affiliation{New York University Abu Dhabi, PO Box 129188, Abu Dhabi, UAE}
\affiliation{Center for Astrophysics and Space Science (CASS), New York University Abu Dhabi, PO Box 129188, Abu Dhabi, UAE}
\affiliation{Sternberg Astronomical Institute, Lomonosov Moscow State University, Universitetskij pr., 13,  Moscow, 119234, Russia}

\author[0000-0001-8427-0240]{Evgenii V. Rubtsov}
\affiliation{Sternberg Astronomical Institute, Lomonosov Moscow State University, Universitetskij pr., 13,  Moscow, 119234, Russia}

\author[0000-0002-4342-9312]{Anna S. Saburova}
\affiliation{Sternberg Astronomical Institute, Lomonosov Moscow State University, Universitetskij pr., 13,  Moscow, 119234, Russia}

\author[0000-0001-8646-0419]{Alexei Yu. Kniazev}
\affiliation{South African Astronomical Observatory, PO Box 9, 7935 Observatory, Cape Town, South Africa}
\affiliation{Southern African Large Telescope Foundation, PO Box 9, 7935 Observatory, Cape Town, South Africa}
\affiliation{Special Astrophysical Observatory of the Russian Academy of Sciences, Nizhnij Arkhyz, 369167 Russia}
\affiliation{Sternberg Astronomical Institute, Lomonosov Moscow State University, Universitetskij pr., 13,  Moscow, 119234, Russia}

\author[0000-0003-4679-1058]{Joseph D. Gelfand}
\affiliation{New York University Abu Dhabi, PO Box 129188, Abu Dhabi, UAE}
\affiliation{Center for Astrophysics and Space Science (CASS), New York University Abu Dhabi, PO Box 129188, Abu Dhabi, UAE}
\affiliation{Center for Cosmology and Particle Physics, New York University, 726 Broadway, room 958, New York, NY 10003}

\author[0000-0003-4946-794X]{Olga K. Sil'chenko}
\affiliation{Sternberg Astronomical Institute, Lomonosov Moscow State University, Universitetskij pr., 13,  Moscow, 119234, Russia}

\author[0000-0002-7924-3253]{Igor V. Chilingarian}
\affiliation{Center for Astrophysics --- Harvard and Smithsonian, 60 Garden St. MS09, Cambridge, MA 02138, USA}
\affiliation{Sternberg Astronomical Institute, Lomonosov Moscow State University, Universitetskij pr., 13,  Moscow, 119234, Russia}

\author[0000-0002-0507-9307]{Alexei V. Moiseev}
\affiliation{Special Astrophysical Observatory of the Russian Academy of Sciences, Nizhnij Arkhyz, 369167 Russia}
\affiliation{Sternberg Astronomical Institute, Lomonosov Moscow State University, Universitetskij pr., 13,  Moscow, 119234, Russia}

\author[0000-0002-1091-5146]{Anastasia V. Kasparova}
\affiliation{Sternberg Astronomical Institute, Lomonosov Moscow State University, Universitetskij pr., 13,  Moscow, 119234, Russia}

\author[0000-0001-9914-4466]{Anatoly Zasov}
\affiliation{Sternberg Astronomical Institute, Lomonosov Moscow State University, Universitetskij pr., 13,  Moscow, 119234, Russia}
\affiliation{Faculty of Physics, Moscow M.V. Lomonosov State University, Leninskie gory 1,  Moscow, 119991, Russia}



\begin{abstract}
Stellar counterrotating (CR) galaxies are systems hosting two large-scale stellar components rotating in opposite directions --- a main, preexisting galaxy body with an older stellar population and a younger CR stellar disk likely formed from externally accreted gas. 
Such systems offer a unique opportunity to study disk assembly by analyzing the stellar populations of each component. 
Using integral field spectroscopic data from the SDSS-IV Mapping Nearby Galaxies at Apache Point Observatory survey, we identified a sample of 120 CR disk galaxies (65 reliable and 55 probable systems) by inspecting their kinematic maps and analyzing the shape of the stellar line-of-sight velocity distribution, which was recovered nonparametrically. Of these, 74 CR galaxies have not been reported in previous studies. 
For one-third of our sample, we further derived the ages and metallicities of stars for both disks via a spectral decomposition technique. 
We show that the observed spatial bimodality --- where the CR disk either is concentrated in the central region (inner counterrotation) or dominates the outer part of the galaxy (outer counterrotation) --- is driven by differences in the stellar mass and angular momentum of the CR disk. 
The wide range of stellar metallicities observed in CR disks suggests that no single source of external material is solely responsible for the formation of counterrotation in all galaxies; instead, proposed mechanisms such as merger with gas-rich satellites, accretion from cosmic filaments, and exchange of gas between neighboring galaxies can dominate in individual cases.

\end{abstract}



\section{Introduction} \label{sec:intro}

Understanding the detailed processes by which galaxies build and evolve their stellar disks remains an important area of study in galaxy formation theory.
However, observationally probing the details of this process is challenging because observed galaxies have already assembled disks, and reconstructing their formation history from spectra integrated along the line-of-sight is nontrivial.
One of the main challenges is that galaxies are not closed, isolated systems; they experience mergers with other galaxies \citep{Toomre1972ApJ...178..623T, Barnes1988ApJ...331..699B} and accrete external material \citep{Sancisi2008A&ARv..15..189S, Combes2014ASPC..480..211C, Putman2017ASSL..430....1P}.
The acquisition of a substantial amount of material, particularly if this material has angular momentum opposite to that of the host, can lead to the formation of distinct kinematic components.
Galaxies with these features, termed ``multi-spin'' galaxies \citep{Rubin1994AJ....108..456R}, include remarkable examples where two stellar components -- typically disks -- rotate in the same plane but in opposite directions, a phenomenon known as stellar counter-rotation (CR).
CR in disks is particularly intriguing because the high rotational amplitude generates pronounced kinematic separation, which facilitates disentangling the stellar populations in the two disks and provides insights into the formation and evolution of each.

Co-spatial stellar CR disks were first discovered and studied in NGC~4550 by \citet{Rubin1992ApJ...394L...9R} and \citet{Rix1992ApJ...400L...5R}.
Notably, even before this discovery, evidence for stellar CR had been found in elliptical galaxies through the detection of kinematically decoupled cores and nuclear disks \citep{Franx1988ApJ...327L..55F, Bender1988A&A...202L...5B, Jedrzejewski1988ApJ...330L..87J, Franx1989ApJ...344..613F, Bertola1992ApJ...401L..79B}.
Subsequent studies identified CR stellar disks in NGC~7217 \citep{Merrifield1994ApJ...432..575M}, NGC~3593 \citep{Bertola1996ApJ...458L..67B}, and NGC~4138 \citep{Jore1996AJ....112..438J}, confirming that the CR phenomenon is not unique to NGC~4550.

The next significant advance in the field came with the development and application of spectral decomposition techniques \citep{Chilingarian2011MNRAS.412.1627C, Coccato2011MNRAS.412L.113C, Johnston2013MNRAS.428.1296J, Katkov2013ApJ...769..105K}, which over the past decade have enabled the assessment of stellar population properties in many CR disk galaxies (Tab.~\ref{tab:cr_galaxies}).
These studies indicate that the CR disk generally contains a younger stellar population that co-rotates with the ionized gas, while the main stellar disk is typically older.
This suggests that CR stellar disks likely form in situ from externally acquired gas rather than through the accretion of ex situ stars during a merger.
The metallicity of CR stars varies widely among galaxies, ranging from lower to comparable or even higher than that of the main disk.
Several scenarios have been proposed to explain the formation of CR disks, including accretion from cosmic filaments \citep{2014MNRAS.437.3596A}, ``wet'' mergers with gas-rich dwarf galaxies \citep{1996ApJ...461...55T}, tidal gas exchange between neighboring galaxies \citep{Khim2021ApJS..254...27K, Sil'chenko2023Galax..11..119S}, and major merger events with finely tuned orbits that preserve the disk structure \citep{2001Ap&SS.276..909P, 2009MNRAS.393.1255C}.
Lower metallicity is expected in CR disks formed from pristine gas accreted via cosmic filaments, whereas other mechanisms tend to result in higher metallicity since the externally accreted gas is already pre-enriched. 
However, the exact metallicity depends on both star formation and metal enrichment history.
Although bar dissolution via separatrix crossing \citep{Evans1994ApJ...420L..67E} has also been proposed as an internal mechanism for the formation of CR disks, it cannot account for the observed differences in stellar populations -- particularly the significant age differences between stellar components in many CR galaxies -- and is therefore not considered here.

\begin{deluxetable}{ll}
\tablecaption{CR disk galaxies studied by means of spectral decomposition technique \label{tab:cr_galaxies}}
\tablehead{
\colhead{Studied CR Galaxy} & \colhead{Reference}
}
\startdata
NGC~5719     & \citet{Coccato2011MNRAS.412L.113C} \\
\multirow{2}{*}{IC~719}       & \citet{Katkov2013ApJ...769..105K} \\ 
 & \citet{Pizzella2018A\string&A...616A..22P} \\
\multirow{2}{*}{NGC~4550}     & \citet{Coccato2013A\string&A...549A...3C} \\ 
 & \citet{Johnston2013MNRAS.428.1296J} \\
NGC~3593     & \citet{Coccato2013A\string&A...549A...3C} \\
NGC~4138     & \citet{Pizzella2014A\string&A...570A..79P} \\
NGC~4191     & \citet{Coccato2015A\string&A...581A..65C} \\
NGC~448      & \citet{Katkov2016MNRAS.461.2068K} \\
NGC~1366     & \citet{Morelli2017A\string&A...600A..76M} \\
NGC~5102     & \citet{Mitzkus2017MNRAS.464.4789M} \\
SDSS~J0748+4441 & \multirow{2}{*}{\citet{Bao2024MNRAS.528.2643B}} \\
(PGC~21856) & \\
PGC~66551    & \citet{Katkov2024ApJ...962...27K} \\
\enddata
\end{deluxetable}

Massive spectroscopic IFU surveys, particularly SDSS-IV MaNGA \citep{Bundy2015ApJ...798....7B}, which observed $\sim$10,000 galaxies, have greatly advanced the studies of stellar CR and demonstrated that the phenomenon is not particularly rare.
Several research teams identified numerous CR galaxies in MaNGA, suggesting that $\approx 1\%$ of its galaxies exhibit stellar CR \citep{Graham2018MNRAS.477.4711G, Bao2022ApJ...926L..13B, Bevacqua2022MNRAS.511..139B}. A detailed comparison between our sample and presented in previous works is provided in App.~\ref{sec:App_A}.
In this work, we perform a detailed search for, and analysis of, CR galaxies within the MaNGA survey using a different approach than past work.
Our final selection of CR galaxies is based not only on the stellar and gas kinematic maps derived from MaNGA data but also on an analysis of the non-parametrically recovered stellar Line-Of-Sight Velocity Distribution (LOSVD), which allowed us to reliably detect CR galaxies even when the kinematic maps are unclear.
For $\approx1/3$ of the identified CR galaxies, we have also carried out spectral decomposition and assessed the stellar population properties in both the main and CR disks.

This paper is organized as follows.
In Section~\ref{sec:sample}, we describe the data used in this work and our visual inspection procedure.
In Section~\ref{sec:analysis}, we present the workflow for analyzing the MaNGA data.
In Section~\ref{sec:discussion}, we discuss the dichotomy and possible sources of accreted material in CR-galaxies.
Throughout the paper, we adopt the WMAP9 cosmology with $H_0~=~69.3$~km s$^{-1}$ Mpc$^{-1}$ and $\Omega_0 = 0.2865$ \citep{Hinshaw2013ApJS..208...19H}.

\section{Sample construction}
\label{sec:sample}

\subsection{Data}

We performed a search for stellar counter-rotating galaxies using data collected by the MaNGA survey \citep{Abdurro'uf2022ApJS..259...35A} based on Integral Field Unit (IFU) spectroscopic \citep{Drory2015AJ....149...77D} observations,conducted at the Sloan 2.5-m telescope \citep{Gunn2006AJ....131.2332G}, of some 10,000 galaxies in the nearby Universe ($0.01~\lesssim~z~\lesssim~0.15$) covering a wide range of stellar mass and colors \citep{Wake2017AJ....154...86W}.
In the survey, each galaxy was observed using  2~arcsec fibers packed in bundles that vary in diameter from 12~arcsec (19 fibers) to 32~arcsec (127 fibers) covering $(1.5-2.5)\times$ the effective diameter of the target.
Fiber bundles feed the BOSS spectrographs \citep{Smee2013AJ....146...32S}, which provide wavelength coverage of $\lambda\lambda$3600-10300~\AA\ with spectral resolving power $R\sim2000$, corresponding to an instrumental dispersion $\sigma_\mathrm{inst.}\approx75$~\kms\ at 5100~\AA\ \citep{Law2016AJ....152...83L}.
To fill in the gaps between the round fibers in a bundle, each galaxy was observed using 3 spatially dithered exposures \citep{Law2015AJ....150...19L, Yan2016AJ....152..197Y}.

In our work, we also used the quantities from MaNGA value added catalogs (VACs) and external galaxy catalogs: 
\begin{itemize}
    \item MaNGA-HI DR4 \citep{Masters2019MNRAS.488.3396M, Stark2021MNRAS.503.1345S} provides atomic hydrogen measurements based on 21~cm observations from the Green Bank Telescope (GBT), complemented by the ALFALFA survey \citep{Haynes2018ApJ...861...49H}. It includes data for 7290 galaxies in MaNGA survey.
    \item Galaxy Environment for MaNGA (GEMA-VAC) (Argudo-Fernández M. et al. in prep.) includes environmental properties for all MaNGA galaxies, based on the methods described by \cite{Argudo-Fernandez2015A&A...578A.110A, Etherington2015MNRAS.451..660E, Wang2016ApJ...831..164W}.
    \item MaNGA Visual Morphology Catalogue (MVM-VAC) \citep{Vazquez-Mata2022MNRAS.512.2222V} provides visual morphological classification for MaNGA galaxies. Reliable classifications are available for 9361 galaxies;
    \item NASA-Sloan Atlas (NSA) \citep{Blanton2011AJ....142...31B, Wake2017AJ....154...86W} contains photometric and structural parameters for nearby galaxies based on SDSS imaging;
    \item GALEX-SDSS-WISE Legacy Catalog (GSWLC) \citep{Salim2016ApJS..227....2S} combines photometric data from the GALEX, SDSS, and WISE surveys, providing multi-wavelength properties for a broad range of galaxies.
    \item Reference Catalog of Spectral Energy Distributions (RCSED) \citep{Chilingarian2017ApJS..228...14C} offers homogenized UV-NIR photometry and SDSS spectroscopic measurements and analysis for $\sim$800,000 galaxies.
\end{itemize}

\subsection{Visual inspection}
\label{sec:vis_inspec}

\startlongtable
\begin{deluxetable*}{clrrcccccccc}
\label{Tab:Rel_samp}
\tablecaption{Sample of \textbf{reliable} galaxies with stellar counter-rotation.}
\tablewidth{700pt}
\tabletypesize{\tiny}
\tablehead{
\colhead{No.} & 
\colhead{ID} & 
\colhead{R.A.} & 
\colhead{Dec.} & 
\colhead{Type} & 
\colhead{$\log$ (M$_\star$/M$_\odot$)} & 
\colhead{z} & 
\colhead{Features} & 
\colhead{CR config.} &
\colhead{Env.} & 
\colhead{Morph.}& 
\colhead{Ref.}} 
\decimalcolnumbers
\startdata
1 & \href{https://manga.voxastro.org/visualiser/9195-3704}{1-42660}$^{\dagger}$ & 01:52:44.4 & 13:11:33.18 & SABab &10.27 & 0.026 & 2$\sigma$ & outer & \textbf{V}, \textbf{G} & \textbf{LSB} & 2 \\
2 & \href{https://manga.voxastro.org/visualiser/9514-1902}{1-42247}$^{\dagger}$ & 02:14:21.1 & 14:07:06.73 & Sab &9.71 & 0.029 & 2$\sigma$, Srot, CR-GS & inner & \textbf{V}, \textbf{G} & \ldots & \ldots \\
3 & \href{https://manga.voxastro.org/visualiser/8077-6103}{\textbf{1-109056}}$^{\dagger}$ & 02:37:47.2 & 00:24:18.31 & SABa &10.41 & 0.047 & 2$\sigma$ & outer & \textbf{V}, \textbf{P}, \textbf{G} & \ldots & \ldots \\
4 & \href{https://manga.voxastro.org/visualiser/8077-3702}{1-37062}$^{\dagger}$ & 02:47:23.1 & 00:03:31.52 & SAB0a &10.19 & 0.025 & 2$\sigma$, CR-GS, Srot & inner & \textbf{V}, \textbf{G} & \textbf{T}, \textbf{Sh} & 3 \\
5 & \href{https://manga.voxastro.org/visualiser/8077-3704}{1-37068}$^{}$ & 02:48:47.6 & -00:06:33.07 & SAB0a &10.34 & 0.025 & 2$\sigma$, CR-GS, NRR & inner & \textbf{V}, \textbf{G} & \ldots & \ldots \\
6 & \href{https://manga.voxastro.org/visualiser/8154-3703}{\textbf{1-37494}}$^{\dagger}$ & 02:58:30.1 & 00:47:39.24 & S0a &10.18 & 0.043 & 2$\sigma$ & outer & \textbf{V}, \textbf{P}, \textbf{G} & \ldots & 2 \\
7 & \href{https://manga.voxastro.org/visualiser/8154-1902}{\textbf{1-37478}}$^{\dagger}_{\star}$ & 02:59:23.2 & -01:10:00.07 & S0 &10.46 & 0.028 & $\sigma$-elong. & inner & \textbf{V}, \textbf{G} & \ldots & \ldots \\
8 & \href{https://manga.voxastro.org/visualiser/9192-3703}{\textbf{1-109275}}$^{\dagger}$ & 03:04:17.0 & 00:31:11.59 & SBa &10.05 & 0.044 & 2$\sigma$, CR-GS, Srot & inner & \textbf{V}, \textbf{G} & \textbf{B} & 3 \\
9 & \href{https://manga.voxastro.org/visualiser/8155-3702}{1-38543}$^{\dagger}$ & 03:36:07.8 & -00:35:47.18 & S0a &10.36 & 0.023 & 2$\sigma$, CR-GS & inner & \textbf{V}, \textbf{G} & \textbf{E} & 2 \\
10 & \href{https://manga.voxastro.org/visualiser/10216-1902}{\textbf{1-297863}}$^{\dagger}_{\star}$ & 07:47:04.1 & 18:08:18.03 & SAB0 &10.25 & 0.045 & 2$\sigma$, NRR & inner & \textbf{Cl}, \textbf{G} & \ldots & \ldots \\
11 & \href{https://manga.voxastro.org/visualiser/8138-6102}{1-339061}$^{\dagger}$ & 07:48:34.6 & 44:41:17.83 & SAB0a &10.32 & 0.020 & 2$\sigma$, CR-GS, Srot & inner & \textbf{Sh}, \textbf{G}, \textbf{O} & \textbf{T}, \textbf{Sh} & 1,2,4 \\
12 & \href{https://manga.voxastro.org/visualiser/8710-12702}{\textbf{1-378729}}$^{\dagger}$ & 07:52:04.2 & 50:07:50.96 & S0 &10.20 & 0.024 & Gas-mis, 2$\sigma$ & undef & \textbf{Cl}, \textbf{G} & \ldots & \ldots \\
13 & \href{https://manga.voxastro.org/visualiser/8143-1902}{1-44047}$^{\dagger}$ & 07:58:34.4 & 41:34:42.27 & S0 &10.22 & 0.041 & 2$\sigma$, Srot & outer & \textbf{F}, \textbf{G}, \textbf{O} & \ldots & 1,2 \\
14 & \href{https://manga.voxastro.org/visualiser/8143-3702}{1-44483}$^{\dagger}$ & 07:59:20.4 & 42:03:25.46 & S0a &10.12 & 0.025 & 2$\sigma$, Srot & outer & \textbf{Sh}, \textbf{G}, \textbf{O} & \textbf{E}, \textbf{T} & 1,2,3 \\
15 & \href{https://manga.voxastro.org/visualiser/8149-1901}{1-145913}$^{}$ & 08:00:53.2 & 28:07:49.42 & Edc &9.85 & 0.017 & Srot, CR-GS, 2$\sigma$? & inner & \textbf{F}, \textbf{G}, \textbf{O} & \ldots & \ldots \\
16 & \href{https://manga.voxastro.org/visualiser/9496-3701}{1-382889}$^{\dagger}$ & 08:02:28.9 & 20:30:50.33 & Sa &10.37 & 0.029 & 2$\sigma$, CR-GS, Srot & inner & \textbf{Sh} & \textbf{E} & \ldots \\
17 & \href{https://manga.voxastro.org/visualiser/9485-3703}{\textbf{1-121871}}$^{\dagger}$ & 08:03:04.0 & 36:15:52.16 & S0 &10.11 & 0.044 & 2$\sigma$ & inner & \textbf{F}, \textbf{G}, \textbf{O} & \ldots & 2 \\
18 & \href{https://manga.voxastro.org/visualiser/10511-3704}{1-76425}$^{}$ & 08:47:18.6 & 02:45:44.20 & Sab &10.61 & 0.028 & 2$\sigma$, CR-GS & outer & \textbf{Cl}, \textbf{G} & \textbf{T}, \textbf{LSB} & \ldots \\
19 & \href{https://manga.voxastro.org/visualiser/9506-3703}{1-386322}$^{\dagger}$ & 08:56:51.3 & 27:16:12.37 & S0a &9.94 & 0.019 & 2$\sigma$, Srot & outer & \textbf{Sh}, \textbf{G}, \textbf{O} & \textbf{E} & 2 \\
20 & \href{https://manga.voxastro.org/visualiser/12488-6101}{1-300092}$^{}$ & 09:00:47.5 & 29:59:25.02 & Sa &10.74 & 0.054 & CR-GS, Srot & outer & \textbf{F}, \textbf{G} & \textbf{T}, \textbf{Sh} & \ldots \\
21 & \href{https://manga.voxastro.org/visualiser/8249-1901}{1-137890}$^{\dagger}$ & 09:08:52.6 & 44:55:56.16 & Sb &9.89 & 0.027 & 2$\sigma$, CR-GS, Srot & inner & \textbf{F}, \textbf{G}, \textbf{O} & \textbf{E} & 1,2,3 \\
22 & \href{https://manga.voxastro.org/visualiser/12487-1902}{1-386932}$^{\dagger}$ & 09:11:22.7 & 30:05:47.15 & Sa &9.87 & 0.026 & 2$\sigma$, CR-GS, NRR & inner & \textbf{F}, \textbf{G} & \ldots & 3 \\
23 & \href{https://manga.voxastro.org/visualiser/8152-3703}{1-232087}$^{\dagger}$ & 09:30:53.9 & 35:26:23.23 & SABa &10.28 & 0.028 & 2$\sigma$, Srot & outer & \textbf{F}, \textbf{G} & \ldots & \ldots \\
24 & \href{https://manga.voxastro.org/visualiser/10515-12702}{1-78381}$^{\dagger}$ & 09:52:00.7 & 04:09:02.77 & Sa &10.96 & 0.031 & 2$\sigma$, Srot & outer & \textbf{F}, \textbf{G} & \textbf{Sh} & \ldots \\
25 & \href{https://manga.voxastro.org/visualiser/10519-3704}{\textbf{1-182730}}$^{\dagger}_{\star}$ & 10:18:23.6 & 05:44:10.06 & S0 &11.05 & 0.074 & Srot, $\sigma$-elong. & inner & \textbf{V}, \textbf{G} & \ldots & \ldots \\
26 & \href{https://manga.voxastro.org/visualiser/11838-3704}{1-586725}$^{\dagger}$ & 10:24:07.7 & 00:08:30.96 & SABa &10.37 & 0.023 & 2$\sigma$, CR-GS & inner & \textbf{Sh}, \textbf{P}, \textbf{G} & \textbf{T}, \textbf{LSB}, \textbf{C} & \ldots \\
27 & \href{https://manga.voxastro.org/visualiser/8455-3704}{1-274545}$^{}$ & 10:30:44.6 & 40:03:26.79 & S0a &9.69 & 0.023 & 2$\sigma$, Srot, CR-GS & unclear & \textbf{F}, \textbf{G} & \ldots & 2 \\
28 & \href{https://manga.voxastro.org/visualiser/8253-1902}{1-255220}$^{}$ & 10:31:23.0 & 42:16:37.85 & S0a &9.87 & 0.022 & 2$\sigma$, CR-GS, Srot & inner & \textbf{F}, \textbf{G}, \textbf{O} & \textbf{E} & 1,2 \\
29 & \href{https://manga.voxastro.org/visualiser/8999-6102}{1-148987}$^{\dagger}$ & 10:58:12.4 & 50:12:48.44 & Sa &10.34 & 0.024 & 2$\sigma$ & inner & \textbf{Cl}, \textbf{G}, \textbf{O} & \textbf{E} & \ldots \\
30 & \href{https://manga.voxastro.org/visualiser/8995-3703}{\textbf{1-188530}}$^{\dagger}$ & 11:43:16.3 & 55:16:39.74 & S0a &10.66 & 0.055 & CR-GS, Srot & inner & \textbf{F}, \textbf{O} & \textbf{E} & 2,3 \\
31 & \href{https://manga.voxastro.org/visualiser/8989-9101}{1-174947}$^{\dagger}$ & 11:45:26.1 & 49:52:44.54 & SBab &10.68 & 0.033 & 2$\sigma$, Srot & outer & \textbf{F}, \textbf{G}, \textbf{O} & \textbf{B}, \textbf{Sp} & 2,3 \\
32 & \href{https://manga.voxastro.org/visualiser/12489-3703}{1-139814}$^{}$ & 11:57:23.5 & 04:32:35.01 & Sa &10.17 & 0.020 & 2$\sigma$ & unclear & \textbf{F}, \textbf{G} & \ldots & \ldots \\
33 & \href{https://manga.voxastro.org/visualiser/8310-1901}{1-519907}$^{\dagger}$ & 11:58:34.7 & 24:07:01.77 & Sb &10.09 & 0.030 & 2$\sigma$, gas-polar, NRR & outer & \textbf{Sh}, \textbf{G} & \textbf{E}, \textbf{C} & \ldots \\
34 & \href{https://manga.voxastro.org/visualiser/10510-3704}{1-189376}$^{}$ & 12:04:03.4 & 55:12:17.69 & S0 &9.88 & 0.019 & CR-GS, 2$\sigma$, Srot & inner & \textbf{F}, \textbf{G} & \ldots & \ldots \\
35 & \href{https://manga.voxastro.org/visualiser/9873-12703}{1-623416}$^{\dagger}_{\star}$ & 12:57:53.9 & 28:29:59.25 & S0a &10.16 & 0.024 & 2$\sigma$, Srot & outer & \textbf{Cl}, \textbf{G} & \ldots & \ldots \\
36 & \href{https://manga.voxastro.org/visualiser/9875-9101}{1-456884}$^{\dagger}_{\star}$ & 12:58:19.2 & 27:45:43.54 & Sa &9.54 & 0.018 & 2$\sigma$, Srot & inner & \textbf{Cl}, \textbf{G} & \textbf{E} & \ldots \\
37 & \href{https://manga.voxastro.org/visualiser/9874-12705}{1-456672}$^{\dagger}_{\star}$ & 13:00:19.1 & 27:33:13.37 & SB0 &10.13 & 0.020 & 2$\sigma$, Srot & outer & \textbf{Cl}, \textbf{G} & \textbf{B}, \textbf{T} & \ldots \\
38 & \href{https://manga.voxastro.org/visualiser/11009-12703}{1-456645}$^{\dagger}$ & 13:03:05.9 & 26:31:52.02 & SABa &9.96 & 0.019 & 2$\sigma$, NRR & outer & \textbf{Cl}, \textbf{G} & \textbf{LSB} & \ldots \\
39 & \href{https://manga.voxastro.org/visualiser/8444-3701}{1-419257}$^{\dagger}$ & 13:24:33.7 & 31:32:46.96 & SB0a &10.31 & 0.023 & 2$\sigma$, CR-GS, Srot & inner & \textbf{F}, \textbf{G}, \textbf{O} & \textbf{B} & 2 \\
40 & \href{https://manga.voxastro.org/visualiser/8982-1902}{1-457547}$^{}$ & 13:27:41.0 & 26:03:41.88 & SABab &9.86 & 0.024 & 2$\sigma$ & unclear & \textbf{F}, \textbf{G}, \textbf{O} & \ldots & 2 \\
41 & \href{https://manga.voxastro.org/visualiser/11020-1901}{1-244115}$^{\dagger}$ & 13:38:01.5 & 55:11:15.52 & Sa &10.14 & 0.025 & 2$\sigma$, CR-GS, NRR & inner & \textbf{Cl}, \textbf{G} & \textbf{E} & \ldots \\
42 & \href{https://manga.voxastro.org/visualiser/8447-3703}{\textbf{1-260823}}$^{\dagger}$ & 13:53:10.0 & 40:07:09.11 & S0 &10.47 & 0.051 & 2$\sigma$, NRR & inner & \textbf{Sh}, \textbf{G}, \textbf{O} & \ldots & \ldots \\
43 & \href{https://manga.voxastro.org/visualiser/11012-3702}{1-245235}$^{\dagger}$ & 14:05:41.9 & 55:46:36.83 & Sab &10.15 & 0.041 & 2$\sigma$, CR-GS, Srot & inner & \textbf{F}, \textbf{G} & \ldots & \ldots \\
44 & \href{https://manga.voxastro.org/visualiser/11012-6104}{1-245301}$^{\dagger}$ & 14:10:07.7 & 54:55:11.75 & SABa &10.75 & 0.040 & 2$\sigma$, CR-GS & outer & \textbf{Cl}, \textbf{G} & \textbf{LSB}, \textbf{Sh} & \ldots \\
45 & \href{https://manga.voxastro.org/visualiser/9868-12704}{1-593328}$^{\dagger}$ & 14:38:37.7 & 46:39:47.23 & Sa &11.20 & 0.037 & 2$\sigma$, CR-GS, Srot & inner & \textbf{Cl}, \textbf{G} & \ldots & 2 \\
46 & \href{https://manga.voxastro.org/visualiser/11834-1902}{\textbf{1-13736}}$^{\dagger}$ & 14:52:19.4 & -00:16:09.14 & S0 &10.08 & 0.044 & 2$\sigma$, CR-GS, Srot & inner & \textbf{Cl}, \textbf{G} & \ldots & \ldots \\
47 & \href{https://manga.voxastro.org/visualiser/11962-3701}{1-566335}$^{}$ & 15:05:07.9 & 08:06:26.47 & SABa &10.33 & 0.047 & 2$\sigma$, Srot & outer & \textbf{F} & \ldots & 3 \\
48 & \href{https://manga.voxastro.org/visualiser/11962-6103}{1-333770}$^{}_{\star}$ & 15:05:13.6 & 08:47:47.79 & Edc &11.06 & 0.045 & 2$\sigma$, Srot & unclear & \textbf{F}, \textbf{G} & \textbf{T}, \textbf{Sh} & \ldots \\
49 & \href{https://manga.voxastro.org/visualiser/9872-3701}{1-633000}$^{\dagger}$ & 15:32:55.7 & 42:26:17.72 & SAB0 &10.06 & 0.020 & 2$\sigma$, CR-GS, Srot & inner & \textbf{Cl}, \textbf{G}, \textbf{O} & \ldots & 2,3 \\
50 & \href{https://manga.voxastro.org/visualiser/8314-3702}{1-199775}$^{}$ & 16:03:23.5 & 39:59:08.68 & S0 &10.22 & 0.030 & 2$\sigma$, CR-GS, NRR & inner & \textbf{F}, \textbf{G} & \ldots & 3 \\
51 & \href{https://manga.voxastro.org/visualiser/8600-1902}{\textbf{1-210611}}$^{\dagger}$ & 16:17:33.7 & 41:41:18.76 & SAB0 &10.30 & 0.027 & CR-GS, $\sigma$-elong. & inner & \textbf{F}, \textbf{G}, \textbf{O} & \ldots & \ldots \\
52 & \href{https://manga.voxastro.org/visualiser/8604-6103}{1-248869}$^{\dagger}$ & 16:18:45.8 & 39:20:04.01 & S0a &10.85 & 0.032 & 2$\sigma$, CR-GS & outer & \textbf{F}, \textbf{G}, \textbf{O} & \ldots & 2 \\
53 & \href{https://manga.voxastro.org/visualiser/9027-3703}{1-323764}$^{}$ & 16:24:13.0 & 31:52:33.09 & Sbc &9.71 & 0.021 & 2$\sigma$, NRR & unclear & \textbf{V}, \textbf{G}, \textbf{O} & \ldots & 2 \\
54 & \href{https://manga.voxastro.org/visualiser/11942-3702}{\textbf{1-135190}}$^{\dagger}$ & 16:27:00.8 & 40:48:37.89 & S0 &10.47 & 0.030 & CR-GS, $\sigma$-elong. & inner & \textbf{Cl}, \textbf{P}, \textbf{G} & \ldots & \ldots \\
55 & \href{https://manga.voxastro.org/visualiser/8604-3703}{1-135503}$^{}_{\star}$ & 16:31:31.7 & 39:53:51.55 & Sa &10.72 & 0.030 & 2$\sigma$, star-polar & undef & \textbf{Cl}, \textbf{G}, \textbf{O} & \textbf{Sh}, \textbf{P} & \ldots \\
56 & \href{https://manga.voxastro.org/visualiser/9026-1902}{\textbf{1-94773}}$^{\dagger}$ & 16:37:12.6 & 44:05:07.84 & SB0a &10.18 & 0.033 & 2$\sigma$, CR-GS & inner & \textbf{F}, \textbf{G}, \textbf{O} & \textbf{B} & 1,2 \\
57 & \href{https://manga.voxastro.org/visualiser/11946-3701}{1-211561}$^{\dagger}$ & 16:40:46.8 & 37:02:31.32 & Sa &10.27 & 0.030 & 2$\sigma$, CR-GS, NRR & inner & \textbf{F}, \textbf{G} & \ldots & 3 \\
58 & \href{https://manga.voxastro.org/visualiser/11946-12703}{1-135767}$^{\dagger}$ & 16:40:55.1 & 37:54:15.54 & Sab &10.55 & 0.031 & 2$\sigma$, Srot & outer & \textbf{F}, \textbf{G} & \textbf{LSB}, \textbf{Sh}, \textbf{Sp} & \ldots \\
59 & \href{https://manga.voxastro.org/visualiser/9026-3704}{1-94690}$^{\dagger}$ & 16:45:30.7 & 43:34:53.91 & S0a &10.22 & 0.031 & 2$\sigma$, CR-GS & inner & \textbf{F}, \textbf{G}, \textbf{O} & \textbf{E} & 2 \\
60 & \href{https://manga.voxastro.org/visualiser/8606-3702}{1-136248}$^{\dagger}$ & 16:55:10.5 & 36:54:22.72 & Sa &10.52 & 0.024 & 2$\sigma$, Srot & outer & \textbf{F}, \textbf{O} & \ldots & 1,2,3 \\
61 & \href{https://manga.voxastro.org/visualiser/9186-3704}{\textbf{1-178027}}$^{\dagger}$ & 17:14:58.2 & 29:08:18.57 & SABab &10.44 & 0.045 & 2$\sigma$, CR-GS & outer & \textbf{F}, \textbf{G} & \textbf{LSB}, \textbf{C}, \textbf{Sp} & \ldots \\
62 & \href{https://manga.voxastro.org/visualiser/8618-6102}{1-635506}$^{\dagger}$ & 21:16:24.8 & 10:16:24.19 & SB0 &10.21 & 0.017 & CR-GS, 2$\sigma$ & inner & \textbf{V}, \textbf{G} & \textbf{B} & \ldots \\
63 & \href{https://manga.voxastro.org/visualiser/8615-1902}{1-179561}$^{\dagger}$ & 21:19:00.3 & -00:57:50.37 & S0 &10.09 & 0.020 & 2$\sigma$, CR-GS, Srot & inner & \textbf{V}, \textbf{P}, \textbf{G} & \textbf{C} & 1,2,3,5 \\
64 & \href{https://manga.voxastro.org/visualiser/8652-6102}{\textbf{1-26197}}$^{\dagger}$ & 22:02:53.7 & -00:47:04.57 & S0a &10.93 & 0.064 & 2$\sigma$, CR-GS, Srot & outer & \textbf{V}, \textbf{G} & \ldots & 3 \\
65 & \href{https://manga.voxastro.org/visualiser/7977-3701}{1-115097}$^{\dagger}$ & 22:08:48.8 & 13:21:53.06 & Sab &10.23 & 0.027 & 2$\sigma$, CR-GS, Srot & inner & \textbf{V}, \textbf{G} & \textbf{E} & 2,3 \\
\enddata
\begin{tablenotes}
\small
\item \textsc{NOTE:} Column (1) lists the sequential number of galaxy.
(2) is the MaNGA ID of the galaxy.
In the e-version, the ID contains a hyperlink to the galaxy page in the interactive visualizer service \url{https://manga.voxastro.org}.
Symbol $\dagger$ marks galaxies with a clear X-shape in the non-parametric LOSVD (Sec.~\ref{sec:np_losvd}), $\star$ marks galaxies without emission lines. Galaxies with bold IDs were upgraded from the probable to the reliable category after LOSVD inspection.
(3) and (4) are sky coordinates.
(5) morphological type from MVM-VAC.
(6) contains the stellar mass from NSA, corrected to the adopted $H_0^{2}$.
(7) is the redshift from MaNGA-DAP.
(8) is the categories assigned during our visual inspection (Sec.~\ref{sec:vis_inspec}).
(9) an inner or outer spatial configuration of the CR-disk (Sec.~\ref{sec:in_out_CR}), ``unclear'' cases refer to galaxies where classification is not possible based on kinematic maps or LOSVD analysis, and ``undef'' where the CR disk is out of the main galactic plane.
(10) provides environmental galaxy properties from the GEMA-VAC
(\textbf{P}~---~Pair,
\textbf{G}~---~Group,
\textbf{O}~---~Overdensity,
\textbf{Cl}~---~Cluster,
\textbf{Sh}~---~Sheet,
\textbf{F}~---~Filament,
\textbf{V}~---~Void).
(11) includes the morphological features from the MVM-VAC
(\textbf{B}~---~Bar,
\textbf{E}~---~Edge-on,
\textbf{T}~---~Tidal)
and from our visual inspection
(\textbf{LSB}~---~LSB outskirts,
\textbf{Sh}~---~Shells/tails,
\textbf{C}~---~Near companion/cluster,
\textbf{Sp}~---~Spirals,
\textbf{P}~---~Polar structure).
(12) indicates the presence of the galaxy in previous studies of galaxies with counter-rotation (App.~\ref{sec:App_A}):
[1]~\cite{Graham2018MNRAS.477.4711G},
[2]~\cite{Bevacqua2022MNRAS.511..139B},
[3]~\cite{Beom2024AJ....168..197B},
[4]~\cite{Bao2024MNRAS.528.2643B},
[5]~\cite{Katkov2024ApJ...962...27K}.

\end{tablenotes}
\end{deluxetable*}

\begin{deluxetable*}{clrrcccccccc}
\label{Tab:Prob_samp}
\tablecaption{Sample of \textbf{{probable}} galaxies with stellar counter-rotation. The column descriptions are provided in the Tab.~\ref{Tab:Rel_samp}.}
\tablewidth{700pt}
\tabletypesize{\tiny}
\tablehead{
\colhead{No.} & 
\colhead{ID} & 
\colhead{R.A.} & 
\colhead{Dec.} & 
\colhead{Type} & 
\colhead{$\log$ (M$_\star$/M$_\odot$)} & 
\colhead{z} & 
\colhead{Features} & 
\colhead{CR config.} &
\colhead{Env.} & 
\colhead{Morph.}& 
\colhead{Ref.}} 
\decimalcolnumbers
\startdata
1 & \href{https://manga.voxastro.org/visualiser/8656-6101}{1-600853}$^{}$ & 00:31:12.1 & -00:24:26.48 & Sa &10.35 & 0.019 & 2$\sigma$?, CR-GS, KDC & inner & \textbf{V}, \textbf{G} & \ldots & 3 \\
2 & \href{https://manga.voxastro.org/visualiser/12093-3701}{1-40771}$^{}$ & 01:11:08.3 & 14:13:22.11 & Sa &11.20 & 0.054 & 2$\sigma$?, CR-GS, Srot & unclear & \textbf{V}, \textbf{G} & \textbf{T}, \textbf{Sh} & \ldots \\
3 & \href{https://manga.voxastro.org/visualiser/8092-12705}{1-41072}$^{}$ & 01:21:36.3 & 14:27:10.34 & SABa &10.65 & 0.052 & CR-GS, Srot & unclear & \textbf{V}, \textbf{G} & \textbf{T}, \textbf{Sh} & \ldots \\
4 & \href{https://manga.voxastro.org/visualiser/12075-12705}{1-34036}$^{}$ & 01:25:29.1 & -00:03:26.30 & S0 &9.84 & 0.021 & CR-GS, Srot, 2$\sigma$? & inner & \textbf{V}, \textbf{G} & \ldots & \ldots \\
5 & \href{https://manga.voxastro.org/visualiser/7993-9101}{1-41752}$^{}$ & 02:07:48.7 & 13:36:34.54 & Sa &10.87 & 0.065 & 2$\sigma$, CR-GS & inner & \textbf{V}, \textbf{G} & \textbf{T}, \textbf{Sh} & \ldots \\
6 & \href{https://manga.voxastro.org/visualiser/8080-3702}{1-38062}$^{}$ & 03:16:54.9 & -00:02:31.23 & S0 &10.20 & 0.023 & 2$\sigma$, Srot & unclear & \textbf{V}, \textbf{G} & \ldots & \ldots \\
7 & \href{https://manga.voxastro.org/visualiser/9498-6103}{1-217842}$^{}$ & 07:50:40.8 & 23:29:39.55 & S0 &10.35 & 0.019 & CR-GS, 2$\sigma$ & inner & \textbf{Sh}, \textbf{G} & \ldots & \ldots \\
8 & \href{https://manga.voxastro.org/visualiser/10218-1902}{1-382697}$^{}$ & 07:56:49.8 & 17:22:49.09 & S0 &10.56 & 0.036 & CR-GS, 2$\sigma$? & inner & \textbf{Sh}, \textbf{G} & \textbf{E} & \ldots \\
9 & \href{https://manga.voxastro.org/visualiser/9491-6102}{1-382818}$^{}$ & 07:59:43.3 & 18:28:03.84 & Sb &9.95 & 0.038 & 2$\sigma$, CR-GS & inner & \textbf{Sh}, \textbf{P}, \textbf{G}, \textbf{O} & \textbf{E} & \ldots \\
10 & \href{https://manga.voxastro.org/visualiser/10220-1901}{1-153127}$^{}$ & 08:03:13.3 & 33:07:56.42 & Sab &9.81 & 0.018 & 2$\sigma$, CR-GS & unclear & \textbf{Cl}, \textbf{G} & \textbf{LSB} & \ldots \\
11 & \href{https://manga.voxastro.org/visualiser/9495-6101}{1-298482}$^{}$ & 08:10:12.2 & 22:46:20.22 & SABa &9.57 & 0.016 & CR-GS, 2$\sigma$? & inner & \textbf{Cl}, \textbf{P}, \textbf{G} & \ldots & \ldots \\
12 & \href{https://manga.voxastro.org/visualiser/9506-1902}{1-386154}$^{}$ & 08:54:18.0 & 26:30:22.03 & SB0a &9.93 & 0.027 & 2$\sigma$, CR-GS & inner & \textbf{Sh}, \textbf{G}, \textbf{O} & \textbf{B} & \ldots \\
13 & \href{https://manga.voxastro.org/visualiser/10512-1902}{1-77006}$^{}$ & 09:02:21.2 & 03:36:56.66 & S0 &10.03 & 0.024 & 2$\sigma$, CR-GS & unclear & \textbf{V}, \textbf{G} & \ldots & \ldots \\
14 & \href{https://manga.voxastro.org/visualiser/12488-1901}{1-386394}$^{}$ & 09:02:46.6 & 29:17:17.66 & Sa &9.94 & 0.042 & CR-GS, 2$\sigma$? & unclear & \textbf{Sh}, \textbf{P}, \textbf{G} & \ldots & \ldots \\
15 & \href{https://manga.voxastro.org/visualiser/8461-3701}{1-166613}$^{}_{\star}$ & 09:38:03.8 & 42:58:27.56 & S0a &11.07 & 0.047 & Srot & unclear & \textbf{F}, \textbf{G}, \textbf{O} & \ldots & 1,2 \\
16 & \href{https://manga.voxastro.org/visualiser/10844-6103}{1-181720}$^{}$ & 09:44:21.9 & 06:04:31.71 & Sa &10.90 & 0.062 & Srot, $\sigma$-elong. & unclear & \textbf{Cl}, \textbf{G} & \textbf{T}, \textbf{C} & \ldots \\
17 & \href{https://manga.voxastro.org/visualiser/8459-1901}{1-167079}$^{}$ & 09:51:12.4 & 44:00:33.51 & Sab &9.51 & 0.016 & 2$\sigma$ & unclear & \textbf{V}, \textbf{G}, \textbf{O} & \ldots & \ldots \\
18 & \href{https://manga.voxastro.org/visualiser/10517-9102}{1-78858}$^{}$ & 10:06:02.2 & 04:31:19.69 & S0a &10.20 & 0.023 & CR-GS, 2$\sigma$? & inner & \textbf{F}, \textbf{G} & \ldots & \ldots \\
19 & \href{https://manga.voxastro.org/visualiser/8455-1902}{1-274440}$^{}$ & 10:22:50.1 & 39:22:08.19 & S0 &9.55 & 0.026 & 2$\sigma$ & unclear & \textbf{F}, \textbf{G}, \textbf{O} & \ldots & 2 \\
20 & \href{https://manga.voxastro.org/visualiser/8254-1902}{1-255878}$^{}$ & 10:52:52.6 & 43:25:42.26 & S0 &10.02 & 0.024 & CR-GS, NRR?, 2$\sigma$?? & inner & \textbf{Cl}, \textbf{P}, \textbf{G}, \textbf{O} & \textbf{T}, \textbf{LSB} & 3 \\
21 & \href{https://manga.voxastro.org/visualiser/8255-3704}{1-277289}$^{}$ & 11:01:06.0 & 45:17:35.91 & S0a &9.86 & 0.037 & 2$\sigma$ & unclear & \textbf{Cl}, \textbf{G}, \textbf{O} & \textbf{E} & \ldots \\
22 & \href{https://manga.voxastro.org/visualiser/8945-3701}{1-279554}$^{}_{\star}$ & 11:26:30.8 & 46:59:54.36 & SB0a &10.43 & 0.025 & 2$\sigma$? & unclear & \textbf{F}, \textbf{G}, \textbf{O} & \textbf{B}, \textbf{E} & \ldots \\
23 & \href{https://manga.voxastro.org/visualiser/8990-12701}{1-174429}$^{}$ & 11:33:23.0 & 49:02:16.96 & Sc &10.11 & 0.031 & 2$\sigma$ & outer & \textbf{Cl}, \textbf{G}, \textbf{O} & \textbf{E} & \ldots \\
24 & \href{https://manga.voxastro.org/visualiser/8991-6103}{1-188177}$^{}$ & 11:44:31.1 & 53:40:57.79 & S0a &10.30 & 0.027 & CR-GS, 2$\sigma$ & inner & \textbf{F}, \textbf{P}, \textbf{G}, \textbf{O} & \ldots & 2 \\
25 & \href{https://manga.voxastro.org/visualiser/8995-1901}{1-189197}$^{}$ & 11:47:46.6 & 55:23:24.07 & S0 &10.15 & 0.026 & 2$\sigma$, NRR & unclear & \textbf{Sh}, \textbf{G}, \textbf{O} & \ldots & \ldots \\
26 & \href{https://manga.voxastro.org/visualiser/10507-3703}{1-280716}$^{}_{\star}$ & 11:49:35.2 & 47:49:52.40 & S0 &10.39 & 0.025 & Srot, 2$\sigma$ & unclear & \textbf{F}, \textbf{G} & \ldots & \ldots \\
27 & \href{https://manga.voxastro.org/visualiser/8260-3703}{1-281935}$^{}$ & 12:12:31.6 & 43:22:01.34 & Sbc &9.68 & 0.024 & 2$\sigma$??, NRR?? & unclear & \textbf{F}, \textbf{G} & \ldots & \ldots \\
28 & \href{https://manga.voxastro.org/visualiser/11956-1901}{\textbf{1-151418}}$^{\dagger}$ & 12:27:58.0 & 53:02:34.21 & Sab &10.04 & 0.030 & CR-GS, 2$\sigma$? & unclear & \textbf{F}, \textbf{G} & \textbf{LSB}, \textbf{Sh} & \ldots \\
29 & \href{https://manga.voxastro.org/visualiser/11953-3701}{1-283130}$^{}$ & 12:36:14.4 & 42:38:12.27 & S0a &9.83 & 0.028 & 2$\sigma$, Srot & unclear & \textbf{Sh}, \textbf{G} & \ldots & \ldots \\
30 & \href{https://manga.voxastro.org/visualiser/11014-3702}{1-456775}$^{}_{\star}$ & 12:56:06.4 & 27:38:52.89 & S0a &10.05 & 0.025 & 2$\sigma$? & unclear & \textbf{Cl}, \textbf{G} & \textbf{E} & \ldots \\
31 & \href{https://manga.voxastro.org/visualiser/11761-1901}{1-152010}$^{}$ & 12:57:48.8 & 52:46:45.36 & Sa &9.72 & 0.028 & CR-GS, 2$\sigma$ & inner & \textbf{F}, \textbf{G} & \ldots & 3 \\
32 & \href{https://manga.voxastro.org/visualiser/11007-6102}{1-456731}$^{}_{\star}$ & 12:58:00.8 & 27:27:14.29 & S0 &9.65 & 0.026 & Srot, KDC? & unclear & \textbf{Cl}, \textbf{G} & \ldots & \ldots \\
33 & \href{https://manga.voxastro.org/visualiser/8318-6103}{1-284335}$^{}$ & 13:07:17.0 & 45:43:41.35 & S0a &9.71 & 0.035 & 2$\sigma$, CR-GS & inner & \textbf{F}, \textbf{G}, \textbf{O} & \textbf{E} & \ldots \\
34 & \href{https://manga.voxastro.org/visualiser/8322-3703}{1-421145}$^{}_{\star}$ & 13:23:42.8 & 31:30:55.71 & S0 &10.10 & 0.018 & 2$\sigma$, Srot, KDC? & unclear & \textbf{F}, \textbf{P}, \textbf{G}, \textbf{O} & \textbf{E} & \ldots \\
35 & \href{https://manga.voxastro.org/visualiser/8311-1901}{1-523154}$^{}$ & 13:38:23.3 & 23:58:19.72 & S0a &10.20 & 0.028 & CR-GS, 2$\sigma$ & inner & \textbf{F}, \textbf{G}, \textbf{O} & \ldots & \ldots \\
36 & \href{https://manga.voxastro.org/visualiser/8446-1902}{1-418253}$^{}$ & 13:44:37.0 & 37:10:17.87 & SB0 &10.50 & 0.027 & CR-GS, $\sigma$-elong. & unclear & \textbf{F}, \textbf{G}, \textbf{O} & \textbf{B}, \textbf{T}, \textbf{Sh} & \ldots \\
37 & \href{https://manga.voxastro.org/visualiser/9882-3702}{1-537081}$^{}$ & 13:47:47.7 & 23:22:02.05 & S0a &11.03 & 0.084 & 2$\sigma$, NRR & unclear & \textbf{Cl}, \textbf{G} & \textbf{T}, \textbf{Sh} & \ldots \\
38 & \href{https://manga.voxastro.org/visualiser/8332-3704}{1-251198}$^{}$ & 13:58:33.5 & 41:43:27.06 & Sab &10.10 & 0.043 & 2$\sigma$, CR-GS & inner & \textbf{V}, \textbf{G}, \textbf{O} & \textbf{E} & 2 \\
39 & \href{https://manga.voxastro.org/visualiser/11013-1902}{1-245176}$^{}$ & 13:59:35.4 & 55:35:33.12 & Sa &10.16 & 0.041 & 2$\sigma$, NRR & unclear & \textbf{F}, \textbf{G} & \textbf{Sh} & \ldots \\
40 & \href{https://manga.voxastro.org/visualiser/8335-1901}{1-251783}$^{}_{\star}$ & 14:21:14.8 & 39:39:09.12 & E &10.18 & 0.026 & Srot, 2$\sigma$?? & unclear & \textbf{F}, \textbf{G}, \textbf{O} & \ldots & 1,2 \\
41 & \href{https://manga.voxastro.org/visualiser/8333-1902}{1-266244}$^{}$ & 14:26:11.2 & 42:42:40.47 & S0a &9.22 & 0.017 & 2$\sigma$?, Srot & unclear & \textbf{F}, \textbf{G}, \textbf{O} & \textbf{T}, \textbf{LSB} & 1,2 \\
42 & \href{https://manga.voxastro.org/visualiser/8980-3701}{1-235983}$^{}$ & 14:54:32.8 & 42:23:07.60 & S0 &9.79 & 0.018 & gas-polar, Srot, KDC? & unclear & \textbf{F}, \textbf{G}, \textbf{O} & \ldots & 1,2 \\
43 & \href{https://manga.voxastro.org/visualiser/9865-1902}{1-246484}$^{}$ & 14:55:27.9 & 51:02:51.85 & S0a &9.72 & 0.030 & gas-polar, 2$\sigma$ & unclear & \textbf{F}, \textbf{G}, \textbf{O} & \textbf{E} & \ldots \\
44 & \href{https://manga.voxastro.org/visualiser/8487-1902}{\textbf{1-134361}}$^{\dagger}$ & 16:00:53.7 & 47:11:49.26 & Sa &10.42 & 0.046 & 2$\sigma$ & outer & \textbf{Sh}, \textbf{G} & \textbf{E} & \ldots \\
45 & \href{https://manga.voxastro.org/visualiser/8979-3701}{1-248410}$^{}$ & 16:04:50.0 & 42:02:08.91 & Sa &9.59 & 0.025 & 2$\sigma$, NRR? & outer & \textbf{F}, \textbf{G}, \textbf{O} & \textbf{E} & 2 \\
46 & \href{https://manga.voxastro.org/visualiser/9048-6101}{1-318513}$^{}$ & 16:16:08.6 & 24:04:46.30 & S0a &9.84 & 0.032 & 2$\sigma$ & outer & \textbf{Cl}, \textbf{G}, \textbf{O} & \textbf{E} & 2 \\
47 & \href{https://manga.voxastro.org/visualiser/9027-3702}{1-323886}$^{}$ & 16:20:34.0 & 32:56:05.74 & Sab &10.03 & 0.041 & 2$\sigma$ & outer & \textbf{Sh}, \textbf{G}, \textbf{O} & \textbf{E} & \ldots \\
48 & \href{https://manga.voxastro.org/visualiser/9027-1902}{1-323766}$^{}$ & 16:24:42.2 & 32:03:51.74 & E &9.88 & 0.022 & CR-GS, Srot, $\sigma$-elong. & inner & \textbf{V}, \textbf{G}, \textbf{O} & \ldots & 2,3 \\
49 & \href{https://manga.voxastro.org/visualiser/11975-6104}{\textbf{1-543599}}$^{\dagger}$ & 16:41:07.3 & 26:03:39.05 & Sab &10.60 & 0.055 & 2$\sigma$? & outer & \textbf{F}, \textbf{G} & \ldots & \ldots \\
50 & \href{https://manga.voxastro.org/visualiser/7972-3704}{1-113405}$^{}$ & 21:07:21.9 & 11:03:59.11 & Sa &10.11 & 0.042 & 2$\sigma$, CR-GS & inner & \textbf{V}, \textbf{G} & \textbf{E} & \ldots \\
51 & \href{https://manga.voxastro.org/visualiser/8615-3702}{\textbf{1-635590}}$^{\dagger}_{\star}$ & 21:24:13.1 & 01:07:06.18 & S0 &11.22 & 0.049 & 2$\sigma$, NRR? & unclear & \textbf{V}, \textbf{G} & \ldots & 2 \\
52 & \href{https://manga.voxastro.org/visualiser/8617-3703}{1-180900}$^{}_{\star}$ & 21:41:28.6 & 00:36:29.67 & Edc &10.73 & 0.055 & 2$\sigma$, Srot & unclear & \textbf{V}, \textbf{G} & \ldots & \ldots \\
53 & \href{https://manga.voxastro.org/visualiser/12071-3702}{1-106664}$^{}$ & 23:09:02.9 & 00:16:01.05 & SBbc &10.16 & 0.033 & 2$\sigma$ & unclear & \textbf{V}, \textbf{G} & \textbf{B}, \textbf{E}, \textbf{T}, \textbf{Sh} & \ldots \\
54 & \href{https://manga.voxastro.org/visualiser/8655-6103}{1-29728}$^{}$ & 23:43:18.0 & 00:26:32.91 & Sab &9.87 & 0.037 & CR-GS, 2$\sigma$ & unclear & \textbf{V}, \textbf{G} & \textbf{E} & \ldots \\
55 & \href{https://manga.voxastro.org/visualiser/8655-1902}{1-29809}$^{}$ & 23:53:52.5 & -00:05:55.43 & S0a &9.56 & 0.022 & 2$\sigma$?, CR-GS, Srot? & inner & \textbf{V}, \textbf{G} & \ldots & 3 \\
\enddata
\end{deluxetable*}

\begin{figure*}
    \centering
        \includegraphics[width=0.48\textwidth]{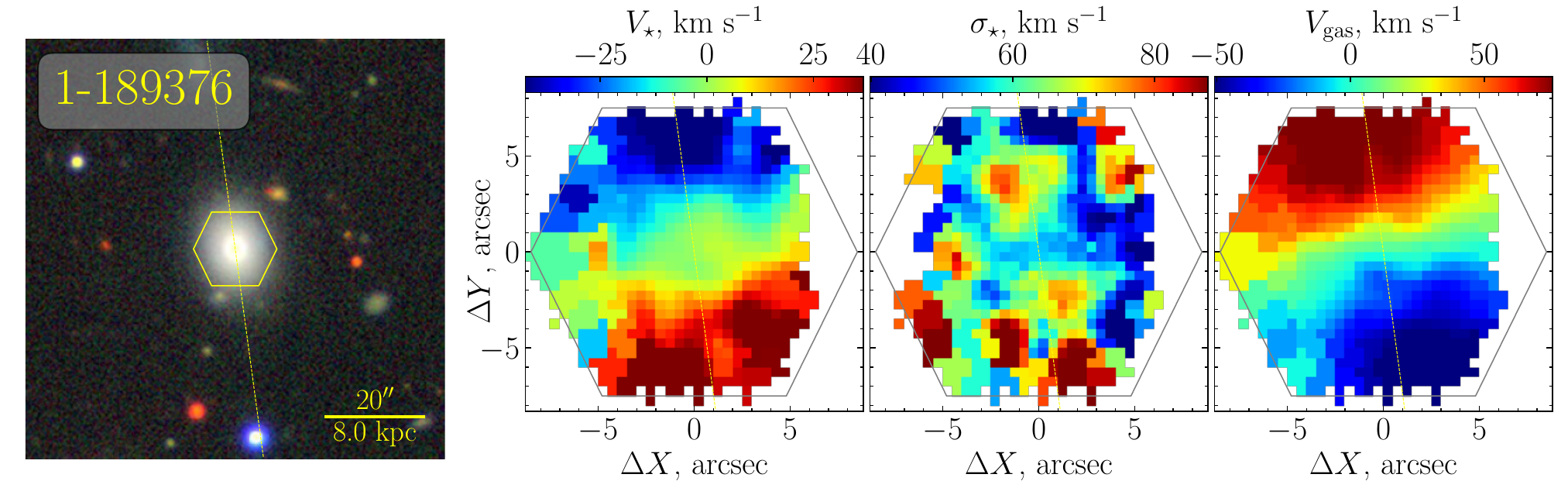}
	\includegraphics[width=0.48\textwidth]{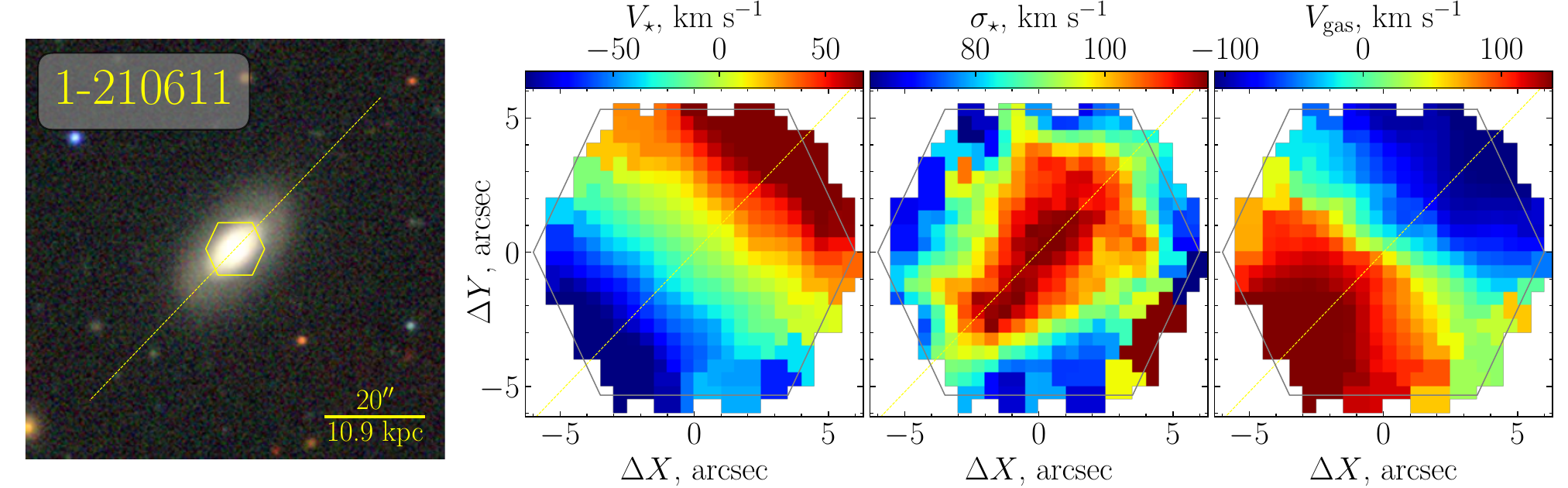}
	\includegraphics[width=0.48\textwidth]{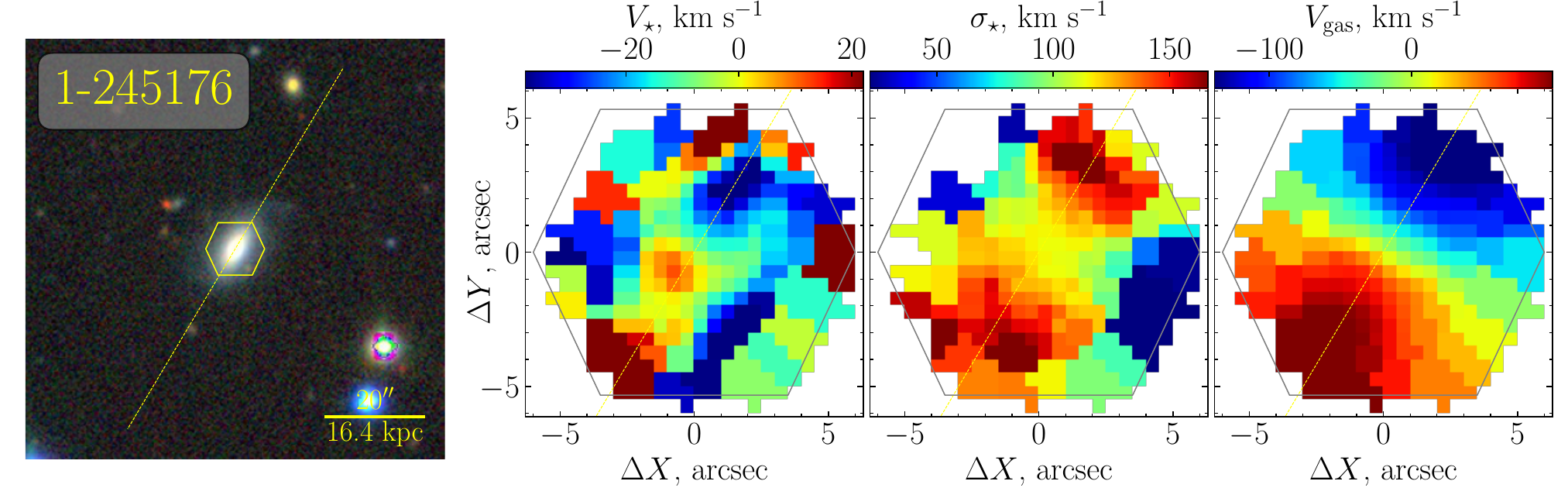}
	\includegraphics[width=0.48\textwidth]{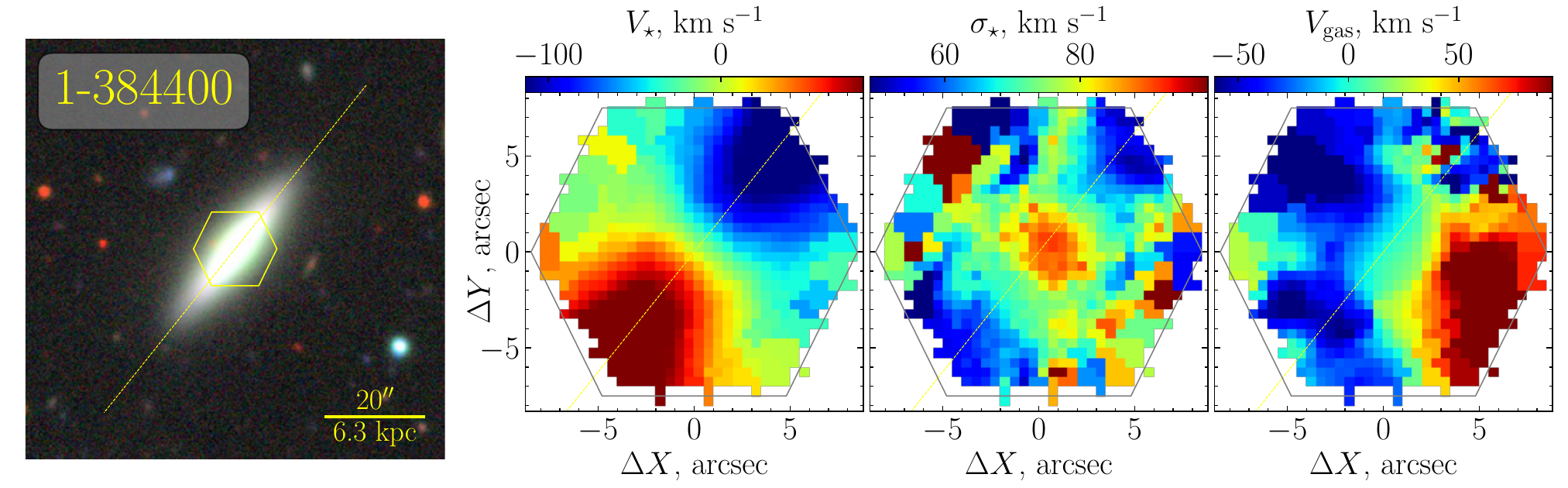}
	\includegraphics[width=0.48\textwidth]{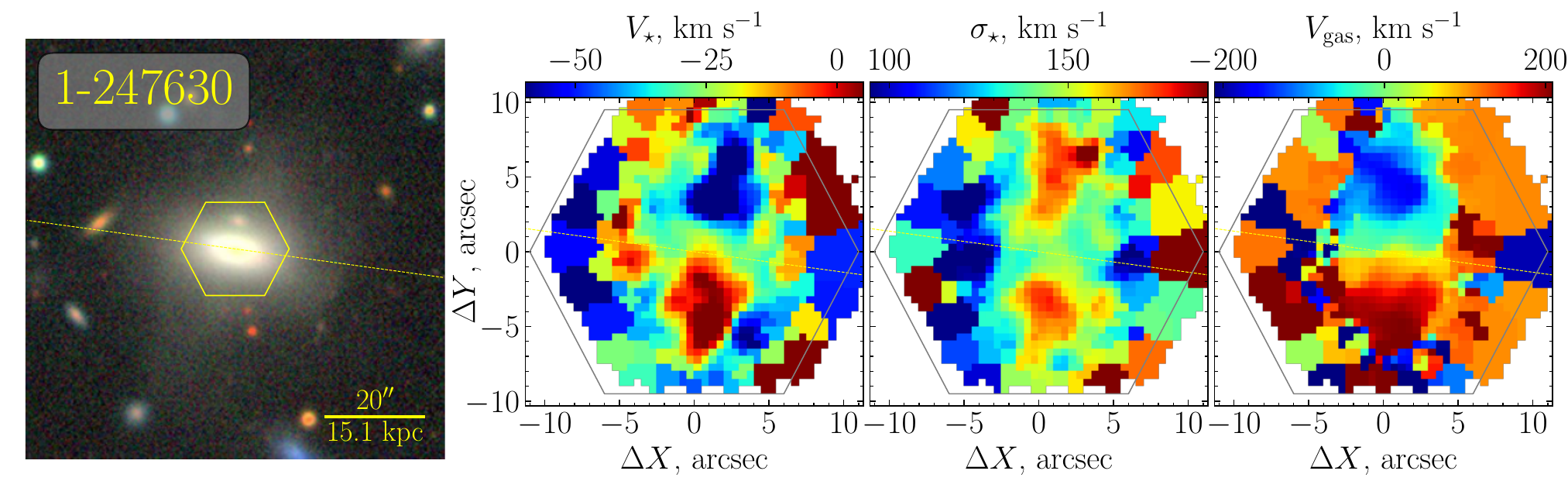}
	\includegraphics[width=0.48\textwidth]{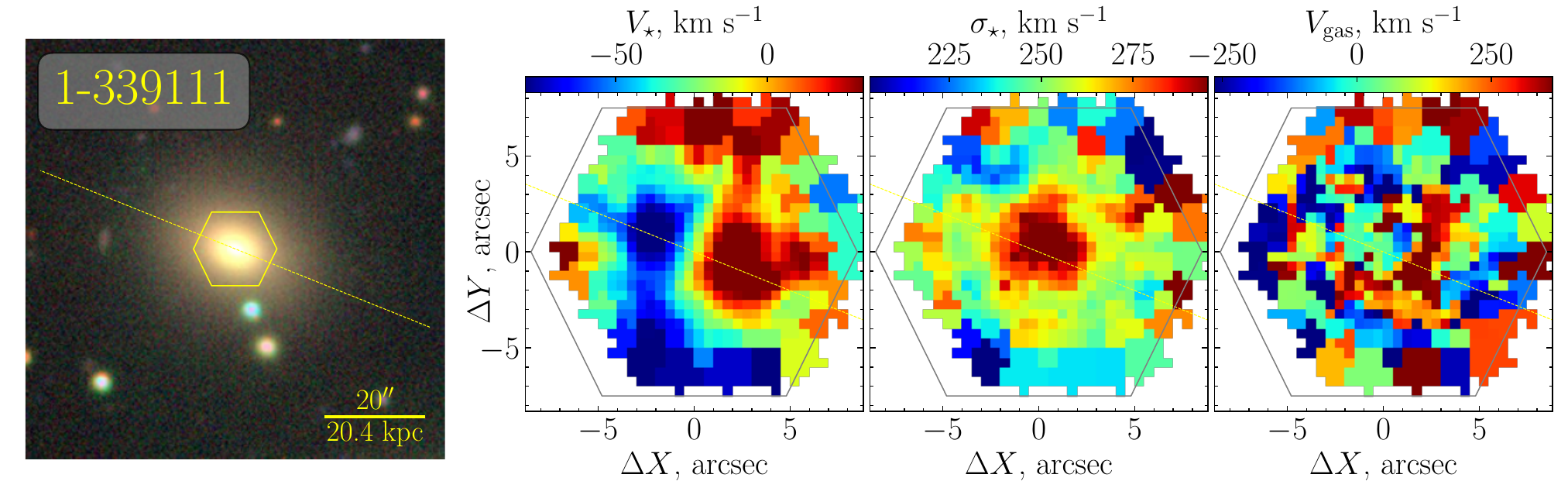}
	\includegraphics[width=0.48\textwidth]{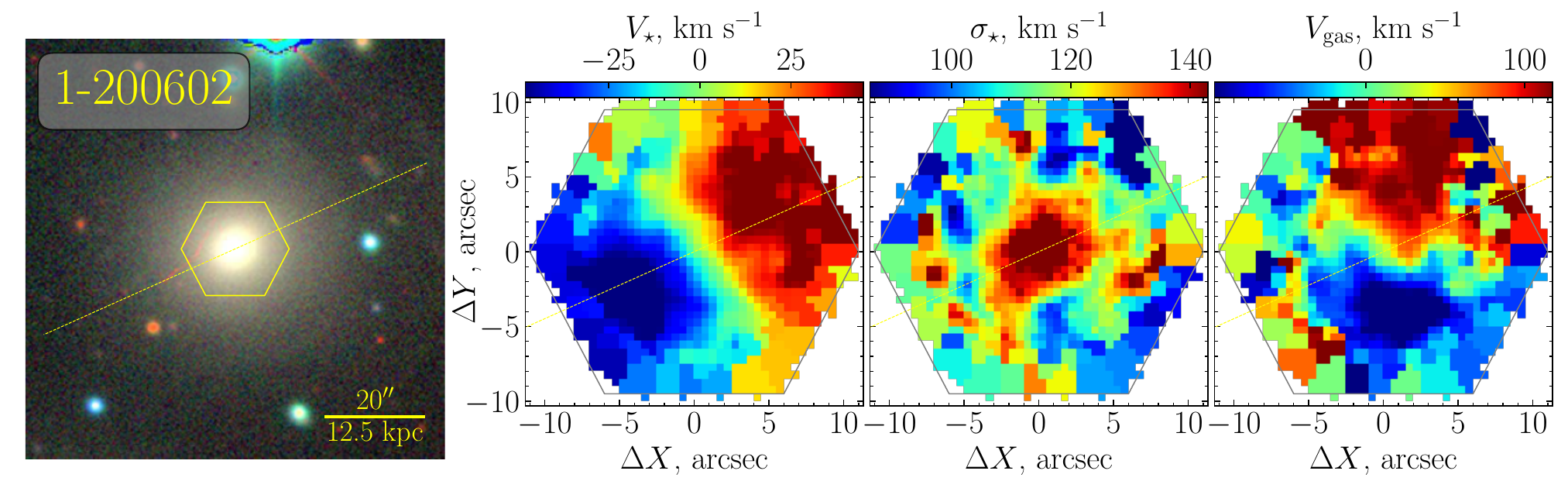}
	\includegraphics[width=0.48\textwidth]{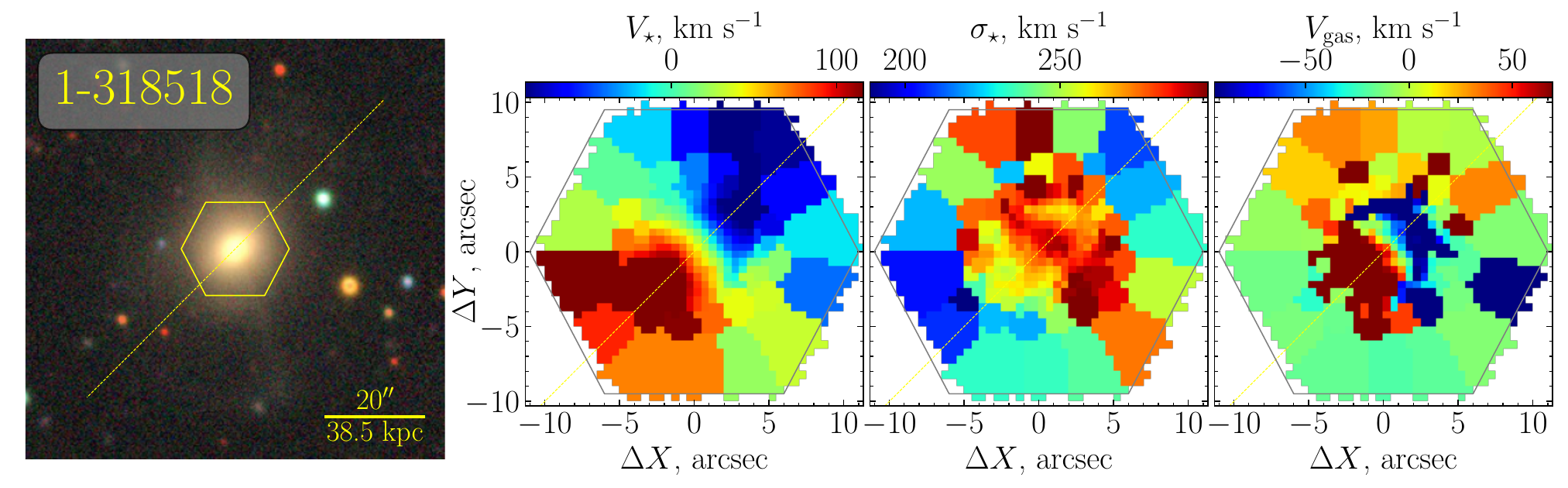}
\caption{Eight examples of different types of kinematic misalignment. Each individual panel shows a composite RGB image of the galaxy taken from the Legacy Imaging Surveys (\citet{Dey2019_legacysurveys}, \href{http://legacysurvey.org/}{legacysurvey.org}), kinematic maps are taken from MaNGA-DAP (stellar velocity $V_\star$, dispersion of stellar velocity $\sigma_\star$, and \Ha\ gas velocity $V_\mathrm{gas}$), hexagon shows MaNGA FoV, dashed line lies along the major axis (by the position angle from the NSA catalog). From left to right, from top to bottom: \textbf{1-189376} --- 2$\sigma$, Srot, CR-GS; \textbf{1-210611} --- CR-GS, $\sigma$-elong.; \textbf{1-245176} --- 2$\sigma$, NRR; \textbf{1-384400} --- gas-polar; \textbf{1-247630} --- 2$\sigma$, star-polar, gas-polar; \textbf{1-339111} --- KDC; \textbf{1-200602} --- Gas-mis; \textbf{1-318518} --- other.\label{fig:im-prob}}
\end{figure*}

To identify galaxies with CR disks, we visually inspected the Data Analysis Products (DAP) maps \citep{Westfall2019AJ....158..231W,Belfiore2019AJ....158..160B,Law2021AJ....161...52L} of the stellar velocity, gas (\Ha) velocity, and stellar velocity dispersion of all MaNGA galaxies. Galaxies possibly exhibiting kinematic misalignment features associated with CR disks were identified and attributed to one of the following types (Fig.~\ref{fig:im-prob}) by the following criteria:
\begin{itemize}
    \item \textit{2$\sigma$} (132 galaxies) --- characterized by two symmetrically positioned off-centered peaks in the stellar velocity dispersion map aligned along the major axis of the galaxy. This apparent structure results from modeling two stellar populations with distinct kinematics with a single spectral component \citep{Krajnovic2011MNRAS.414.2923K, Rubino2021A&A...654A..30R};

    \item $\sigma$-elongated \textit{$\sigma$-elong.} (11 galaxies) --- analogous to the \textit{2$\sigma$} type, but characterized by an elongated peak along the major axis observed in the stellar velocity dispersion map, formed by the merging of the two velocity dispersion peaks;
    
    \item S-rotation \textit{Srot} (58 galaxies) --- characterized by a change in the direction of stellar rotation between the inner and outer regions of the galaxy. So named because the stellar velocity profile along the major axis has a ``S'' shape;

    \item Non Regular (irregular) Rotation \textit{NRR} (24 galaxies) --- disk galaxies with stellar velocity maps that do not indicate regular rotation;

    \item Gas / Star Counter Rotation \textit{CR-GS} (207 galaxies) --- galaxies in which the gas and stars have different rotation patterns, as indicated by comparing their velocity maps. For galaxies with a prominent S-shaped rotation, the gas velocity map is compared to the outer regions of the stellar velocity map;

    \item \textit{Gas-polar} (119 galaxies) --- characterized by the major kinematic axis of the stellar velocity field being perpendicular to that of the gas velocity field;

    \item Gas misalignment \textit{Gas-mis} (173 galaxies) --- where the kinematic major axes of stellar and gas velocity fields are misaligned but neither perpendicular nor counter-rotating;
    
    \item \textit{Star-polar} (2 galaxies) --- the rare case where the stellar velocity map reveals a polar structure, typically associated with a morphological feature in the polar direction relative to the main disk, as detected in DECaLS images;
    
    \item Kinematically Decoupled Core \textit{KDC} (8 galaxies) --- galaxies with a physically small, kinematically distinct region in the central part of the galaxy \citep{Efstathiou1982MNRAS.201..975E};  and 
    
    \item \textit{Other} (7 galaxies) --- galaxies that exhibit irregular kinematics that does not fit into the above categories.
\end{itemize}

All the types mentioned above were initially identified through visual inspection without supplementary validation via non-parametric LOSVD analysis (Sec.~\ref{sec:np_losvd}).
Therefore, this preliminary classification may not be perfect and minor variations in types may be present.
In total, we identified 588 unique galaxies exhibiting some form of kinematical misalignment, including some galaxies that could fit in multiple categories.
We omitted clear cases of violent ongoing mergers involving decoupled stellar and gaseous subsystems in an unrelaxed state.
While these systems could be studied with our methods \citep{MazzilliCiraulo2021A&A...653A..47M}, they are beyond the scope of this work.

Our primary goal during the inspection was to identify galaxies with stellar CR.
Therefore, we focused on the galaxies exhibiting 2$\sigma$/$\sigma$-elongated, Srot, and CR-GS (a subsample of 294 unique galaxies).
Among these, only 115 galaxies exhibited prominent kinematic signatures in their kinematic maps indicative of stellar CR components.
This subset constitutes the initial and primary sample for further analysis.
However, clear identification of stellar CR was not always possible due to observational limitations such as the limited field of view of MaNGA observations, occasionally poor signal-to-noise ratio (SNR),  effective spatial resolution of $\approx$2.5~arcsec, and additional complications from projection effects, dust lanes, or overlapping galactic components. 
In such cases, where the kinematic features were obscured, we flagged the corresponding feature types with a ``?'' and classified these galaxies as probable.
Based on this visual classification, the initial sample of 115 galaxies was divided into two categories: the 49 galaxies with strong and unambiguous kinematic signatures of stellar CR were classified as \textit{reliable}, while the remaining 66 galaxies with less certain features were considered as \textit{probable} CR galaxies.
During the inspection process, we extensively used our web-based interactive service for the visualization of MaNGA data (\url{https://manga.voxastro.org}\footnote{\url{https://manga.voxastro.org} is our first prototype service for the visualization of IFU data. \url{https://ifu.voxastro.org} \citep{Katkov2024ASPC..535..239K} is the next generation of the service in the early stage of development which includes data from other spectroscopic surveys (SAMI, Califa, Atlas3D).}).

We then carried out a non-parametric analysis of the stellar LOSVD (see details in Sec.~\ref{sec:np_losvd}) for the complete sample of kinematically misaligned galaxies (588 galaxies).
As expected, most of the reliable CR galaxies exhibit clear X-shaped structures in their position-velocity diagrams (see the beginning of Sec.~\ref{sec:np_losvd}), suggesting two counter-rotating stellar disks (35 galaxies).
Furthermore, we identified clear X-shaped structures in 15 probable CR candidates, which were subsequently reclassified into the reliable category (bold IDs in Tab.~\ref{Tab:Rel_samp}).
Moreover, we found five additional galaxies within the total sample of kinematically misaligned galaxies that did not show obvious features in the kinematic maps but exhibited possible X-shaped LOSVDs.
Of these, we added four to the probable CR category and reclassified one galaxy with the most noticeable X-shaped LOSVD (1-37478) into the reliable category.

\begin{figure}
    \centering
    \includegraphics[width=0.49\textwidth]{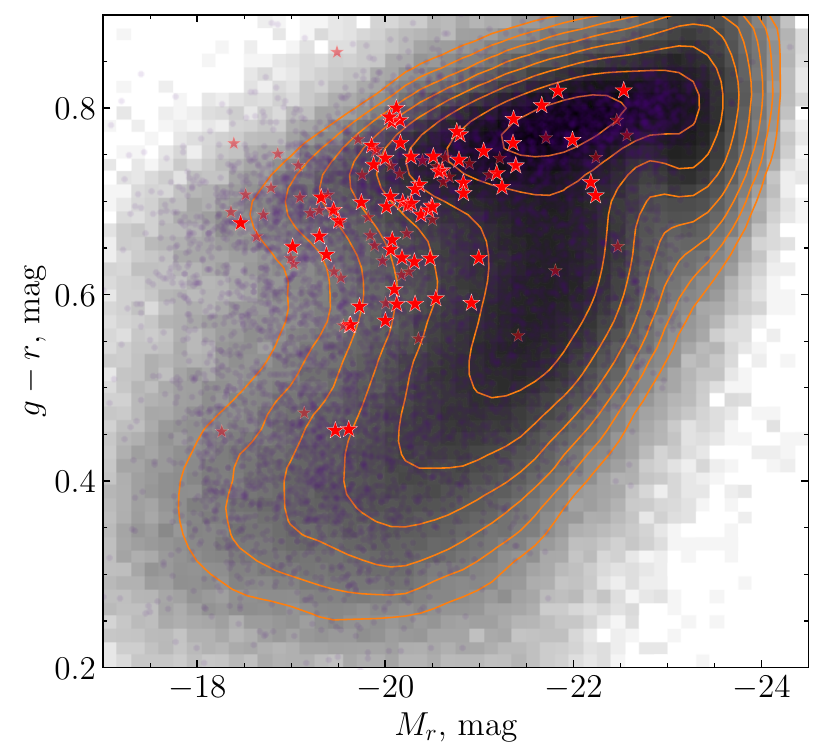}
	\includegraphics[width=0.49\textwidth]{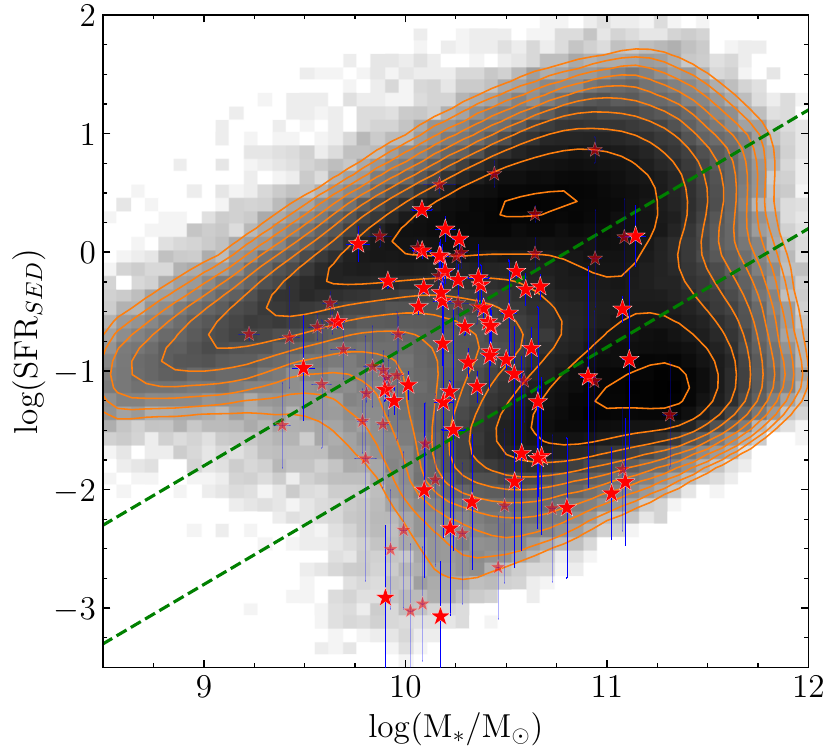}
\caption{\textbf{Top:} Color-magnitude diagram for CR sample overlaid on a background computed from the RCSED catalog \citep[\url{rcsed.sai.msu.ru},][]{Chilingarian2017ApJS..228...14C}.
All MaNGA galaxies were cross-matched with the RCSED and are plotted as transparent purple dots.
\textbf{Bottom:} M$_\star$--SFR diagram with the CR galaxies.
GSWLC catalog \citep{Salim2016ApJS..227....2S} was used to construct the grey background.
The top and bottom green dashed lines correspond to specific star formation rates (sSFR) of $\log \textrm{sSFR} = -10.8$ yr$^{-1}$ and $\log \textrm{sSFR} = -11.8$ yr$^{-1}$, respectively.
For both panels, the bright red stars denote the reliable CR sample (Tab.~\ref{Tab:Rel_samp}), while small red stars indicate the probable sample (Tab.~\ref{Tab:Prob_samp}).
\label{fig:cmd_M_SFR}}
\end{figure}

\begin{figure*}
\centering
\includegraphics[width=\textwidth]{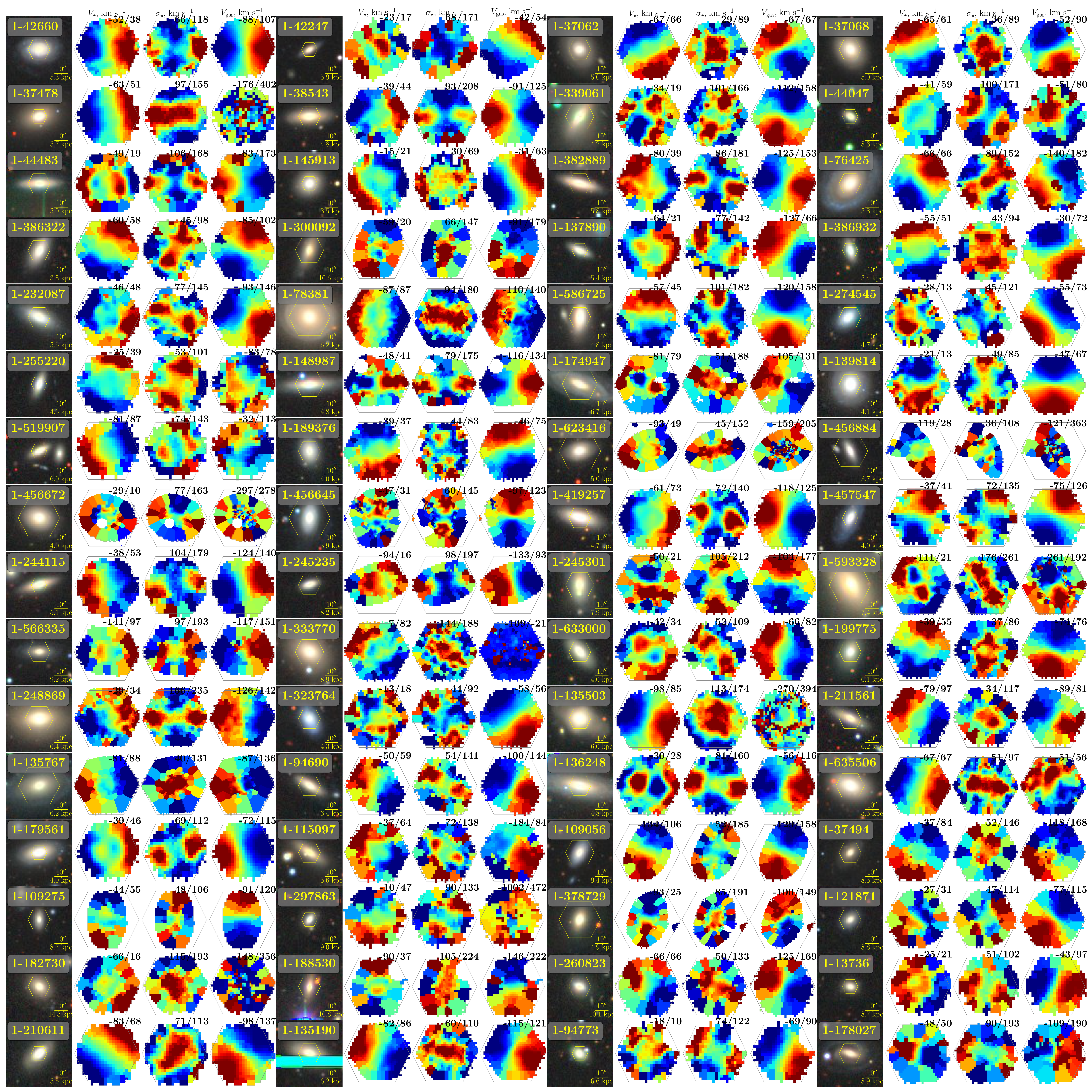}
\caption{
Sample of galaxies with reliable CR.
The panel description is similar to Fig.~\ref{fig:im-prob}. 
Numbers in the upper right corner of the maps indicate the minimum (blue) and maximum (red) values of the corresponding parameter. \label{fig:im_kin_maps_rel}}
\end{figure*}

\begin{figure*}
\centering
\includegraphics[width=\textwidth]{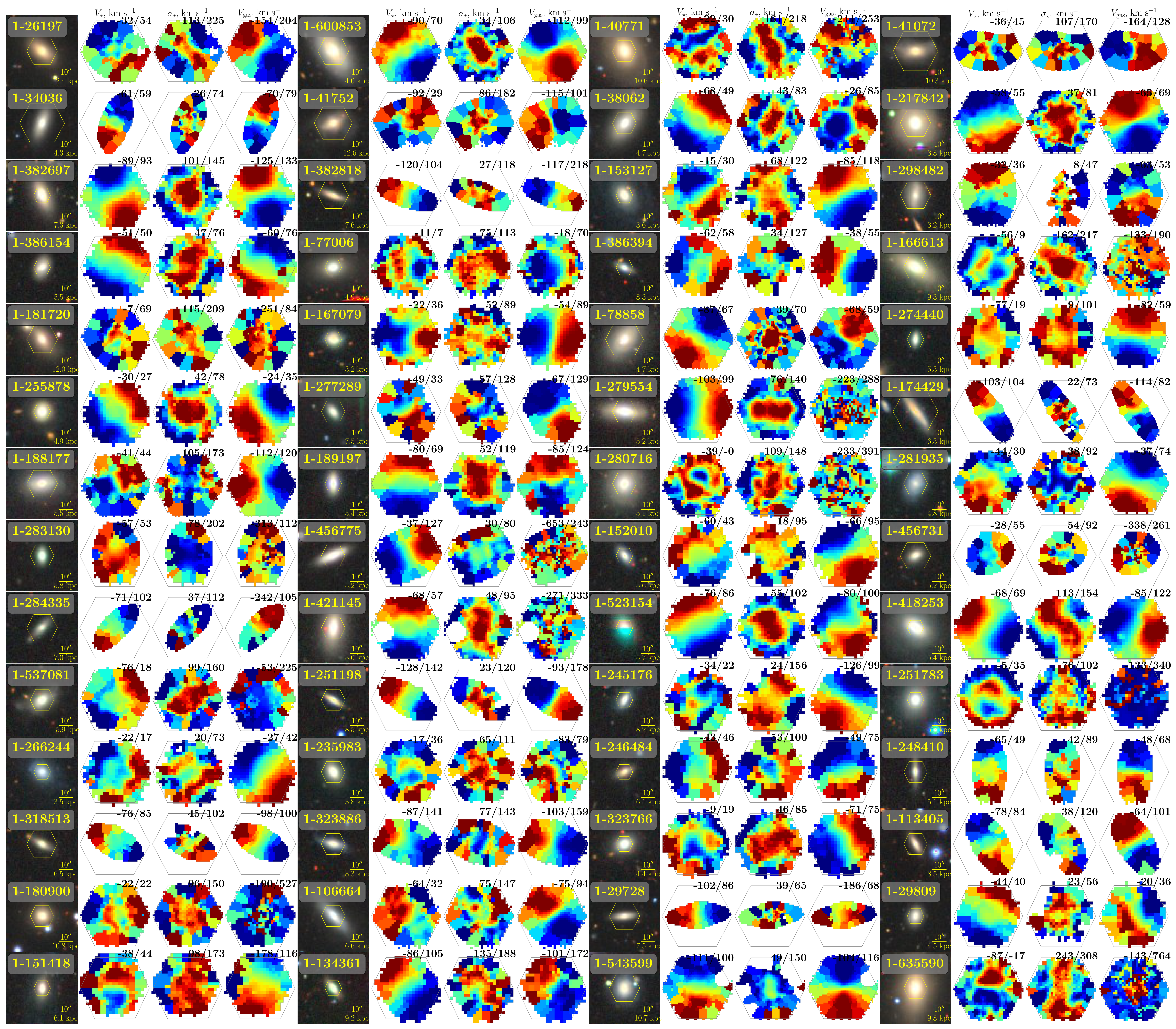}
\caption{Same as Fig.~\ref{fig:im_kin_maps_rel}, but for the probable sample of CR galaxies.\label{fig:im_kin_maps_prob}}
\end{figure*}

In summary, our final sample consists of 120 stellar CR galaxies, including 65 reliable and 55 probable cases. 
As noted above, the reliable class includes galaxies with strong kinematic features in the kinematic maps, typically reinforced by an X-shaped LOSVD. The non-parametric LOSVD analysis provides additional confirmation for most -- but not all -- reliable cases, as it is more sensitive to SNR and spectral resolution.
The main properties of these galaxies are presented in Tab.~\ref{Tab:Rel_samp} and \ref{Tab:Prob_samp}.
Fig.~\ref{fig:cmd_M_SFR} shows the distributions of these galaxies across the Color-Magnitude Diagram (CMD) and star formation rate (SFR) --- stellar mass diagram.
Fig.~\ref{fig:im_kin_maps_rel} and \ref{fig:im_kin_maps_prob} present color images and kinematic maps from MaNGA-DAP for our sample of CR galaxies.

\begin{deluxetable*}{ccccccccc}
\tablecaption{
Fraction of galaxies with stellar counter-rotation (``CR'') and with any type of kinematic misalignment (``All'') across different morphological types.
The ``MaNGA'' columns show raw fractions with respect to the total number of galaxies of each morphological type in the MaNGA survey.
The ``Corr.’’ columns give the corresponding volume-corrected fractions as derived in the MVM-VAC (V\'azquez-Mata et al. submitted).
The numbers in parentheses indicate the number of galaxies with corresponding morphology passed T-type and classification quality filters (targets with ``Unsure’’ classification were filtered out).
Note that lenticular S0/S0a galaxies form a subclass of early-type galaxies (ETGs), which also include E and Edc types (ellipticals with embedded discs).  
Galaxies with irregular or ``no classification'' morphology were excluded.
\label{Tab:frac_gals}
}
\tablehead{
     & \multicolumn{2}{c}{S0/S0a (1797)}    & \multicolumn{2}{c}{ETG (3576)} & \multicolumn{2}{c}{LTG (5720)} & \multicolumn{2}{c}{Total (9296)} \\
     & MaNGA & Corr. & MaNGA & Corr. & MaNGA & Corr. & MaNGA & Corr. 
}
\startdata
CR (120)   &  3.1\% (56) &  2.9\%  &  1.7\% (61)  &  1.5\% & 0.8\% (48)  &  0.8\% & 1.2\% (109) & 1.0\% \\
All (588) &  12.9\% (231) &  11.9\%  &  10.6\% (379)  &  7.9\% & 3.1\% (178)  &  3.0\%  & 6.0\% (557) & 4.9\% \\
\enddata
\end{deluxetable*}

The MVM-VAC includes 3576 early-type galaxies (ETGs; classified as E, S0, and S0a) and 5720 late-type galaxies (LTGs; Sa--Sm).
Among the final sample of 120 CR galaxies, 61 are classified as ETGs, accounting for $\approx$1.7\% of all ETGs in the MaNGA survey and $\approx$3.1\% of S0/S0a galaxies (a morphological subclass within ETGs).
Another 48 CR galaxies are LTGs, representing $\approx$ 0.8\% of all LTGs.
The relatively high number of LTGs with stellar CR is not surprising, as many of them are Sa/Sab types and may be misclassified and share properties with ETGs.
To calculate these statistics, we filtered the galaxy sample using the T-type classification and morphological quality flags (\textit{unsure}) provided in MVM-VAC, excluding uncertain or irregular systems (T-type code over 10).  
This explains why the total numbers of galaxies in each morphological bin (shown in parentheses in Tab.~\ref{Tab:frac_gals}) do not exactly match the raw totals of 120 CR galaxies or 588 galaxies in the total sample of galaxies with kinematic misalignment.
Considering this total sample (588 galaxies), there are 379 ETGs ($\approx$ 10.6\%, $\approx$ 12.9\% of only lenticular S0/S0a galaxies), and 178 LTGs ($\approx$ 3.1\%, primarily Sa/Sab/Sb types).
These fractions, along with those adjusted for volume correction following V\'azquez-Mata et al. (submitted, priv. comm.) are listed in Tab.~\ref{Tab:frac_gals}.
Our measured fraction of CR galaxies among ETGs ($\approx 1.5\%$) is consistent with previous studies, the estimate of $(2.3 \pm 0.3)$ by \cite{Bevacqua2022MNRAS.511..139B} and $(1.5 \pm 0.8)$ by \cite{Krajnovic2011MNRAS.414.2923K}. 
Our measured fraction of galaxies with kinematic misalignments (7.9\% among ETGs and 4.9\% of the total galaxy population) is in general agreement with the estimate by \citep{Krajnovic2011MNRAS.414.2923K}: about $9 \%$ and $< 3\%$, respectively.

Most galaxies in our sample are disky, meaning that both the pre-existing main component and the counter-rotating one are co-planar stellar disks. However, there are a few unusual cases where the main stellar body may be a rotating spheroid (1-248869 or with Edc photometric classification, ellipticals with embedded discs), or where the two components appear to be non-co-planar (1-378729). We retain these systems in the sample, as the formation of the secondary (CR) component likely follows the same external accretion scenario. Given their small number, they do not significantly affect our overall conclusions.
Several galaxies in the CR sample exhibit both stellar CR disks and bars.
Some of these (1-178027, 1-245301, 1-298482, and 1-94773) have close companions that may induce bar-forming instabilities \citep{2022MNRAS.514.1006I}.
In other cases (1-174947, 1-386154, 1-633000, and 1-635506), the mechanism responsible for bar formation remains unclear.

\section{Analysis}
\label{sec:analysis}
We examined the MaNGA spectra of all identified CR galaxies using a workflow similar to the one we previously used for detailed investigations of CR galaxies IC~719 \citep{Katkov2013ApJ...769..105K}, NGC~448 \citep{Katkov2016MNRAS.461.2068K}, and PGC~66551 \citep{Katkov2024ApJ...962...27K}.
Hereafter, we present the details of our analysis.

\subsection{Single-component full-spectrum fitting}
\label{sec:1comp_nb}

The first step in analyzing the spectra of CR galaxies is single-component full-spectrum fitting using simple stellar population (SSP) models.
We used the full spectral fitting package \nb\ \citep{nburst_a, nburst_b}, which models the entire galaxy spectrum by effectively extracting the information encoded within an absorption-line spectrum. We used a grid of synthetic SSP models \textsc{E-MILES} \citep{Vazdekis2016MNRAS.463.3409V}. 
The model spectrum was broadened by convolving it with the Gaussian stellar LOSVD.
To account for discrepancies between the model spectrum and the observed spectrum due to imperfect calibration and dust extinction, we applied a multiplicative polynomial continuum represented by Legendre polynomials of the 19th degree.
The model also includes a set of strong emission lines (\Hg, \Hb, [O\iii], [O\oi], [N\ii], \Ha, [S\ii]) represented by a common parametric Gaussian LOSVD profile independent of stellar kinematics. These emission lines are treated as additive components, and their fluxes are determined by solving a constrained linear problem at each step of a non-linear minimization loop. Both emission line templates and SSP models were pre-convolved with the wavelength-dependent instrumental resolution prior to the main minimization process.

We used these results to construct maps of the gas and stellar kinematics that include velocity and velocity dispersion. Additionally, we built maps of SSP-equivalent estimates of stellar population ages and metallicities, along with emission line flux maps for all galaxies in the sample that exhibit kinematic misalignment. To ensure consistency, we compared the derived kinematic maps with those provided in the DAP and confirmed the presence of kinematic misalignment in all selected galaxies.

\subsection{Non-parametric stellar LOSVD analysis}
\label{sec:np_losvd}

\begin{figure*}
\centering
\includegraphics[width=0.84\textwidth]{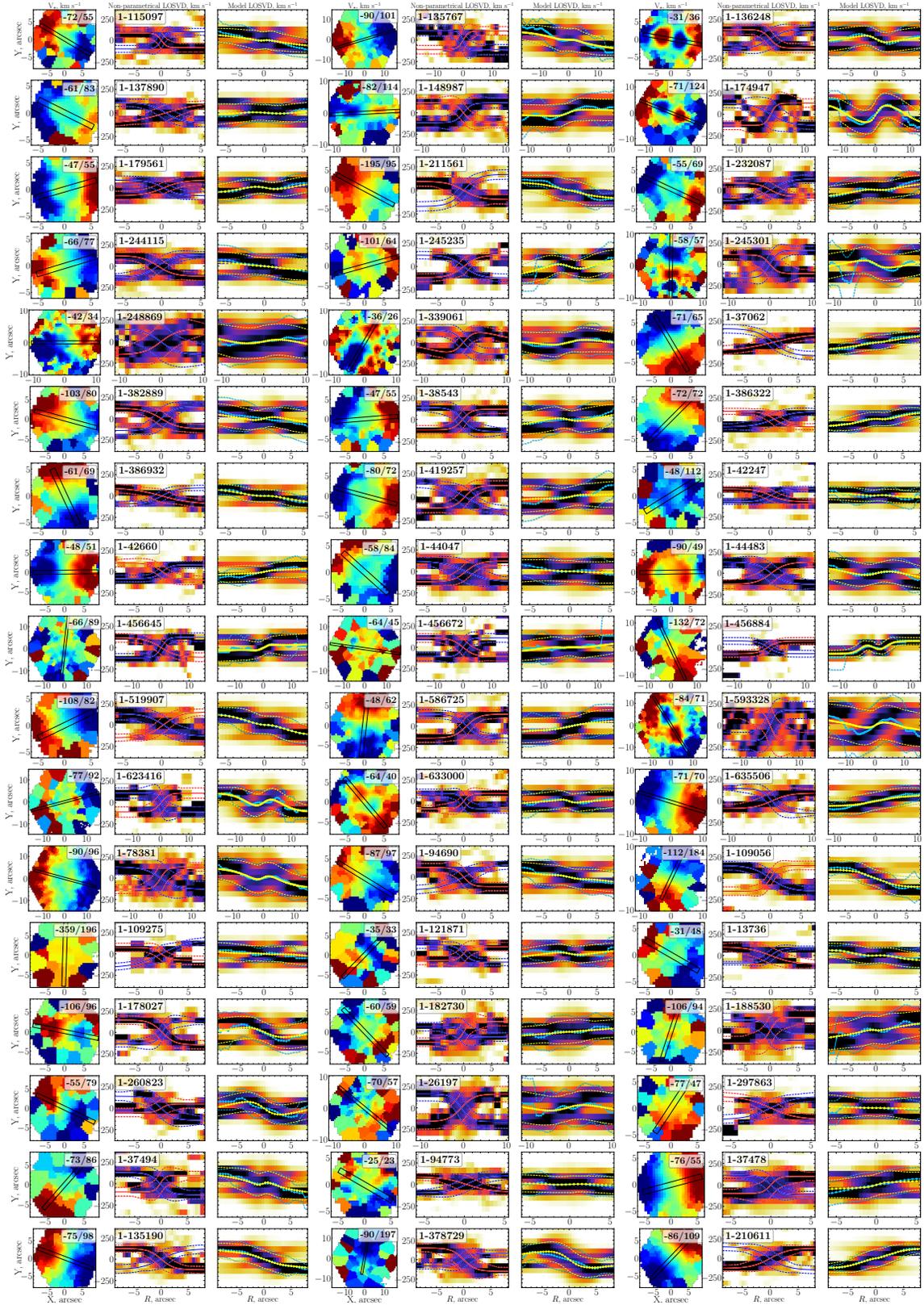}
\caption{
Subset of CR galaxies with distinct X-shaped structures in the non-parametrically recovered stellar LOSVD.
The left panel displays the stellar velocity field derived from MaNGA spectra using a single-component \nb\ analysis, with the pseudo-slit oriented along the major axis.
The central panel presents the LOSVD extracted along the pseudo-slit. The solid red and blue lines represent the modeled rotation velocities $v_i$ (Eq.~\ref{eq:v_prof}), of the main and CR disks, respectively. The dashed lines above and below each $v_i$ profiles indicate $v_i \pm \sigma_i$, where $\sigma_i$ is the velocity dispersion of the corresponding disk.
A negative radius corresponds to the left side of the map (the region with negative X-coordinates).
The right panel displays a 2D representation of the LOSVD model.
Cyan lines indicate $v_i$ and $v_i \pm \sigma_i$ from the single-component \nb\ analysis of the MaNGA spectrum, while the yellow lines are derived from the LOSVD model.\label{fig:np_losvd}}
\end{figure*}

The LOSVD of a galaxy represents a luminosity-weighted sum of the kinematics of all its stellar components. In a case of counter-rotating stellar disk, this results in two oppositely directed rotation curves, each broadened by stellar velocity dispersion. When visualized, such a LOSVD often exhibits a characteristic ``X'' shape (Fig.~\ref{fig:np_losvd} and~\ref{fig:full_analysis_example}), which we refer to as the X-shaped LOSVD. To reveal this feature, we applied the non-parametric LOSVD recovery approach successfully used in our previous studies \citep[e.g.,][]{Katkov2016MNRAS.461.2068K,Kasparova2020MNRAS.493.5464K,Katkov2024ApJ...962...27K} which is described in detail by \citet{Gasymov2024ASPC..535..279G}.
The properties of these X-shape LOSVDs serve as a reliable initial estimate for the kinematical parameters required for spectral decomposition.

The stellar spectrum of a galaxy, obtained as the difference between the observed and the emission line spectra, logarithmically binned in wavelength, can be modeled as the convolution of a high-resolution stellar population template with the stellar LOSVD.
In our approach, we treat the deconvolution problem as an ill-posed linear problem whose solution is the desired LOSVD.
We solve this problem using a linear least-squares method with regularization, as the solution is sensitive to noise in the spectra. The objective function to be minimized is: 
\begin{equation}
\left\lVert Y - A \cdot \mathfrak{L} \right\rVert + \lambda \left\lVert R \cdot \mathfrak{L} \right\rVert \mathop{\longrightarrow}\limits_{\mathfrak{L}} \min,
\end{equation}
where $\mathfrak{L}$ is the vector of the LOSVD (i.e., the solution), $Y$ is the vector of the observed spectrum (i.e., the data), $A$ is the design matrix composed of velocity-shifted unbroadened stellar population spectra, $R$ is the regularization matrix, and $\lambda$ is the regularization strength parameter that controls the contribution of the regularization term to the objection function. Different choices of R influence different aspects of the solution shape. We use L2 (Tikhonov, or ridge) regularization \citep{tikhonov1977solutions} where $R = I$, which penalizes the overall amplitude of the solution and effectively enforces smoothness. Additionally, we apply ``edge'' regularization where $R$ is constructed to suppress non-physical high-velocity wings in the solution by penalizing large deviations from zero velocity.

We used regularization with different regularization coefficient settings: ``weak'' ($\lambda \sim 0.01$), ``medium'' ($\lambda \sim 0.2-0.3$), and ``strong'' ($\lambda \sim 4-5$), which encompass models with varying degrees of the influence of the regularization on the solution.
As a template, we used the unbroadened best-fit model of stellar population from the single-component fitting described in Sec.~\ref{sec:1comp_nb}.

Instead of directly computing the full LOSVD, we reformulate the problem by expressing the solution as the sum of a predefined, peak-like LOSVD component and a minor correction vector.
We then perform a linear least-squares fit to determine the correction vector, applying regularization to ensure its smoothness.
In our iterative procedure, we start with an initial Gaussian LOSVD (obtained from single-component fitting) and, in subsequent iterations, use the full solution from the previous step as the fixed, predefined component.

Our method is provided as a Python package \textsc{sla}\footnote{\url{https://pypi.org/project/sla/}, \url{https://github.com/gasymovdf/sla}}.
Additionally, a similar approach based on the Bayesian method for non-parametric stellar LOSVD recovery, \textsc{bayes-losvd} by \citet{Falcon-Barroso2021A&A...646A..31F}, is also publicly accessible.

Following this analysis, we then determined the LOSVD for each spatial bin, yielding a three-dimensional model of the galaxy’s stellar kinematics (X-Y-V) from the IFU spectral cube.
To facilitate subsequent analyses, we extracted a position-velocity (PV) diagram along the galaxy’s major axis in a 1~arcsec pseudo-slit using \textsc{pvextractor}\footnote{\url{https://pvextractor.readthedocs.io/}} \citep{Ginsburg2016ascl.soft08010G}.
We then modeled the LOSVD within the position (radius)-velocity coordinate frame.

Applying this analysis to the entire sample of 588 galaxies with kinematical misalignment, we identified 51 galaxies exhibiting a clear X-shaped structure in their LOSVDs (Fig.~\ref{fig:np_losvd}), and four galaxies with unclear LOSVDs. 
One galaxy (1-37478) shows no obvious CR features in its kinematic maps (first column in the second row of Fig.~\ref{fig:im_kin_maps_rel}), but it displays a well-defined X-shape in the recovered LOSVD (right column in the penultimate row of Fig.~\ref{fig:np_losvd}). 
Additionally, 15 galaxies were upranked from the ``probable'' to the ``reliable'' sample status based on their recovered X-shape LOSVDs. 
Finally, we identified four galaxies (1-151418, 1-134361, 1-543599, and 1-635590) whose LOSVDs are noisy but resemble an X-shape without strong symmetry. These galaxies were assigned to the ``probable'' sample with a note on possible CR disk presence.

Following this inspection, 51 out of the 120 galaxies in our CR sample exhibited clear X-shaped recovered LOSVDs within the MaNGA Field-of-View (the central panels of Fig.~\ref{fig:np_losvd}).
These galaxies were chosen for further analysis of the recovered LOSVD to obtain kinematics and relative luminosity weights of both disks. We assumed that at any given position $R$ the LOSVD could be parameterized by two Gaussians.
These Gaussians are characterized by velocity dispersion ($\sigma$), intensity ($I$), and the velocity ($v$), which depend on the radial position $R$.
Velocity $v$ and velocity dispersion $\sigma$ profiles are parametrized as follows:
\begin{equation}
    v(R)  =  v_\mathrm{max} \left( \tanh(\pi |R|/R_0) + c |R|/R_0\right),
    \label{eq:v_prof}
\end{equation}
where $v_\mathrm{max}$ is the maximum rotation velocity, $R_0$ is the scale radius at which the velocity curve flattens, and $c$ is a linear slope term that modifies the shape of the outer velocity profile (typically $c=0$ for a flat plateau) proposed by \cite{Chung2021ApJS..257...66C}.
\begin{equation}
    \sigma(R) = \max\{\sigma_0,~\sigma_1 - k |R|\},
\end{equation}
where $\sigma_1$ is the central velocity dispersion, $\sigma_0$ is the minimum (``disk'') dispersion at large radii, and $k$ is the radial gradient controlling the decline in dispersion between center and outer regions ($k>0$).
The fluxes of both disks were described by the standard disk exponential law \citep{Freeman1970ApJ...160..811F}:

\begin{equation}
    \label{eq:exp_prof}
    I_i(R) = I_{0,i} \exp( -|R|/h_i).
\end{equation}
By definition, the LOSVD is normalized to unity within each radial bin $R$.
Therefore, we encountered a challenge due to degeneracy between the disk scale lengths $h_i$ and their central intensities $I_{0,i}$.
To break this degeneracy, we simplified the analysis to fit only two parameters: the relative intensity $I_0 = I_{02} / I_{01}$ and the relative scale $h_0 = \frac{h_1 h_2}{h_1 - h_2}$. Utilizing these parameters, we computed the relative weight profiles within each radial bin of the LOSVD:
\begin{equation}
    W_1(R) = \frac{I_1(R)}{I_1(R) + I_2(R)} = \frac{1}{I_0 \exp(-|R|/h_0) + 1},
\end{equation}
\begin{equation}
    W_2 (R) = 1 - W_1(R).
\end{equation}

However, using only these two parameters does not allow us to determine the total weight of the counter-rotating disk $W_{\textrm{CR}}$.
To compute this parameter, we introduced the total weight of the disk as defined by the following equation:
\begin{equation}
    W_{\textrm{CR}} = \frac{\int I_\textrm{CR} R dR}{\int I_\textrm{CR} R dR + \int I_\textrm{m} R dR}  = \frac{1}{1 + I_0 (h_{\textrm{CR}} / h_{\textrm{m}})^2}.
\end{equation}

This formulation implies that we need to determine the relative intensity $I_0$ and the scale $h_0$, along with at least one physical scale, either $h_{\textrm{CR}}$ or $h_{\textrm{m}}$.
We multiplied the model profiles $W_i(R)$ by the flux measured along the slice in the SDSS $g$-band MaNGA intensity map and fitted the resulting profiles with exponential laws $I_i(R)$ (Eq.~\ref{eq:exp_prof}) using the $I_0$ and $h_0$ parameters obtained from kinematic analysis.
Finally, we recalculated the $I_{0,i}$ and $h_i$ parameters to derive $W_{\textrm{CR}}$. These steps are illustrated in Fig.~\ref{fig:full_analysis_example}.

One limitation of our LOSVD reconstruction method is its reliance on a single stellar template to recover the complex LOSVD shape formed by two distinct stellar disks.
The analysis assumes that both stellar components have similar stellar population properties, allowing them to be accurately represented by the same template.
However, if this assumption does not hold, it may introduce distortions in the recovered LOSVD.
We estimate that our analysis has a detection sensitivity threshold of approximately 10\% in terms of luminosity weight, consistent with the detection limits reported by \citet{Kuijken1996MNRAS.283..543K, Pizzella2004A&A...424..447P, Rubino2021A&A...654A..30R}.

\subsection{Two component full-spectrum fitting}
\label{sec:2comp_nb}

LOSVD analysis provided accurate estimates of the kinematics and weights for both disks, which we used as priors for a two-component SSP fitting using the \nb\ technique.
This method, known as spectral decomposition \citep{Chilingarian2011MNRAS.412.1627C, Coccato2011MNRAS.412L.113C, Johnston2013MNRAS.428.1296J, Katkov2013ApJ...769..105K}, was previously applied for several CR galaxies shown in Tab.~\ref{tab:cr_galaxies}.
The essence of this approach is that the observed galaxy spectrum is modeled as a combination of two SSPs with individual properties, each convolved with an independent Gaussian LOSVD and assigned a free relative weight.

\begin{figure*}
\centering
\includegraphics[width=0.7\textwidth]{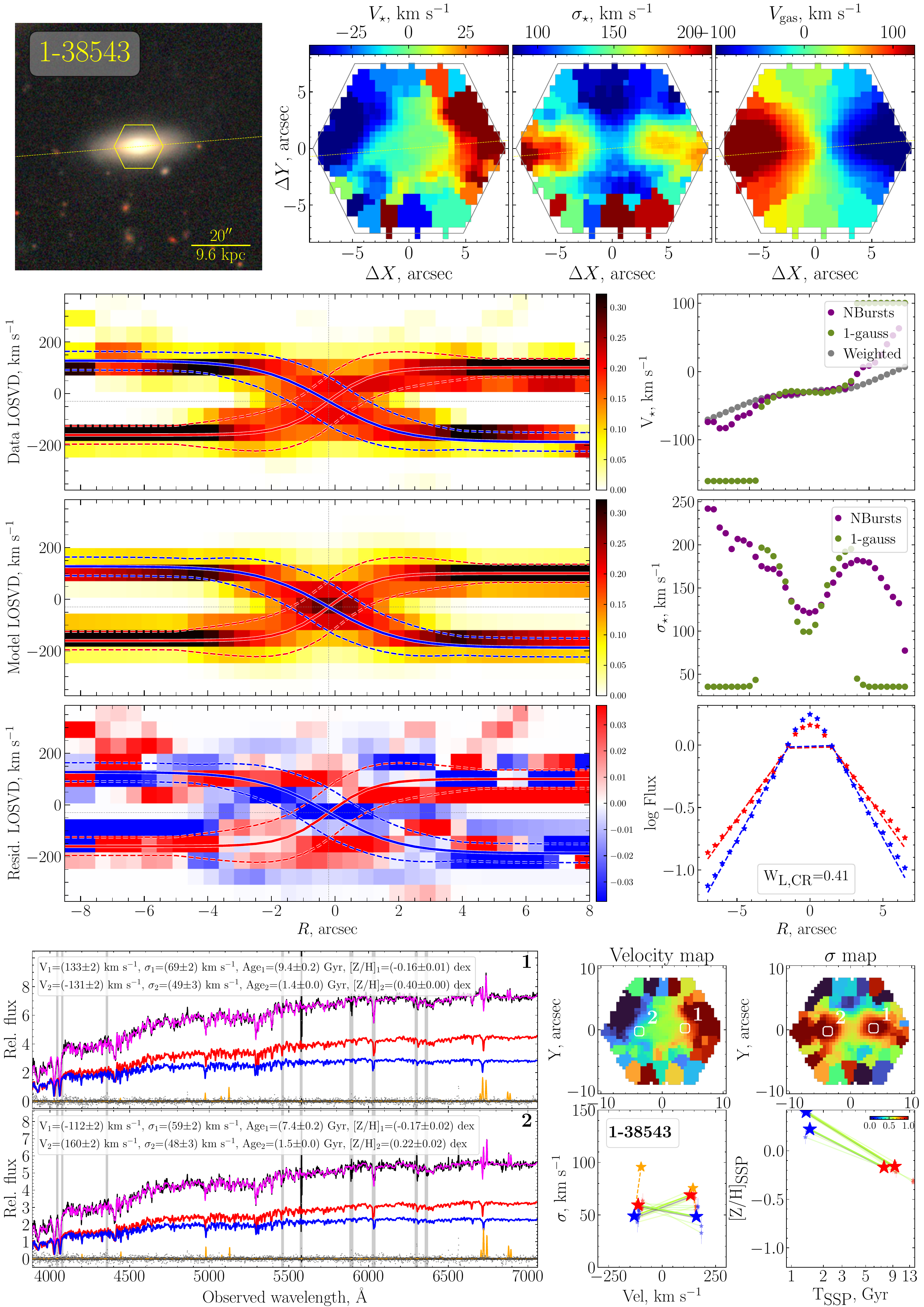}
\caption{
Example of all analysis steps for \textbf{1-38543}. We prepared similar plots for all CR galaxies (Sec.~\ref{sec:app_C}).
\textbf{Top part:} Description is similar to Fig.~\ref{fig:im-prob}.
\textbf{Middle part:} Non-parametrically recovered stellar LOSVD and its analysis. 
The first column shows the extracted non-parametric stellar LOSVD along the major axis, along with its model and the residuals between them.
The second column displays the weighted stellar velocity from the model ($W_1 \times V_1 + (1 - W_1) \times V_2$), compared to a single-component \nb\ fit and a pure Gaussian fitting for each spatial bin.
A similar comparison is provided for the velocity dispersion $\sigma_\star$.
The bottom plot shows the extracted stellar intensity in the SDSS g-band along the major axis, with each disk weighted according to its exponential shape in LOSVD modeling.
\textbf{Bottom part:} Two-component spectral decomposition.
The left two plots display the decomposition of bins \textbf{1} and \textbf{2}, whose positions are shown on the stellar velocity and velocity dispersion maps derived from single-component fitting (top central and right panels).
Black lines represent the observed spectra and thin black lines show the error of these spectra.
The magenta line is the sum of two stellar components and emission lines, red indicates the old main component, and blue -- the young CR component.
Emission lines are shown in orange, while grey dots indicate the residuals between the observed spectrum and the full model.
The bottom middle and right panels show the kinematic and stellar population parameters for the old (red marks) and young (blue) components, with orange stars representing gas kinematics correlated with the young component.
Transparent symbols show obtained parameters for each unbinned spaxel within the large bins.
The color of the connecting marks reflects the mean weight of the old component at $\lambda=5100 \pm 50$\AA. 
The color bar of this weight is shown in the top right angle of the  stellar population plot.
\label{fig:full_analysis_example}}
\end{figure*}

The quality of spectral decomposition depends on SNR, spectral resolution, and the physical separation of the LOS velocities between components.
With MaNGA spectra, distinguishing two stellar components is challenging in the galaxy center (minimal kinematic separation) and near the IFU edge (low SNR).
Therefore, we performed spectral decomposition in two optimal Voronoi bins selected from both sides of the galaxy along the major kinematic axis at the radius where the rotation curve reaches a plateau (largest $\Delta V$) and the component weights are roughly equal.
Typically, these bins correspond to the location of the 2$\sigma$ peaks.
Fig.~\ref{fig:full_analysis_example} demonstrates an example of spectral decomposition for 1-38543.
Consistent parameters from two independent bins confirm the reliability of the fit, while the discrepancies between them allow us to estimate the output errors.

In Tab.~\ref{tab:np_losvd_2comp}, we report the ages and metallicities as averages over the two bins, with uncertainties
calculated by combining the standard deviation (for two bins, half of the difference between the values) and the individual errors of the bins.
The kinematics and relative weights of the components fitted within selected bins are consistent with estimates obtained from non-parametric LOSVD analyses. 
All components co-rotating with the ionized gas consistently exhibit younger ages than the main disk.
Additionally, we computed the stellar mass-to-light ratio $(M/L)_\star$, denoted as $\mu$, for the derived SSP parameters using photometric models calculated with the \textsc{PEGASE.2} code \citep{Fioc1997A&A...326..950F} based on the low-resolution BaSeL synthetic stellar library \citep{Lejeune1997A&AS..125..229L} and assuming a Kroupa Initial Mass Function \citep{Kroupa2001MNRAS.322..231K}.
We then recalibrated the total luminosity weight of the CR disk $W_\mathrm{CR}$ into its corresponding total mass weight $W_\mathrm{m,CR}$ using the following equation:
\begin{equation}
\label{eq:Wm_from_W}
W_\mathrm{m,CR} = \dfrac{1}{1 + \left( 1/W_\mathrm{CR} - 1 \right) \mu_\mathrm{m}/\mu_\mathrm{CR}},
\end{equation}
where $\mu_m$ and $\mu_{CR}$ are the $(M/L)_\star$ ratios for the main and CR components, respectively.

\begin{deluxetable*}{lccccccccccccc}
\label{tab:np_losvd_2comp}
\tablecaption{Table with parameters from non-parametric LOSVD and 2-component \nb~analysis.}
\tabletypesize{\tiny}
\tablehead{
\colhead{ID} & 
\colhead{CR loc.} & 
\colhead{$\log~\dfrac{\textrm{M}_{\star}}{\textrm{M}_\odot}$} & 
\colhead{h$_{\textrm{m}}$, kpc} & 
\colhead{h$_{\textrm{CR}}$, kpc} & 
\colhead{W$_{\textrm{LOSVD}}$} & 
\colhead{T$_{\textrm{SSP,m}}$} & 
\colhead{[Z/H]$_{\textrm{SSP,m}}$} & 
\colhead{T$_{\textrm{SSP,CR}}$} & 
\colhead{[Z/H]$_{\textrm{SSP,CR}}$} & 
\colhead{$\mu_{\textrm{m}}$} & 
\colhead{$\mu_{\textrm{CR}}$} & 
\colhead{W$_{\textrm{mass}}$} & 
\colhead{$\log~\dfrac{\textrm{M}_{\star,\textrm{CR}}}{\textrm{M}_\odot}$}} 
\decimalcolnumbers
\startdata
1-42660 & outer & 10.27 & 1.59 $\pm$ 0.07 & 2.36 $\pm$ 0.11 & (74 $\pm$ 3)\% & $9.6 \pm 0.6$ & $-0.20 \pm 0.03$ & $1.3 \pm 0.2$ & $-0.03 \pm 0.00$ & 4.33 & 0.89 & (37 $\pm$ 4)\% & 9.83 $\pm$ 0.04 \\
1-42247 & inner & 9.71 & 1.20 $\pm$ 0.13 & 0.80 $\pm$ 0.01 & (40 $\pm$ 10)\% & \ldots & \ldots & \ldots & \ldots & \ldots & \ldots & \ldots & \ldots \\
1-37062 & inner & 10.19 & 1.08 $\pm$ 0.02 & 1.02 $\pm$ 0.02 & (9 $\pm$ 2)\% & $3.2 \pm 0.4$ & $-0.01 \pm 0.10$ & $1.7 \pm 0.4$ & $-1.13 \pm 0.15$ & 2.15 & 0.65 & (2.9 $\pm$ 0.7)\% & 8.64 $\pm$ 0.11 \\
1-37478$^{\dagger}$ & inner & 10.46 & 0.93 $\pm$ 0.01 & 0.93 $\pm$ 0.01 & (28.9 $\pm$ 0.1)\% & \ldots & \ldots & \ldots & \ldots & \ldots & \ldots & \ldots & \ldots \\
1-38543 & inner & 10.36 & 1.28 $\pm$ 0.02 & 0.98 $\pm$ 0.03 & (41 $\pm$ 2)\% & $8.4 \pm 1.0$ & $-0.17 \pm 0.03$ & $1.4 \pm 0.1$ & $0.31 \pm 0.09$ & 4.06 & 1.27 & (18 $\pm$ 1)\% & 9.62 $\pm$ 0.03 \\
1-339061 & inner & 10.32 & 1.54 $\pm$ 0.03 & 1.23 $\pm$ 0.02 & (33 $\pm$ 3)\% & $6.8 \pm 0.4$ & $0.00 \pm 0.04$ & $1.7 \pm 0.2$ & $-0.73 \pm 0.10$ & 4.12 & 0.77 & (8.5 $\pm$ 0.9)\% & 9.24 $\pm$ 0.05 \\
1-44047 & outer & 10.22 & 1.22 $\pm$ 0.01 & 1.32 $\pm$ 0.02 & (54 $\pm$ 3)\% & $13.7 \pm 0.9$ & $-0.40 \pm 0.12$ & $1.7 \pm 0.2$ & $0.38 \pm 0.06$ & 4.91 & 1.60 & (28 $\pm$ 2)\% & 9.67 $\pm$ 0.03 \\
1-44483 & outer & 10.12 & 1.30 $\pm$ 0.03 & 2.69 $\pm$ 0.50 & (67 $\pm$ 4)\% & $12.9 \pm 0.7$ & $-0.36 \pm 0.04$ & $1.5 \pm 0.2$ & $0.18 \pm 0.23$ & 4.84 & 1.25 & (35 $\pm$ 4)\% & 9.66 $\pm$ 0.05 \\
1-382889 & inner & 10.37 & 2.86 $\pm$ 0.38 & 1.53 $\pm$ 0.02 & (31 $\pm$ 4)\% & $4.3 \pm 0.4$ & $-0.08 \pm 0.07$ & $1.7 \pm 0.3$ & $-0.83 \pm 0.09$ & 2.69 & 0.73 & (11 $\pm$ 2)\% & 9.40 $\pm$ 0.08 \\
1-386322 & outer & 9.94 & 0.54 $\pm$ 0.09 & 1.15 $\pm$ 0.01 & (82 $\pm$ 2)\% & $13.0 \pm 1.2$ & $-0.16 \pm 0.08$ & $3.8 \pm 0.4$ & $-0.25 \pm 0.05$ & 5.72 & 2.13 & (62 $\pm$ 4)\% & 9.73 $\pm$ 0.03 \\
1-137890$^{\dagger}$ & inner & 9.89 & 1.22 $\pm$ 0.02 & 1.57 $\pm$ 0.05 & (42.2 $\pm$ 0.3)\% & $3.8 \pm 1.6$ & $-0.12 \pm 0.08$ & $1.5 \pm 0.3$ & $-1.29 \pm 0.08$ & 2.37 & 0.59 & (15.5 $\pm$ 0.1)\% & 9.08 $\pm$ 0.00 \\
1-386932 & inner & 9.87 & 1.00 $\pm$ 0.14 & 0.65 $\pm$ 0.01 & (17 $\pm$ 3)\% & $6.0 \pm 0.7$ & $-0.25 \pm 0.09$ & $2.1 \pm 0.1$ & $-0.94 \pm 0.06$ & 2.88 & 0.84 & (6 $\pm$ 1)\% & 8.64 $\pm$ 0.10 \\
1-232087 & outer & 10.28 & 1.68 $\pm$ 0.12 & 3.32 $\pm$ 0.16 & (72 $\pm$ 3)\% & $11.2 \pm 1.6$ & $-0.44 \pm 0.07$ & $1.0 \pm 0.0$ & $-0.25 \pm 0.03$ & 4.02 & 0.64 & (29 $\pm$ 3)\% & 9.73 $\pm$ 0.05 \\
1-78381 & outer & 10.96 & 1.84 $\pm$ 0.04 & 3.41 $\pm$ 0.41 & (78 $\pm$ 3)\% & $14.0 \pm 0.8$ & $0.15 \pm 0.06$ & $11.0 \pm 0.7$ & $-0.57 \pm 0.08$ & 8.08 & 3.63 & (62 $\pm$ 4)\% & 10.76 $\pm$ 0.03 \\
1-586725 & inner & 10.37 & 1.40 $\pm$ 0.03 & 1.22 $\pm$ 0.02 & (34.4 $\pm$ 0.3)\% & $6.4 \pm 1.3$ & $-0.13 \pm 0.07$ & $1.2 \pm 0.1$ & $-0.37 \pm 0.14$ & 3.42 & 0.67 & (9.3 $\pm$ 0.1)\% & 9.34 $\pm$ 0.01 \\
1-148987 & inner & 10.34 & 1.73 $\pm$ 0.04 & 1.47 $\pm$ 0.03 & (65 $\pm$ 4)\% & $9.5 \pm 3.4$ & $-0.32 \pm 0.07$ & $2.3 \pm 0.1$ & $-0.12 \pm 0.04$ & 3.92 & 1.52 & (42 $\pm$ 5)\% & 9.96 $\pm$ 0.05 \\
1-174947 & outer & 10.68 & 1.43 $\pm$ 0.01 & 3.99 $\pm$ 1.89 & (45 $\pm$ 8)\% & $12.5 \pm 1.8$ & $-0.46 \pm 0.20$ & $4.6 \pm 2.9$ & $-0.84 \pm 1.04$ & 4.36 & 1.66 & (24 $\pm$ 6)\% & 10.07 $\pm$ 0.11 \\
1-519907 & outer & 10.09 & 0.72 $\pm$ 0.01 & 1.74 $\pm$ 1.04 & (85 $\pm$ 2)\% & $14.0 \pm 1.5$ & $-0.23 \pm 0.11$ & $8.2 \pm 0.2$ & $-0.17 \pm 0.03$ & 5.75 & 3.98 & (80 $\pm$ 3)\% & 9.99 $\pm$ 0.02 \\
1-623416 & outer & 10.16 & 0.87 $\pm$ 0.01 & 1.91 $\pm$ 0.62 & (42 $\pm$ 5)\% & $13.7 \pm 0.7$ & $-0.32 \pm 0.06$ & $2.8 \pm 0.3$ & $-0.40 \pm 0.17$ & 5.27 & 1.44 & (16 $\pm$ 3)\% & 9.37 $\pm$ 0.07 \\
1-456884 & inner & 9.54 & 0.93 $\pm$ 0.01 & 0.41 $\pm$ 0.10 & 20$^*$\% & $10.0 \pm 2.5$ & $-0.69 \pm 0.09$ & $6.5 \pm 0.6$ & $0.01 \pm 0.09$ & 3.10 & 3.95 & 20$^*$\% & 8.84 $\pm$ 0.78 \\
1-456672 & outer & 10.13 & 1.08 $\pm$ 0.03 & 1.48 $\pm$ 0.02 & (48.5 $\pm$ 0.3)\% & \ldots & \ldots & \ldots & \ldots & \ldots & \ldots & \ldots & \ldots \\
1-456645 & outer & 9.96 & 1.19 $\pm$ 0.02 & 1.29 $\pm$ 0.02 & (41.6 $\pm$ 0.1)\% & $4.3 \pm 1.3$ & $-0.19 \pm 0.25$ & $1.9 \pm 0.4$ & $-0.97 \pm 0.12$ & 2.49 & 0.77 & (18.03 $\pm$ 0.08)\% & 9.21 $\pm$ 0.00 \\
1-419257 & inner & 10.31 & 1.95 $\pm$ 0.04 & 1.38 $\pm$ 0.04 & (35.4 $\pm$ 1.0)\% & $6.4 \pm 2.3$ & $-0.02 \pm 0.08$ & $1.3 \pm 0.1$ & $-0.58 \pm 0.09$ & 3.78 & 0.63 & (8.3 $\pm$ 0.3)\% & 9.23 $\pm$ 0.02 \\
1-244115 & inner & 10.14 & 1.46 $\pm$ 0.04 & 1.34 $\pm$ 0.03 & (32.6 $\pm$ 0.2)\% & $8.7 \pm 1.4$ & $-0.09 \pm 0.10$ & $1.4 \pm 0.2$ & $-0.36 \pm 0.14$ & 4.45 & 0.73 & (7.38 $\pm$ 0.07)\% & 9.01 $\pm$ 0.00 \\
1-245235 & inner & 10.15 & 3.46 $\pm$ 0.17 & 1.34 $\pm$ 0.27 & 40$^*$\% & $5.0 \pm 1.0$ & $-0.05 \pm 0.09$ & $2.0 \pm 0.1$ & $-0.84 \pm 0.15$ & 3.06 & 0.84 & 13$^*$\% & 9.27 $\pm$ 0.29 \\
1-245301 & outer & 10.75 & 1.82 $\pm$ 0.12 & 3.02 $\pm$ 0.15 & (58 $\pm$ 1)\% & \ldots & \ldots & \ldots & \ldots & \ldots & \ldots & \ldots & \ldots \\
1-593328 & inner & 11.20 & 4.11 $\pm$ 0.46 & 2.26 $\pm$ 0.05 & (60 $\pm$ 30)\% & $9.8 \pm 0.7$ & $-0.03 \pm 0.10$ & $1.8 \pm 0.6$ & $-0.13 \pm 0.11$ & 5.14 & 1.18 & 30$^*$\% & 10.61 $\pm$ 0.35 \\
1-633000 & inner & 10.06 & 1.03 $\pm$ 0.01 & 0.86 $\pm$ 0.01 & (48 $\pm$ 10)\% & $10.7 \pm 2.5$ & $-0.33 \pm 0.07$ & $1.0 \pm 0.0$ & $-0.18 \pm 0.08$ & 4.25 & 0.66 & (13 $\pm$ 4)\% & 9.16 $\pm$ 0.15 \\
1-248869 & outer & 10.85 & 1.72 $\pm$ 0.02 & 2.45 $\pm$ 0.11 & (50 $\pm$ 4)\% & $13.5 \pm 0.9$ & $-0.15 \pm 0.05$ & $3.8 \pm 0.9$ & $-0.10 \pm 0.12$ & 5.97 & 2.37 & (28 $\pm$ 3)\% & 10.30 $\pm$ 0.05 \\
1-211561 & inner & 10.27 & 1.94 $\pm$ 0.25 & 1.17 $\pm$ 0.03 & (12 $\pm$ 4)\% & $7.3 \pm 0.5$ & $-0.24 \pm 0.06$ & $1.9 \pm 0.4$ & $-0.95 \pm 0.29$ & 3.42 & 0.78 & (3 $\pm$ 1)\% & 8.74 $\pm$ 0.15 \\
1-135767 & outer & 10.55 & 1.23 $\pm$ 0.01 & 3.96 $\pm$ 2.66 & (70 $\pm$ 10)\% & $14.0 \pm 0.7$ & $-0.34 \pm 0.07$ & $3.5 \pm 0.5$ & $-0.05 \pm 0.13$ & 5.24 & 2.29 & (50 $\pm$ 10)\% & 10.25 $\pm$ 0.11 \\
1-94690 & inner & 10.22 & 2.55 $\pm$ 0.61 & 1.24 $\pm$ 0.01 & (29 $\pm$ 7)\% & $7.6 \pm 3.3$ & $-0.18 \pm 0.11$ & $2.1 \pm 0.2$ & $-0.81 \pm 0.10$ & 3.77 & 0.92 & (9 $\pm$ 3)\% & 9.18 $\pm$ 0.13 \\
1-136248 & outer & 10.52 & 1.77 $\pm$ 0.14 & 4.14 $\pm$ 0.23 & (68 $\pm$ 4)\% & $8.9 \pm 1.7$ & $-0.35 \pm 0.06$ & $3.6 \pm 0.3$ & $-0.04 \pm 0.06$ & 3.61 & 2.32 & (58 $\pm$ 5)\% & 10.28 $\pm$ 0.04 \\
1-635506 & inner & 10.21 & 1.35 $\pm$ 0.02 & 1.34 $\pm$ 0.02 & (14.2 $\pm$ 0.4)\% & $6.2 \pm 1.1$ & $-0.08 \pm 0.04$ & $3.0 \pm 0.3$ & $-0.58 \pm 0.08$ & 3.54 & 1.33 & (5.8 $\pm$ 0.2)\% & 8.98 $\pm$ 0.01 \\
1-179561 & inner & 10.09 & 1.11 $\pm$ 0.03 & 0.56 $\pm$ 0.09 & (34 $\pm$ 4)\% & \ldots & \ldots & \ldots & \ldots & \ldots & \ldots & \ldots & \ldots \\
1-115097 & inner & 10.23 & 2.09 $\pm$ 0.62 & 0.98 $\pm$ 0.01 & (22 $\pm$ 8)\% & $8.4 \pm 0.4$ & $-0.41 \pm 0.06$ & $2.8 \pm 0.3$ & $-0.02 \pm 0.07$ & 3.29 & 1.91 & (14 $\pm$ 6)\% & 9.38 $\pm$ 0.18 \\
1-109056 & outer & 10.41 & 1.88 $\pm$ 0.01 & 2.83 $\pm$ 0.21 & (86 $\pm$ 7)\% & \ldots & \ldots & \ldots & \ldots & \ldots & \ldots & \ldots & \ldots \\
1-37494 & outer & 10.18 & 1.49 $\pm$ 0.04 & 2.43 $\pm$ 0.32 & (54 $\pm$ 5)\% & \ldots & \ldots & \ldots & \ldots & \ldots & \ldots & \ldots & \ldots \\
1-109275 & inner & 10.05 & 2.26 $\pm$ 0.31 & 1.39 $\pm$ 0.03 & 50$^*$\% & \ldots & \ldots & \ldots & \ldots & \ldots & \ldots & \ldots & \ldots \\
1-297863 & inner & 10.25 & 1.47 $\pm$ 0.03 & 1.26 $\pm$ 0.01 & (52 $\pm$ 5)\% & $7.7 \pm 1.9$ & $-0.55 \pm 0.10$ & $2.0 \pm 0.3$ & $0.31 \pm 0.14$ & 2.77 & 1.81 & (42 $\pm$ 5)\% & 9.87 $\pm$ 0.05 \\
1-378729 & undef & 10.20 & 0.70 $\pm$ 0.01 & 0.80 $\pm$ 0.01 & (22.7 $\pm$ 0.1)\% & \ldots & \ldots & \ldots & \ldots & \ldots & \ldots & \ldots & \ldots \\
1-121871 & inner & 10.11 & 2.09 $\pm$ 0.05 & 1.44 $\pm$ 0.08 & (48 $\pm$ 6)\% & \ldots & \ldots & \ldots & \ldots & \ldots & \ldots & \ldots & \ldots \\
1-182730 & inner & 11.05 & 3.56 $\pm$ 0.09 & 2.69 $\pm$ 0.09 & (54 $\pm$ 7)\% & $11.9 \pm 2.3$ & $-0.27 \pm 0.16$ & $2.9 \pm 0.5$ & $-0.06 \pm 0.19$ & 4.82 & 1.92 & (32 $\pm$ 6)\% & 10.56 $\pm$ 0.09 \\
1-188530 & inner & 10.66 & 3.77 $\pm$ 0.08 & 2.16 $\pm$ 0.17 & (56 $\pm$ 10)\% & \ldots & \ldots & \ldots & \ldots & \ldots & \ldots & \ldots & \ldots \\
1-260823 & inner & 10.47 & 2.12 $\pm$ 0.04 & 1.42 $\pm$ 0.10 & 60$^*$\% & \ldots & \ldots & \ldots & \ldots & \ldots & \ldots & \ldots & \ldots \\
1-13736 & inner & 10.08 & 1.22 $\pm$ 0.02 & 1.18 $\pm$ 0.01 & (36 $\pm$ 2)\% & $10.6 \pm 3.7$ & $-0.35 \pm 0.13$ & $3.2 \pm 1.2$ & $-0.77 \pm 0.32$ & 4.13 & 1.23 & (14.3 $\pm$ 1.0)\% & 9.24 $\pm$ 0.03 \\
1-210611 & inner & 10.30 & 1.36 $\pm$ 0.02 & 0.73 $\pm$ 0.10 & (22.6 $\pm$ 0.9)\% & $6.5 \pm 0.3$ & $-0.17 \pm 0.04$ & $1.7 \pm 0.4$ & $-0.64 \pm 0.19$ & 3.36 & 0.78 & (6.4 $\pm$ 0.3)\% & 9.10 $\pm$ 0.02 \\
1-135190 & inner & 10.47 & 1.68 $\pm$ 0.14 & 1.11 $\pm$ 0.01 & (20 $\pm$ 3)\% & $6.9 \pm 0.9$ & $-0.02 \pm 0.04$ & $1.4 \pm 0.0$ & $-0.58 \pm 0.13$ & 4.04 & 0.66 & (4.0 $\pm$ 0.8)\% & 9.07 $\pm$ 0.08 \\
1-94773 & inner & 10.18 & 0.92 $\pm$ 0.02 & 0.88 $\pm$ 0.02 & (40.7 $\pm$ 0.9)\% & $6.0 \pm 1.0$ & $-0.20 \pm 0.09$ & $3.5 \pm 1.5$ & $-0.76 \pm 0.21$ & 3.07 & 1.37 & (23.4 $\pm$ 0.7)\% & 9.55 $\pm$ 0.01 \\
1-178027 & outer & 10.44 & 1.68 $\pm$ 0.09 & 2.50 $\pm$ 0.06 & (34 $\pm$ 2)\% & $13.5 \pm 1.1$ & $-0.22 \pm 0.07$ & $3.0 \pm 0.6$ & $0.03 \pm 0.19$ & 5.63 & 2.10 & (16 $\pm$ 1)\% & 9.64 $\pm$ 0.04 \\
1-26197 & outer & 10.93 & 3.48 $\pm$ 0.05 & 5.92 $\pm$ 0.54 & (45 $\pm$ 4)\% & $14.0 \pm 1.3$ & $-0.07 \pm 0.12$ & $2.7 \pm 0.3$ & $-0.34 \pm 0.17$ & 6.53 & 1.48 & (15 $\pm$ 2)\% & 10.12 $\pm$ 0.07 \\
\enddata
\begin{tablenotes}
\small
\item \textsc{NOTE:}
(1) The MaNGA-ID of the galaxy.
(2) The location of the CR disk relative to the main component; $\dagger$ indicates cases where the scale estimates from the LOSVD analysis are inconsistent with the kinematic maps due to a noisy or irregular LOSVD pattern.
(3) The stellar mass from NSA, corrected to $H_0^{2}$.
(4) and (5) are the exponential scales of the main and CR disks in kpc (Eq.~\ref{eq:exp_prof}).
(6) The weight of the CR disk in the total luminosity of the galaxy (Sec.~\ref{sec:np_losvd}), $*$ indicates that the uncertainty of this parameter is too large.
(7) and (9) are the SSP-equivalent ages of the main and CR disks in Gyr, respectively. ``\ldots’’ means failure of the two-component analysis.
(8) and (10) are the SSP-equivalent stellar metallicities of the main and CR disks in dex.
(11) and (12) are the (M/L) ratio in the SDSS $g$-band for the main and CR disks.
(13) CR disk stellar mass fraction  (Eq.~\ref{eq:Wm_from_W}).
(14) The estimated stellar mass of the CR disk.
\end{tablenotes}
\end{deluxetable*}

Additionally, by combining the $(M/L)_\star$ values with our model of the radial structure of the stellar LOSVD (see Sec.~\ref{sec:np_losvd}), we estimated the angular momentum of both the CR and main disks.
We consider a simple model for a disk with velocity profile $v(R)$ (Eq.~\ref{eq:v_prof}) obtained from LOSVD analysis and an exponential density profile $\Sigma(R)$ characterized by scale length $h$ and central density $\Sigma_0$.
In this model, the total angular momentum is given by:
\begin{equation}
\label{eq:disk_momentum}
J_{\textrm{disk}} = \int R v(R) dM = \Sigma_0 \int\limits_0^\infty v(R) \exp\left( -\dfrac{R}{h}\right) 2 \pi R^2 dR.
\end{equation}
Central density $\Sigma_0$ can be calculated from the total mass of the exponential disk: $M_0 = 2\pi \Sigma_0 h^2$.
The parameters $V_\mathrm{max}$, $R_0$, $c$, and $h$ were taken from the LOSVD model (Sec.~\ref{sec:np_losvd}), and the velocity amplitude was corrected for inclination $i$ estimated as follows \citep{Tully1998AJ....115.2264T}:
\begin{equation}
\label{eq:inclination_correction}
\cos^2 i = \dfrac{(1 - e)^2 - (1 - 0.8)^2}{1 - (1 - 0.8)^2},
\end{equation}
where ellipticity $e$ was taken from the MaNGA DAP (parameter \textsc{ecooell}).

\subsection{Ionized gas}
\label{sec:ion_gas}

Most CR galaxies in our sample show strong ionized gas emission lines.
We have already used kinematic maps derived from emission line spectra to supplement our analysis of stellar populations.
In this section, we discuss the sources of ionization and gas-phase metallicity.

\begin{figure*}
    \centering
	\includegraphics[width=\textwidth]{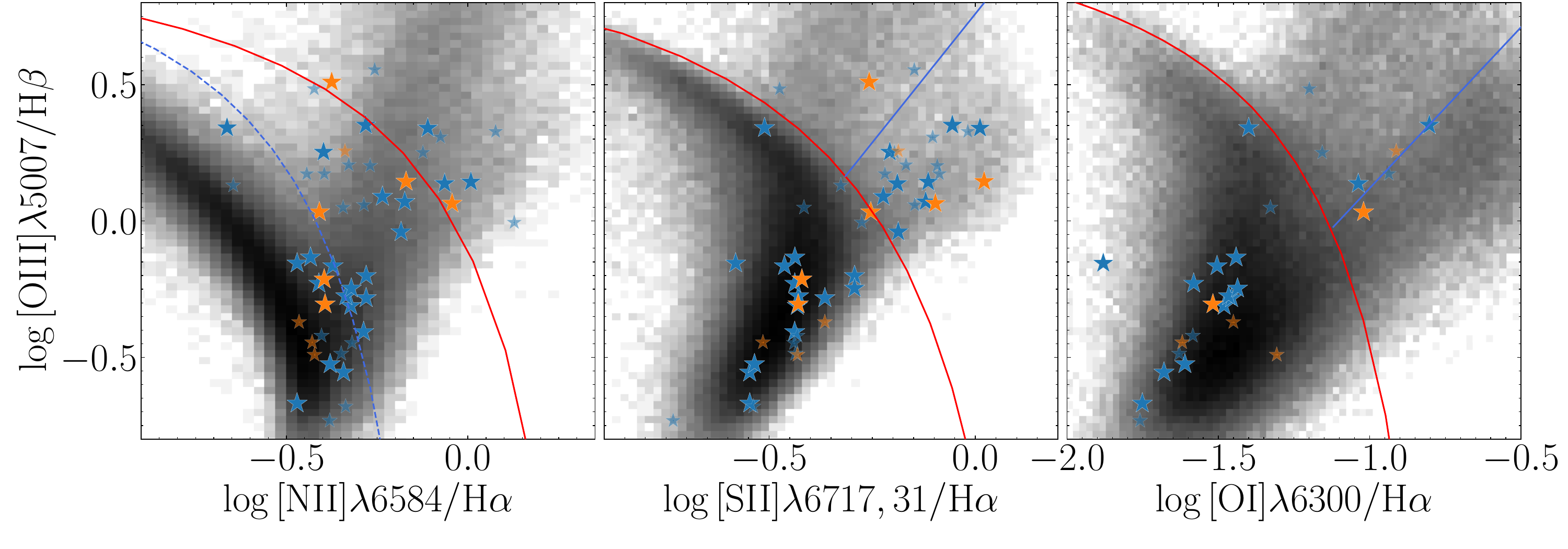}
\caption{BPT-[N\ii], -[S\ii] and -[O\i] diagrams for the sample of CR galaxies. Bright and small stars show \textbf{reliable} and \textbf{probable} CR galaxies. Blue and orange colors show \textbf{inner} and \textbf{outer} configurations of CR disk (Sec.~\ref{sec:in_out_CR}). The blue dashed line in the BPT-[N\ii] plot represents the demarcation line introduced by \citet{kauffmann03}, which distinguishes between the star-forming and the so-called composite regions. The additional demarcation lines taken from \citet{kewley01}. They separate composite and AGN regions on the BPT-[N\ii] diagram and categorize star-forming, Seyfert/LINER regions in the BPT-[O\i] and BPT-[S\ii] diagrams. The distribution of emission line measurements from the RCSED \citep[\url{rcsed.sai.msu.ru},][]{Chilingarian2017ApJS..228...14C} is represented in grey.\label{fig:BPT_sum}}
\end{figure*}

To identify the mechanisms driving gas excitation, we employed Baldwin-Phillips-Terlevich diagrams \citep[BPT,][]{bpt}, incorporating the demarcation lines of \citet{kewley01} and \citet{kauffmann03}.
For each galaxy in our sample, we computed the integrated line flux within the effective radius $R_\mathrm{eff}$ using line flux maps from the MaNGA DAP. 
The $R_\mathrm{eff}$ is the Petrosian effective radius from the NSA.
We then filtered out faint detections by requiring that the mean \textsc{ganr} within $R_\mathrm{eff}$ exceeds 2 and calculated the corresponding line ratios to construct the BPT diagrams (Fig.~\ref{fig:BPT_sum}).
In the DAP, \textsc{ganr} is defined as the amplitude of the Gaussian fit of an emission line relative to the median noise measured in two sidebands near the line.
Our analysis indicates that CR galaxies are evenly distributed across SF, Composite, and AGN/LINER regions (see Tab.~\ref{tab:ioniz_bpt}).

\begin{deluxetable}{ccccc}
\tablecaption{
Percentage of galaxies in different regions of the BPT-[N\ii] by type of kinematic misalignment.
Galaxies with stellar CR were excluded from the pure gas-star misalignment categories (CR-GS, Gas-Polar, Gas-Mis.).
The ``Weak lines'' designates galaxies with too weak emission lines to be used in the BPT-[N\ii] diagram.
The number of galaxies of each type is given in the table head.
}
\label{tab:ioniz_bpt}
\tablehead{
     & \textbf{CR}    & \textbf{CR-GS} & \textbf{Gas-Polar} & \textbf{Gas-Mis.}\\
   & (120) & (145) & (116)     & (172)
}
\startdata
SF    &  29\%  &  21\%  &  8\%   &  9\%   \\
Comp. &  28\%  &  26\%  &  20\%  &  22\%  \\
AGN   &  24\%  &  34\%  &  43\%  &  34\%  \\
Weak lines &  19\% &  19\%  &  29\%  &  35\%  
\enddata
\end{deluxetable}

Interestingly, when moving from stellar CR to pure gas/star CR (excluding cases with stellar CR), the distribution across BPT-[N\ii] classes changes, gas/star CR galaxies tending to populate the AGN region rather than the SF one (see Tab.~\ref{tab:ioniz_bpt}).
This trend is stronger in galaxies with off-plane gas/star misalignment, and particularly in polar gas disks, which we identified during sample construction (see Sec.~\ref{sec:vis_inspec}).
This implies that accretion and collisions with pre-existing gas may amplify the impact of shocks on gas ionization and potentially trigger AGN activity, as it was already demonstrated in previous work on spatially resolved spectroscopy \citep{Egorov2019MNRAS.486.4186E}, numerical simulations \citep{Khrapov2024OAst...3320231K} and statistical study \citep{Smirnov2020AstL...46..501S} of polar-ring galaxies.

\begin{figure}
    \centering
    \includegraphics[width=0.49\textwidth]{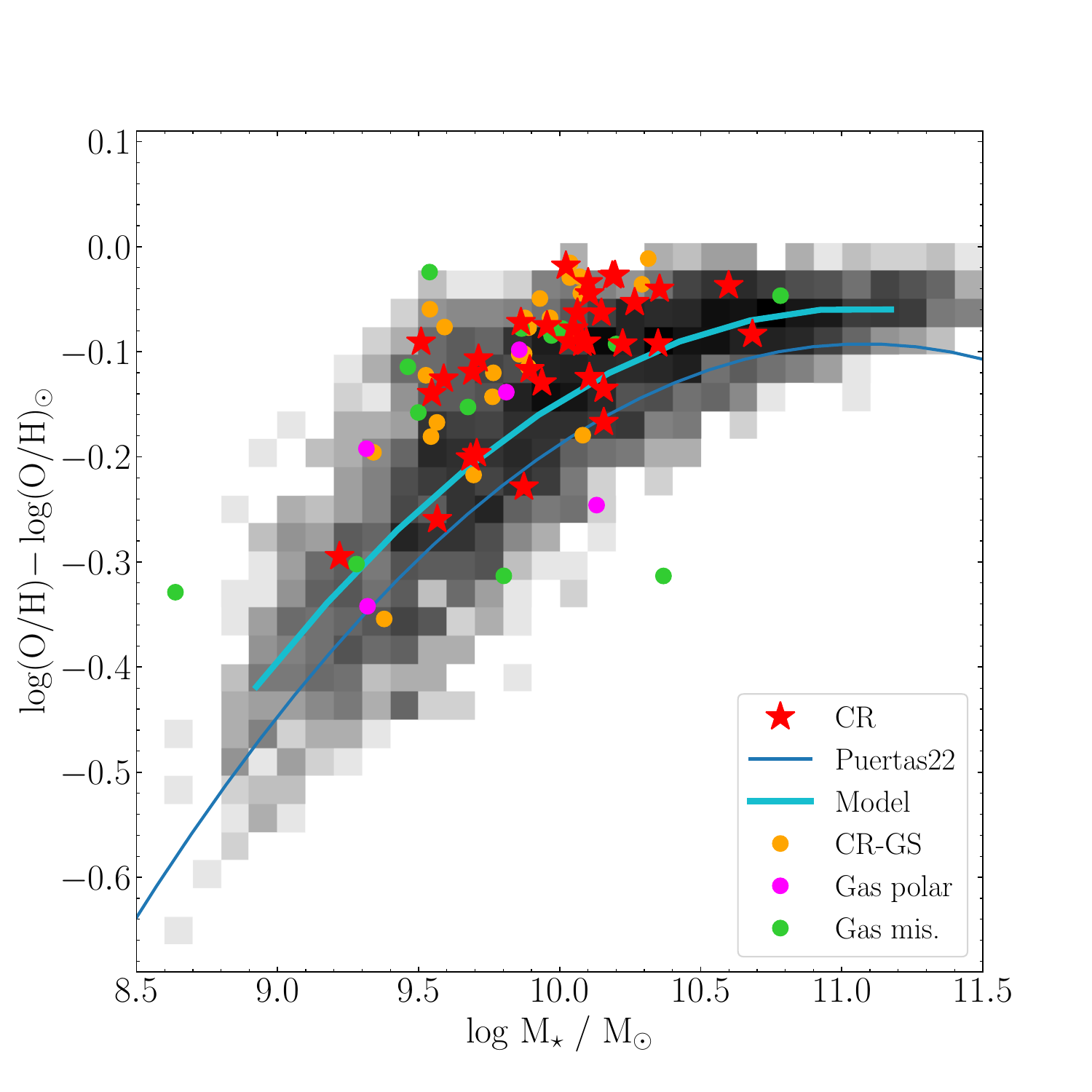}
\caption{
Stellar mass -- gas-phase metallicity relation (MZR).
Metallicity is given relative to $12+\log$(O/H)$_\odot = 8.69$ \citep{Asplund2021A&A...653A.141A}.
The background represents all MaNGA galaxies classified as SF based on the BPT-[N\ii] diagram.
Red stars denote gas in CR galaxies, while orange, purple, and lime dots correspond to galaxies with CR-GS, gas-polar, and gas misalignment, respectively.
The light blue line shows a second-degree polynomial fit to all MaNGA galaxies, while the blue line is taken from \citep{DuartePuertas2022}.
\label{fig:MZR}}
\end{figure}

Gas-phase metallicity is an important observable for discussing the sources of the accreted material that forms CR disks (see discussion in Sec.~\ref{sec:discussion}).
For galaxies with SF dominated excitation, we estimated oxygen abundances using R-calibration \citep{Pilyugin2016MNRAS.457.3678P} after preliminary correcting the fluxes for extinction using the Balmer decrement and \citet{Fitzpatrick1999PASP..111...63F} extinction curve.
The results of our analysis are shown on the Mass-Metallicity (MZR) diagram (Fig.~\ref{fig:MZR}).
For reference, we also analyzed all SF-dominated MaNGA galaxies using the same method, displaying them as grey background in Fig.~\ref{fig:MZR}.
For clarity, we computed a second-degree polynomial fit for these MaNGA galaxies (shown as the thick blue line) that closely matches the MZR shape reported by \citet{DuartePuertas2022} with a deviation of only 0.04 dex.
We observe that the majority of stellar CR galaxies have ionized gas metallicities higher than those predicted by their stellar masses.

We evolutionarily link the ionized gas component to the formation of the CR disk, dismissing any substantial contribution from the pre-existing ISM.
If the CR disk is viewed as an individual galaxy formed apart from its host, we would assign it a much lower stellar mass for the measured gas metallicity, which would increase its deviation from the mean MZR.
The gas-phase metallicity we observe today reflects the cumulative metal enrichment during CR disk formation and depends on both the metallicity of the accreted material and the efficiency of the enrichment process.
The enhanced metallicity in CR galaxies may indicate that the accreted material was already pre-enriched and/or that SF was highly efficient, leading to rapid metal enrichment.
These scenarios can be tested through chemical evolution models, as demonstrated in our study of PGC~66551 \citep{Katkov2024ApJ...962...27K}.

Our approach to use emission lines and oxygen abundance calibration follows the same methodology by \citet{Zinchenko2023A&A...674L...7Z} in their studies of counter-rotating gas in MaNGA galaxies.
They found that red star/gas CR galaxies have lower oxygen abundances for a given mass, while blue ones have typical or higher values.
The findings for our stellar CR galaxies are consistent with those for blue galaxies.
In fact, our targets with reliable gas metallicities are classified as SF on the BPT and exhibit blue colors on the CMD.
However, neither our stellar CR sample nor the larger group of galaxies with gas misalignments (aside from galaxy 1-72211) deviates notably from the MZR toward lower metallicities, as reported by \citet{Zinchenko2023A&A...674L...7Z} for red gas/star CR galaxies.
We suspect that ongoing mergers, excluded during visual inspection and sample construction, may be responsible for this discrepancy.

\section{Discussion}
\label{sec:discussion}

\subsection{Dichotomy between inner and outer CR}
\label{sec:in_out_CR}

\begin{figure*}
\centering
\includegraphics[width=0.98\textwidth]{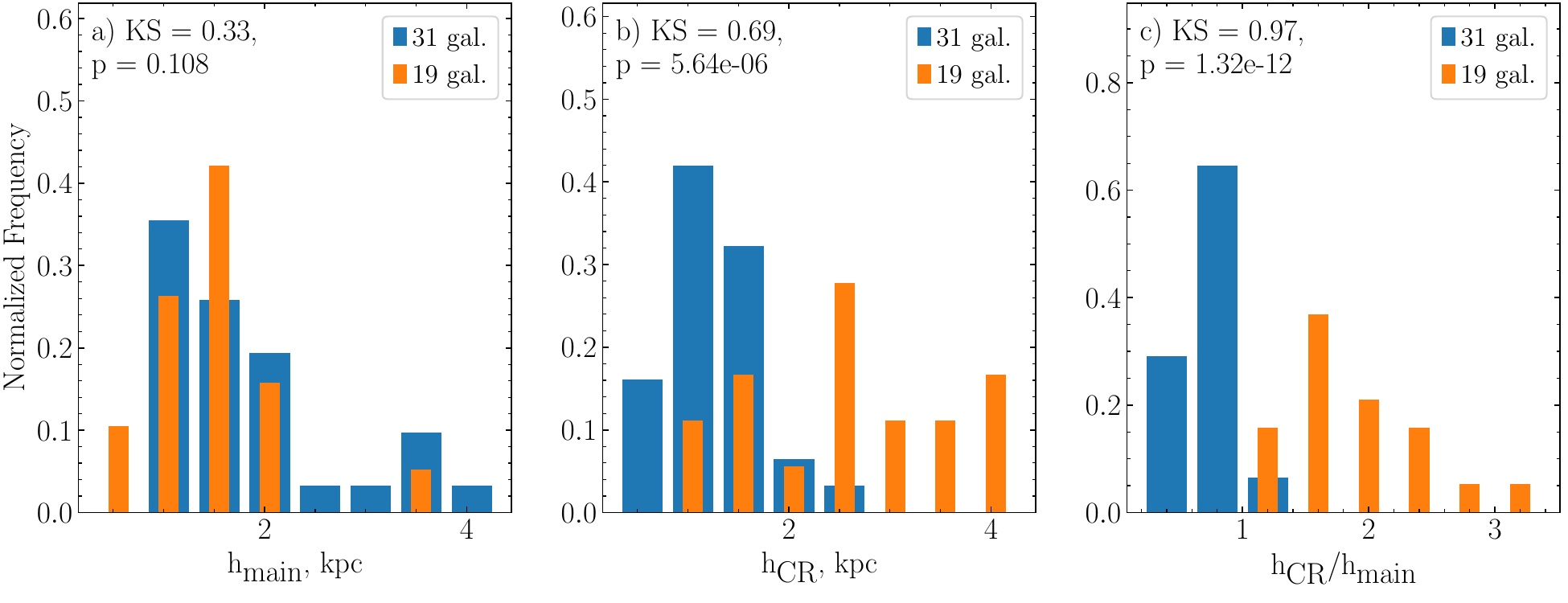}
\caption{Comparison of the disk scale lengths ($h$) of the main and CR components for galaxies with inner and outer CR configurations, as well as their ratio. Blue and orange histograms correspond to inner and outer CR disks, respectively. Each panel includes two numbers in the top-left corner: the Kolmogorov-Smirnov (KS) statistic and the corresponding p-value. The p-value indicates the probability that the two samples are drawn from the same distribution. A value above 0.05 implies no statistically significant difference, while the value below 0.05 suggests that the distributions differ.
\label{fig:in_out_scales}}
\end{figure*}

Due to the large statistical sample, the MaNGA IFU survey has successfully detected numerous CR galaxies.
The growing number of identified CR galaxies, as demonstrated in this paper and previous studies, reveals that the family of CR galaxies is not uniform.
Instead, it comprises two distinct spatial configurations: \textit{inner} and \textit{outer}. We define inner CR disks as those whose contribution to the total luminosity is highest in the central region and declines significantly with radius. Conversely, outer CR disks contribute more prominently at larger radii compare with main stellar body.

We classified galaxies into inner or outer CR categories based on the DAP kinematic maps.
In galaxies exhibiting the ``Srot’’ feature, the stellar rotation direction twists from the central regions toward the outskirts.
By comparing this twist with the rotation of the ionized gas, which we assume traces the CR disk, we can infer whether the CR component dominates in the central region (inner CR) or in the outer parts (outer CR).

Assuming that both components are disk-like, the classification can also be expressed in terms of their spatial scales: inner CR disks have smaller scale lengths than the main component, while outer CR disks have larger ones.
During the LOSVD decomposition, we estimated the exponential scale lengths $h$ of both components (Tab.~\ref{tab:np_losvd_2comp}).
In the vast majority of cases, our classification based on the kinematic maps agrees well with the disk scale lengths derived from the LOSVD analysis (Fig.~\ref{fig:in_out_scales}).
In two cases where this agreement is not observed, the recovered LOSVD is noisy and the resulting decomposition is therefore considered less robust.

\begin{figure*}
\centering
\includegraphics[width=0.98\textwidth]{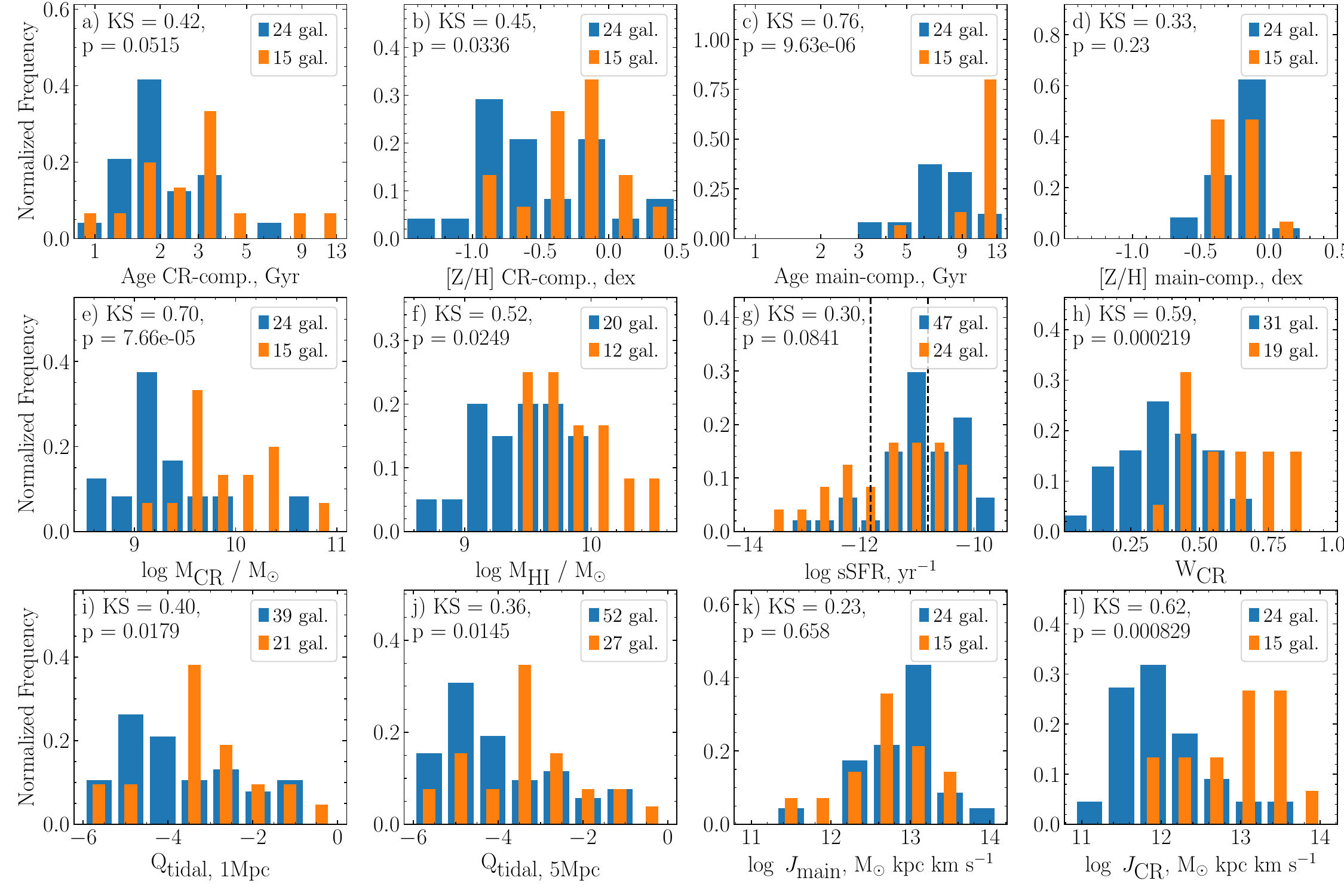}
\caption{Comparison of inner and outer configurations of stellar CR disk. The description is similar to Fig.~\ref{fig:in_out_scales}. The description of parameters can be found in Sec.~\ref{sec:in_out_CR}.
\label{fig:in_out_hists}}
\end{figure*}

Fig.~\ref{fig:in_out_hists} illustrates the differences in the parameters observed between inner and outer stellar CR in galaxies, which can be summarized as follows:
\begin{enumerate}[leftmargin=0em, labelwidth=-0.5em, labelsep=0.5em, align=left]

\item In galaxies with outer CR configuration, the stellar populations of both the CR and main disk are moderately older than those in galaxies with inner CR (see panels \textit{a} and \textit{c}).

\item Panel \textit{b} shows that the metallicity distributions suggest inner CR disks generally have lower metallicity compared to outer CR disks.

\item The separation in stellar metallicities for main disks is weak, if it exists (panel \textit{d}).

\item Panel \textit{h} shows that the luminosity contribution of CR disks is significantly larger for outer configurations, confirming our earlier preliminary conclusion based on a smaller sample of CR galaxies \citet{Gasymov2022muto.confE..17G}.
Nonetheless, our current analysis based on an extended sample does not reveal a significant difference in the total stellar masses of galaxies with inner and outer CR.

\item The parameters $Q_\mathrm{tidal,1Mpc}$ and $Q_\mathrm{tidal,5Mpc}$ (panels \textit{i} and \textit{j}), defined as the logarithm of the ratio between the tidal gravitational strength from neighboring galaxies (within 1 or 5~Mpc) and the galaxy's internal binding forces, were taken from the GEMA-VAC. These parameters do not show a statistically significant difference between inner and outer CR galaxies.
However, there is a weak tendency for inner CR galaxies to be found in sparser environments, particularly in the peaks of the distribution.

\item Galaxies with inner CR tend to exhibit slightly higher specific star formation rates (sSFR, panel \textit{g}) and demonstrate more prominent difference in the atomic phase gas content (panel \textit{f}).

\item Finally, we computed the angular momentum of the stellar main and CR disks (see panels \textit{k} and \textit{l}) and found that the momentum of the CR disk’s inner and outer configurations are clearly distinguished by this parameter, whereas the momentum of the main disk is quite similar in both outer and inner configurations.
\end{enumerate}

We inspected deep images from DECaLS and the Hyper Suprime-Cam Subaru Strategic Program \citep{2018PASJ...70S...1M, hsc2019} and identified another difference between the outer and inner CR configurations.
In general, outer CR galaxies more frequently exhibit morphological peculiarities: low-surface brightness (LSB) structures in galaxy periphery (6 outer vs. 2 inner), tidal tails (4 outer vs. 3 inner).
Also, only outer CR galaxies demonstrate clear spiral structures (1-174947, 1-135767, 1-178027).
Notably, the asymmetrical two-arm spiral pattern in galaxy 1-178027 might be a result of tidal interaction.
Given the lower detection rate of outer CR galaxies in our sample -- 29 outer compared to 53 inner CR galaxies -- the relative frequency of morphological features indicative of possible past interactions is significantly higher among outer CR galaxies.
In Sec.~\ref{sec:discussion_LSB}, we discuss in-depth the connection between CR galaxies and LSB galaxies.

\begin{deluxetable*}{ccc}
\tablecaption{Schema of kinematic diversity of stellar counter-rotation configurations\label{tab:diversity_schema}}
\tablehead{
  \multirow{2}{*}{Dominance of external (E) over pre-existing (P) gas} &
  \multicolumn{2}{c}{Angular momentum of externally accreated material} \\
  \cline{2-3}
  \colhead{} &
  \colhead{Low} &
  \colhead{High}
}
\startdata
E $\gg$ P & inner & outer\\
E $\gtrsim$ P & AGN, central SB / inner & inner / outer\\
E $<$ P & No CR & external CR ring \\
\enddata
\end{deluxetable*}

We propose the following schema for the kinematic diversity of CR galaxies based on two principal parameters: the angular momentum of the externally accreted material and the ratio of mass (or density) of pre-existing and infalling gas.
This schema is summarized in the Tab.~\ref{tab:diversity_schema}.
When external material dominates, it forms either an inner or an outer CR configuration, depending on whether the angular momentum is low or high, respectively.
Such a scenario naturally results in the observed separation between inner and outer CR configurations in the angular momentum histograms shown in Fig.~\ref{fig:in_out_hists}, panel \textit{k} and \textit{l}.
If the pre-existing gas dominates, then no CR formation occurs in the case of low momentum accretion, and possibly an external CR ring can form in the case of high momentum accretion (similar to that formed in NGC~254 \citep{Katkov2022A&A...658A.154K_n254}).
In scenarios where the mass of the external and pre-existing materials is comparable, the picture can be more complex.
With low momentum accretion, the infalling gas collides with the pre-existing gas, losing momentum and spiraling toward the galaxy center, potentially forming an inner CR.
This process is also likely to result in substantial gas deposition at the center, which could trigger AGN activity or a central starburst.
Numerical simulations \citep{Cenci2024ApJ...961L..40C,Peirani2025arXiv250217902P} suggest that, at this stage, a significant fraction of the pre-existing gas can be converted into stars, facilitating subsequent external gas accretion.
In the case of high momentum accretion, both inner and outer CR configurations might emerge.

The accretion scenario underlying our proposed schema aligns with the results of cosmological TNG100 simulations \citep{khoperskov_illustris}.
According to the study of the galaxies with stellar CR found in the results of TNG100, the CR components could be formed by gas accretion from cosmological filaments or gas-rich CGM in the outskirts. 
Also, according to this study, 88 percent of the model galaxies with CR show the CR components as more extended than the host ones, with the CR components consistently younger than the host components across all considered models. In contrast, our sample reveals a significantly lower fraction of outer CR. 
This discrepancy may be partly explained by the selection criterion used in \cite{khoperskov_illustris}, where models were chosen if at least 30\% of their stars rotated in the opposite direction to the main stellar disc.
At the same time, our study suggests that inner CR galaxies tend to have a lower mass for the CR components (see Fig.~\ref{fig:in_out_hists}), therefore, such strict criteria for identifying CR galaxies in simulation might overlook some inner CR.
Another potential explanation for the difference lies in the nature of the data: observations provide luminosity-weighted stellar properties, whereas simulations deal with stellar mass particles.
Additionally, the limited field of view of MaNGA spectra ($\approx 1.5-2.5$ R$_\textrm{eff}$) may result in some unclear cases actually being outer CR.

In a more recent study based on the high-resolution zoom-in simulation \textsc{NewHorizon}, focused on field environments, \citet{Peirani2025arXiv250217902P} identified ten CR galaxies whose CR stars are typically more metal-rich than the co-rotating counterparts and concentrated in the inner regions.
These CR galaxies formed through interactions with satellite galaxies and closely resemble our inner CR systems located at the metal-rich end of the distribution.
Outer CR configurations are also reproduced in these simulations and, in most cases, are associated with major or multiple mergers.

\subsection{On the CR galaxies with LSB structures}
\label{sec:discussion_LSB}

Among the galaxies identified with kinematically misaligned subsystems in our initial inspection (Sec.~\ref{sec:vis_inspec}), we detected extended LSB structures in 20 misaligned galaxies by visual inspection of  the deep images of the Legacy Survey. 
Specifically, among galaxies with stellar CR disks, we observed such structures in 2 inner and 6 outer systems.

Despite having smaller sizes the galaxies with LSB peripheries morphologically resemble giant low-surface brightness galaxies (gLSBGs) --- systems with highly extended LSB disks, reaching radii of up to 130 kpc and containing dynamical masses of up to $10^{12}$~{\Ms} within these disks \citep{Boissier2016}. 
The similarity between gLSBGs and galaxies with CR and LSB structures becomes even more striking, as some gLSBGs also exhibit CR \citep{saburovaetal2021}.
UGC~1382 (1-35391), a well-known gLSBG also identified during the inspection of MaNGA data in this paper, exhibits global counter-rotation of ionized gas and neutral hydrogen relative to its inner stellar component. However, \citet{saburovaetal2021} did not find clear signs of stellar CR in this galaxy, which might indicate  that CR stars in the LSB disk are too faint to be detected in the region dominated by the high surface brightness disk, where spectra were obtained.
UGC~1922 is another gLSBG that hosts kinematically decoupled structures but does not show clear evidence of stellar CR \citep{saburova2018}.

Low star formation efficiency may explain why LSB disks retain their low brightness (even in the presence of cold gas as in UGC~1382 \citep{saburovaetal2021}), and consequently fail to exhibit stellar CR despite the presence of CR gas.
Although measurements of metal content in the ISM of gLSBGs galaxies are limited, the available data indicate a metal deficiency \citep{Saburova2024} supporting this idea.
Recently studied in \citet{Katkov2022A&A...658A.154K_n254} NGC~254, not a gLSBG, but this galaxy hosted an extended outer ring with global gaseous counter-rotation is an example of how low star formation (SFR $\approx 0.02$~\Msperyr) can result in a hidden CR stellar population with a mass fraction of about 1\%, making it nearly impossible to detect purely spectroscopically.

\cite{Saburovaetal2023} proposed that gLSBGs may not represent a distinct class of peculiar systems but rather an extension of normal-sized spirals toward larger radii.
This idea is also supported by \citet{ManceraPina2021}, who found that gLSBGs are not outliers and follow the same relationship between baryonic specific angular momentum, mass and gas fraction as other disk galaxies, despite their notably high angular momentum.
As illustrated in Fig.~\ref{fig:in_out_hists}~(\textit{l}), outer CR components exhibit the highest angular momentum, further emphasizing a possible evolutionary link between gLSBGs and stellar CR galaxies.

\subsection{The origin of accreted material}

\begin{figure}
    \centering
    \includegraphics[width=0.48\textwidth]{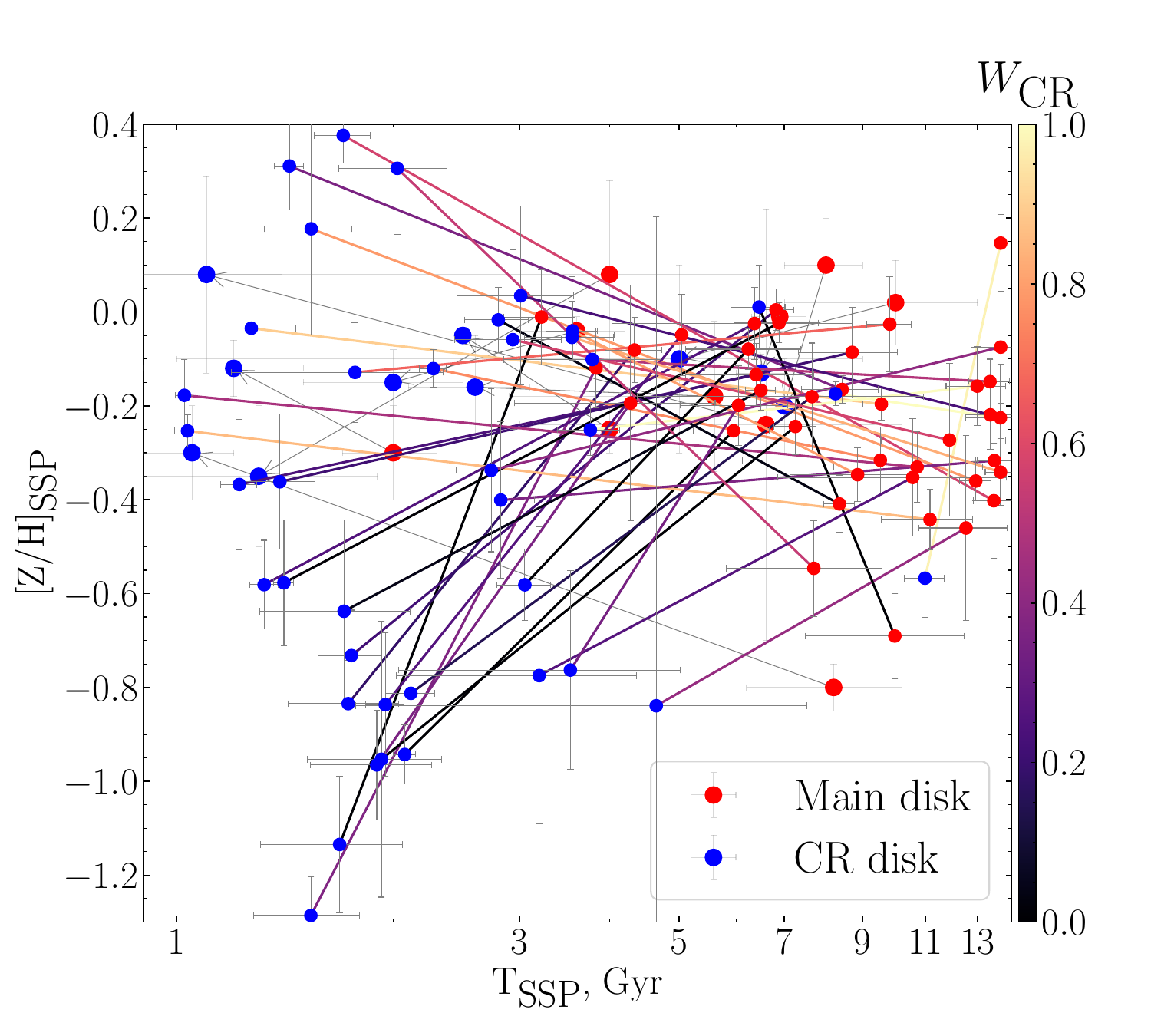}
\caption{SSP-equivalent age --- stellar metallicity diagram for the main (red dots) and CR disks (blue dots). The color of the connecting lines represents the luminosity weight of the CR disk (Tab.~\ref{tab:np_losvd_2comp}), while previously studied CR galaxies (see Tab.~\ref{tab:cr_galaxies}) are shown in the background with grey lines.
\label{fig:age-met}}
\end{figure} 

External accretion of significant amounts of gas is the primary mechanism proposed for the formation of kinematically misaligned components.
This accreted gas may originate from cosmic filaments \citep{Algorry2014MNRAS.437.3596A, khoperskov_illustris}, mergers with gas-rich companions \citep{Thakar1996ApJ...461...55T, Thakar1998ApJ...506...93T, Khim2021ApJS..254...27K, Lu2021MNRAS.503..726L}, or tidal interactions and gas exchange with neighboring galaxies \citep{Khim2021ApJS..254...27K, Sil'chenko2023Galax..11..119S}.
In this section, we aim to determine the most likely origin of the accreted gas responsible for forming the CR components in our sample.

Gas-phase metallicity and stellar metallicity are the key parameters for determining the origin of the accreted gas.
Low stellar metallicity in the CR disk is expected if it formed from gas supplied by cosmic filaments.
These filaments deliver extremely metal-poor, pristine gas to a galaxy.
While the accreting cosmic streams may become slightly enriched through interactions with pre-enriched hot halo gas or outflowing winds \citep{vandeVoort2012MNRAS.423.2991V, Hafen2017MNRAS.469.2292H}, the gas that ultimately settles into the disk and forms the CR disk should largely retain its low metallicity.
The accreted material from other galaxies is expected to be pre-enriched, with gas-rich small companions generally exhibiting low metallicities, while major tidal gas exchanges involve material from massive galaxies that might have approximately solar metallicity.

To constrain the gas source, a detailed chemical evolution model with many parameters is required, such as the toy model applied to galaxy 1-179561 \citep[PGC066551,][]{Katkov2024ApJ...962...27K}.
However, such analysis is beyond the scope of the current study and will be addressed in future papers.
Here, we note that gas-phase metallicity measurement reflects the current chemical abundance of the gas, while the stellar metallicity of the CR disk corresponds to a luminosity-weighted average of the gas-phase metallicity over the disk's star formation history.

\begin{figure}
\centering
\includegraphics[width=0.49\textwidth]{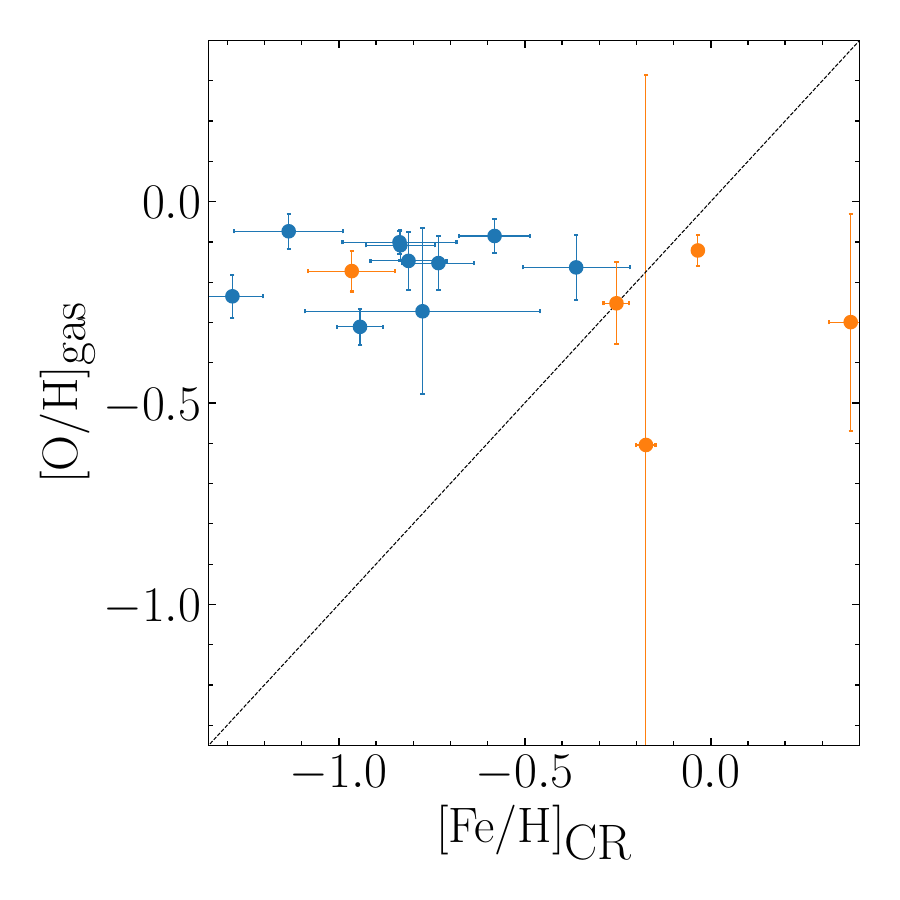}
\caption{
Stellar vs. gas-phase metallicity of CR disks.
Metallicity is given relative to $12+\log$(O/H)$_\odot = 8.69$~dex \citep{Asplund2021A&A...653A.141A}.
Blue and orange symbols represent inner and outer configurations, respectively.
Stellar metallicity is derived from two-component decomposition (Sec.~\ref{sec:2comp_nb}).
For consistency, gas-phase metallicity was calculated using the R-calibration \citep{Pilyugin2016MNRAS.457.3678P} in the same spatial bins where two-component decomposition was conducted, rather than within $R_\mathrm{eff}$ as used in Sec.~\ref{sec:ion_gas}.
\label{fig:MetCR_Metgas}}
\end{figure}

In our sample of CR galaxies, the stellar metallicity of the CR component lies within a broad range, from extremely metal-poor ($\approx~-1.2$~dex) to super-solar ($\approx~0.4$~dex) values (Fig.~\ref{fig:age-met}).
In contrast, the gas-phase metallicity shows a much narrower range, from $\approx-0.3$ dex to $\approx0$ dex (Fig.~\ref{fig:MZR} and Fig.~\ref{fig:MetCR_Metgas}).
For most galaxies, the gas-phase metallicity is noticeably higher than the stellar metallicity in the CR disk.
This configuration is a natural result of the chemical evolution and enrichment of metal-poor gas accreted in the past.
However, for a few galaxies, which turned out to be outer CR galaxies, the gas-phase metallicity is comparable to or even lower than the stellar population metallicity in the CR disk.
This strongly suggests recent or ongoing accretion of low-metallicity gas into these galaxies.
It is important to note that this conclusion is based on a small number of outer CR galaxies with available ionized gas measurements. 
However, in general, the wide range of observed metallicities indicates that no single external gas source can universally explain the accretion history of all studied galaxies.

\subsubsection{Galaxy environments} \label{sec:discussion_environment}

\begin{figure*}
    \centering
	\includegraphics[width=0.32\textwidth]{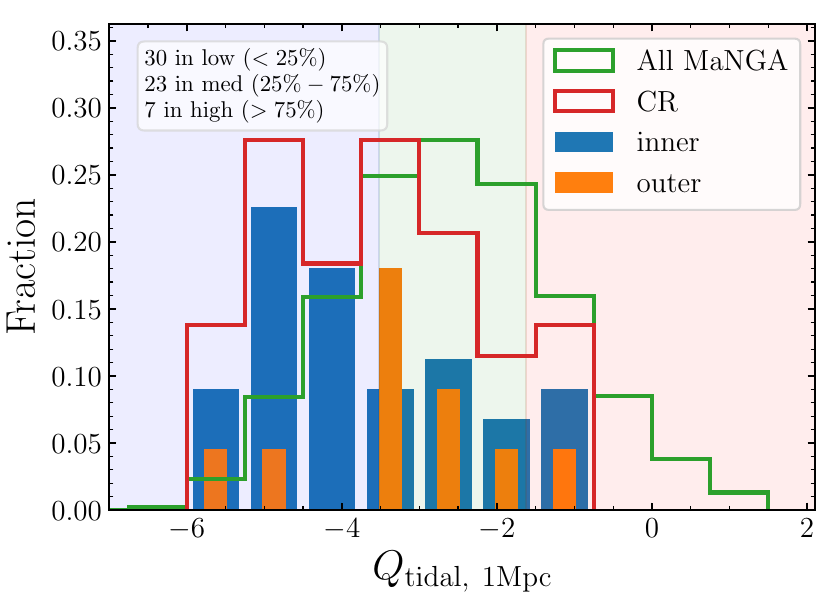}
	\includegraphics[width=0.32\textwidth]{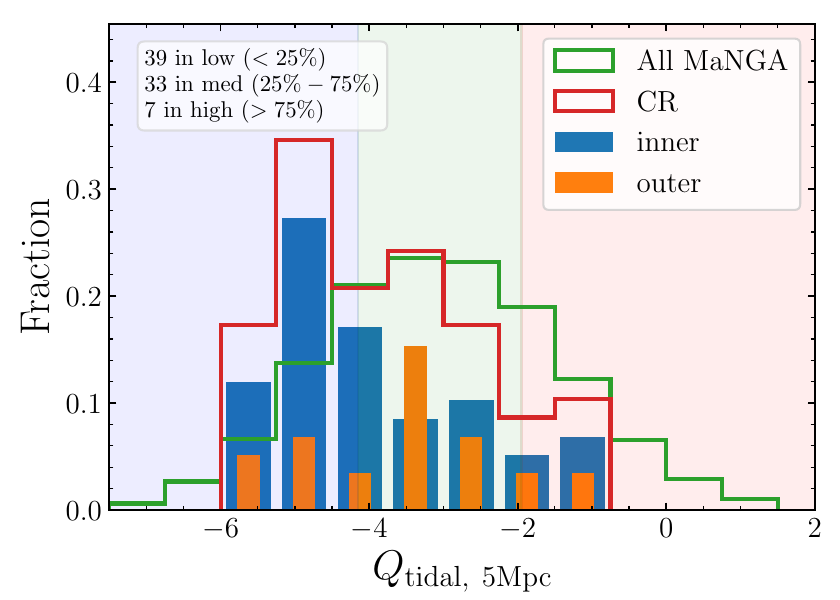}
    \includegraphics[width=0.285\textwidth]{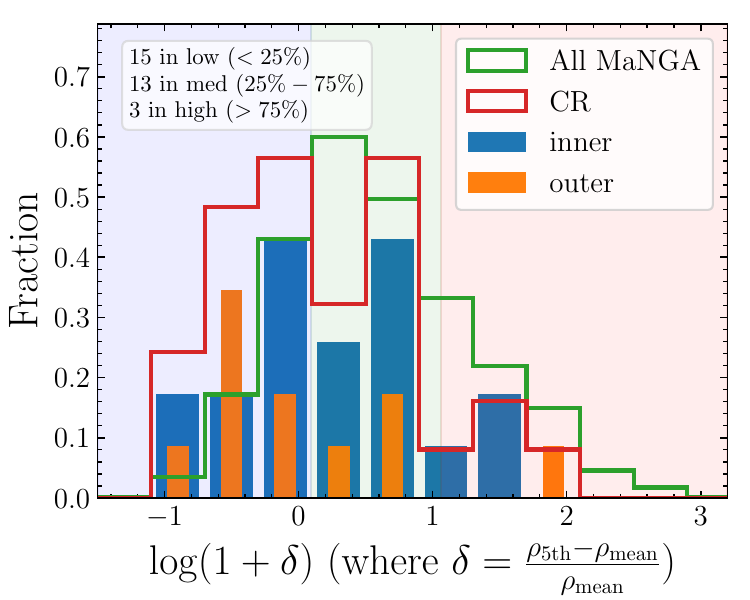}
\caption{Distribution of the environmental parameters from GEMA-VAC (at 1, 5 Mpc, and overdensity). Blue, green, and red zones show sparse, moderate, and dense environments, divided by 25th and 75th percentiles of the parameters. Green histograms represent the properties of all MaNGA galaxies, while red histograms correspond to CR galaxies. Blue and orange histograms show the properties of the inner and outer configurations of the CR disk, respectively. They are normalized to the total number of CR galaxies.\label{fig:Env-1_5Mpc}}
\end{figure*} 

The environmental properties of our galaxies provide additional constraints on possible accretion sources.
Cosmic filaments are viable accretion sources primarily in sparse environments. Conversely, gas exchange with neighboring galaxies requires an intermediate-density environment, as high-density regions are dominated by ram-pressure stripping, which prohibits gas transfer.

We cross-matched our CR sample with the GEMA-VAC, utilizing several parameters to assess environmental density.
These parameters include $Q_{\textrm{1Mpc}}$, $Q_{\textrm{5Mpc}}$ (the description in Sec.~\ref{sec:in_out_CR}), and $\log 1 + \delta$ (where $\delta~=~\dfrac{\rho_{\textrm{5th}} - \rho_{\textrm{m}}}{\rho_{\textrm{m}}}$ represents the difference between the mean density $\rho_{\textrm{m}}$ around the galaxy and the environmental density $\rho_{\textrm{5th}}$ to the 5th nearest neighbor; see details in \cite{Etherington2015MNRAS.451..660E}).
The galaxies in our CR sample are generally found in slightly sparser environments than those in the overall MaNGA sample (see Fig.~\ref{fig:Env-1_5Mpc}), which aligns with our earlier findings of a high incidence of kinematic misalignments in isolated lenticular galaxies \citep{Katkov2014MNRAS.438.2798K, Katkov2015AJ....150...24K}.
However, the broad environmental distribution prevents us from conclusively ruling out any of the proposed above gas sources.

The distributions of $Q_\mathrm{tidal,1Mpc}$ and $Q_\mathrm{tidal,5Mpc}$ (Fig.~\ref{fig:Env-1_5Mpc}) show distinct peaks, with one peak associated with galaxies hosting inner CR disk configurations and the other with outer configurations.
We propose that these peaks reflect different dominant accretion scenarios.
For inner CR disks, accretion of pristine gas from the circumgalactic medium (CGM) in low-density environments could explain the observed peak.
This scenario involves the accretion of low-angular-momentum gas, which facilitates the formation of inner disks and leads to the low stellar metallicities observed in these configurations.
In contrast, some outer CR disks are likely formed through tidal gas exchange with a nearby massive galaxy, a process occurring in intermediate-density environments.
This gas exchange delivers high-angular-momentum material, facilitating the formation of an outer CR disk or a blue star-forming ring.
The accretion of pre-enriched gas leads to the formation of more metal-rich CR disks, which we generally observe in outer CR configurations compared to their inner counterparts.

\vspace{2em}
Summarizing, the sample of CR galaxies displays a wide range of parameters, making it challenging to pinpoint one preferred mechanism for CR disk formation.
While accretion from cosmic filaments and tidal interactions may play important roles in sparse and intermediate-density environments, respectively, we highlight gas-rich companion accretion as a versatile mechanism capable of operating across a wide range of environments -- excluding regions of extreme over-density, which lack CR galaxies.

\section{Conclusions}
\label{sec:conclusions}

In this paper, we present a comprehensive study of stellar counter-rotation in galaxies using the MaNGA spectroscopic survey.
Through a combination of visual inspection and analysis of non-parametric recovered stellar LOSVD, we identified a significantly expanded sample of 120 counter-rotating (CR) galaxies, including 65 \textit{reliable} and 55 \textit{probable} systems.
Of these, 74 galaxies have not been reported in previous studies.
Our sample of CR galaxies represents $\approx 1.5\%$ of all ETGs ($\approx 2.9\%$ of S0 galaxies) and $\approx 0.8\%$ of all LTGs in the Local Universe.
Our conclusions can be summarized as follows:

\begin{enumerate}
    \item In addition to the CR galaxies, we identified 588 galaxies with kinematic misalignments (including 120 galaxies with stellar CR). This entire sample includes 207 galaxies with gas-star counter-rotation (145 excluding stellar CR systems), 119 with polar gas disks (116 excluding stellar CR systems), and 173 with intermediate gas misalignments (172 excluding stellar CR galaxies). Of these kinematically misaligned galaxies, approximately 68\% are early-type galaxies (ETGs), accounting for about 10.6\% of all ETGs in the MaNGA. The remaining 32\% are late-type galaxies (LTGs), corresponding to roughly 3.1\% of all LTGs in MaNGA.
    
    \item Despite the limitations of MaNGA spectral data, we successfully performed spectral decomposition and derived the luminosity-weighted stellar population properties (ages and metallicities) for both disks in 39 out of 120 identified CR galaxies. Our analysis revealed that CR galaxies form a diverse family, exhibiting a wide range of ages and metallicities. However, in all cases, the gas component was found co-rotated with the younger stellar population, reinforcing their evolutionary connection.

    \item We demonstrated the bimodality of spatial configurations in CR galaxies: \textit{inner}, where the CR disk dominates in the central part of the galaxy, and \textit{outer}, where CR stars dominate in the galaxy periphery. Through stellar population analysis, we clearly established a distinction in stellar mass and angular momentum between the inner and outer CR disks.

    \item We detected extended LSB features in 20 kinematically misaligned galaxies (11 with stellar CR). Among them, some systems with stellar CR predominantly showing the outer CR configuration. These galaxies show morphological similarities with the gLSBGs -- systems with very large low-surface brightness disks exhibiting outer configuration in the known cases with counter-rotation. It can indicate that similar processes may drive the formation of both giant disks and CR systems with LSB features.

    \item
    CR galaxies are found in diverse environments but generally tend to reside in sparse environments. The stellar metallicities of CR disks span a broad range. These observational properties make it challenging to identify a single external source of material responsible for CR disk formation. Possible scenarios include accretion from cosmological filaments, mergers with gas-rich satellites, and material exchange with neighboring galaxies. However, mergers with gas-rich satellites may represent a more universal mechanism, as they can occur across different environments and, depending on the prior history of metal enrichment, can produce a wide range of final metallicities in the CR disk.

\end{enumerate}

\section{Acknowledgments}
We thank Andrea Macci\`{o} and Michael Blanton for valuable comments and discussion on this study.
The initial work of DG and IK on the sample construction and the analysis of CR galaxies was supported by the Russian Science Foundation grant No. 21-72-00036.
DG, ER, and AZ research on the analysis, comparison with LSB galaxies, and interpretation of the properties of stellar populations in CR galaxies was supported by the Russian Science Foundation (RSCF) grants No.~23-12-00146.
The authors acknowledge partial support from M.V.Lomonosov Moscow State University Program of Development
Spectral observations reported in this paper were obtained with the Southern African Large Telescope (program 2020-1-SCI-002) supported by the National Research Foundation (NRF) of South Africa.
This material is based upon work supported by Tamkeen under the NYU Abu Dhabi Research Institute grant CASS.
Funding for the Sloan Digital Sky Survey IV has been provided by the Alfred P. Sloan Foundation, the U.S. Department of Energy Office of Science, and the Participating Institutions. SDSS-IV acknowledges support and resources from the Center for High-Performance Computing at the University of Utah. The SDSS website is www.sdss.org.

SDSS-IV is managed by the Astrophysical Research Consortium for the Participating Institutions of the SDSS Collaboration including the Brazilian Participation Group, the Carnegie Institution for Science, Carnegie Mellon University, the Chilean Participation Group, the French Participation Group, Harvard-Smithsonian Center for Astrophysics, Instituto de Astrof\'isica de Canarias, The Johns Hopkins University, Kavli Institute for the Physics and Mathematics of the Universe (IPMU) / University of Tokyo, the Korean Participation Group, Lawrence Berkeley National Laboratory, Leibniz Institut für Astrophysik Potsdam (AIP), Max-Planck-Institut für Astronomie (MPIA Heidelberg), Max-Planck-Institut für Astrophysik (MPA Garching), Max-Planck-Institut für Extraterrestrische Physik (MPE), National Astronomical Observatories of China, New Mexico State University, New York University, University of Notre Dame, Observat\'ario Nacional / MCTI, The Ohio State University, Pennsylvania State University, Shanghai Astronomical Observatory, United Kingdom Participation Group, Universidad Nacional Aut\'onoma de M\'exico, University of Arizona, University of Colorado Boulder, University of Oxford, University of Portsmouth, University of Utah, University of Virginia, University of Washington, University of Wisconsin, Vanderbilt University, and Yale University.

This work made use of Astropy:\footnote{http://www.astropy.org} a community-developed core Python package and an ecosystem of tools and resources for astronomy \citep{astropy:2013, astropy:2018, astropy:2022}.

This research has made use of the NASA/IPAC Extragalactic Database (NED), which is funded by the National Aeronautics and Space Administration and operated by the California Institute of Technology. This research has made use of NASA's Astrophysics Data System Bibliographic Services.

%

\vspace{5mm}
\facilities{SDSS(MaNGA)}


\software{
    Astropy \citep{astropy:2013, astropy:2018, astropy:2022},
    lmfit \citep[v1.0.3,][]{Newville2014zndo.....11813N}
}




\bibliography{bib}{}

@ARTICLE{2018PASJ...70S...1M,
       author = {{Miyazaki}, Satoshi and {Komiyama}, Yutaka and {Kawanomoto}, Satoshi and {Doi}, Yoshiyuki and {Furusawa}, Hisanori and {Hamana}, Takashi and {Hayashi}, Yusuke and {Ikeda}, Hiroyuki and {Kamata}, Yukiko and {Karoji}, Hiroshi and {Koike}, Michitaro and {Kurakami}, Tomio and {Miyama}, Shoken and {Morokuma}, Tomoki and {Nakata}, Fumiaki and {Namikawa}, Kazuhito and {Nakaya}, Hidehiko and {Nariai}, Kyoji and {Obuchi}, Yoshiyuki and {Oishi}, Yukie and {Okada}, Norio and {Okura}, Yuki and {Tait}, Philip and {Takata}, Tadafumi and {Tanaka}, Yoko and {Tanaka}, Masayuki and {Terai}, Tsuyoshi and {Tomono}, Daigo and {Uraguchi}, Fumihiro and {Usuda}, Tomonori and {Utsumi}, Yousuke and {Yamada}, Yoshihiko and {Yamanoi}, Hitomi and {Aihara}, Hiroaki and {Fujimori}, Hiroki and {Mineo}, Sogo and {Miyatake}, Hironao and {Oguri}, Masamune and {Uchida}, Tomohisa and {Tanaka}, Manobu M. and {Yasuda}, Naoki and {Takada}, Masahiro and {Murayama}, Hitoshi and {Nishizawa}, Atsushi J. and {Sugiyama}, Naoshi and {Chiba}, Masashi and {Futamase}, Toshifumi and {Wang}, Shiang-Yu and {Chen}, Hsin-Yo and {Ho}, Paul T.~P. and {Liaw}, Eric J.~Y. and {Chiu}, Chi-Fang and {Ho}, Cheng-Lin and {Lai}, Tsang-Chih and {Lee}, Yao-Cheng and {Jeng}, Dun-Zen and {Iwamura}, Satoru and {Armstrong}, Robert and {Bickerton}, Steve and {Bosch}, James and {Gunn}, James E. and {Lupton}, Robert H. and {Loomis}, Craig and {Price}, Paul and {Smith}, Steward and {Strauss}, Michael A. and {Turner}, Edwin L. and {Suzuki}, Hisanori and {Miyazaki}, Yasuhito and {Muramatsu}, Masaharu and {Yamamoto}, Koei and {Endo}, Makoto and {Ezaki}, Yutaka and {Ito}, Noboru and {Kawaguchi}, Noboru and {Sofuku}, Satoshi and {Taniike}, Tomoaki and {Akutsu}, Kotaro and {Dojo}, Naoto and {Kasumi}, Kazuyuki and {Matsuda}, Toru and {Imoto}, Kohei and {Miwa}, Yoshinori and {Suzuki}, Masayuki and {Takeshi}, Kunio and {Yokota}, Hideo},
        title = "{Hyper Suprime-Cam: System design and verification of image quality}",
      journal = {\pasj},
         year = 2018,
        month = jan,
       volume = {70},
          eid = {S1},
        pages = {S1},
          doi = {10.1093/pasj/psx063},
       adsurl = {https://ui.adsabs.harvard.edu/abs/2018PASJ...70S...1M}
}

@ARTICLE{saburova2018,
       author = {{Saburova}, Anna S. and {Chilingarian}, Igor V. and {Katkov}, Ivan Yu and {Egorov}, Oleg V. and {Kasparova}, Anastasia V. and {Khoperskov}, Sergey A. and {Uklein}, Roman I. and {Vozyakova}, Olga V.},
        title = "{A Malin 1 `cousin' with counter-rotation: internal dynamics and stellar content of the giant low surface brightness galaxy UGC 1922}",
      journal = {\mnras},
         year = 2018,
        month = dec,
       volume = {481},
       number = {3},
        pages = {3534-3547},
          doi = {10.1093/mnras/sty2519},
archivePrefix = {arXiv},
       eprint = {1809.04333},
 primaryClass = {astro-ph.GA},
       adsurl = {https://ui.adsabs.harvard.edu/abs/2018MNRAS.481.3534S}
}

@ARTICLE{saburovaetal2021,
       author = {{Saburova}, Anna S. and {Chilingarian}, Igor V. and {Kasparova}, Anastasia V. and {Sil'chenko}, Olga K. and {Grishin}, Kirill A. and {Katkov}, Ivan Yu and {Uklein}, Roman I.},
        title = "{Observational insights on the origin of giant low surface brightness galaxies}",
      journal = {\mnras},
         year = 2021,
        month = may,
       volume = {503},
       number = {1},
        pages = {830-849},
          doi = {10.1093/mnras/stab374},
archivePrefix = {arXiv},
       eprint = {2011.01238},
 primaryClass = {astro-ph.GA},
       adsurl = {https://ui.adsabs.harvard.edu/abs/2021MNRAS.503..830S}
}

@ARTICLE{Saburovaetal2023,
       author = {{Saburova}, Anna S. and {Chilingarian}, Igor V. and {Kulier}, Andrea and {Galaz}, Gaspar and {Grishin}, Kirill A. and {Kasparova}, Anastasia V. and {Toptun}, Victoria and {Katkov}, Ivan Yu},
        title = "{The volume density of giant low surface brightness galaxies}",
      journal = {\mnras},
         year = 2023,
        month = mar,
       volume = {520},
       number = {1},
        pages = {L85-L90},
          doi = {10.1093/mnrasl/slad005},
archivePrefix = {arXiv},
       eprint = {2209.09906},
 primaryClass = {astro-ph.GA},
       adsurl = {https://ui.adsabs.harvard.edu/abs/2023MNRAS.520L..85S}
}

@ARTICLE{Boissier2016,
   author = {{Boissier}, S. and {Boselli}, A. and {Ferrarese}, L. and {C{\^o}t{\'e}}, P. and
	{Roehlly}, Y. and {Gwyn}, S.~D.~J. and {Cuillandre}, J.-C. and
	{Roediger}, J. and {Koda}, J. and {Mu{\~n}os Mateos}, J.~C. and
	{Gil de Paz}, A. and {Madore}, B.~F.},
    title = "{The properties of the Malin 1 galaxy giant disk. A panchromatic view from the NGVS and GUViCS surveys}",
  journal = {\aap},
archivePrefix = "arXiv",
   eprint = {1610.00918},
     year = 2016,
    month = oct,
   volume = 593,
      eid = {A126},
    pages = {A126},
      doi = {10.1051/0004-6361/201629226},
   adsurl = {http://adsabs.harvard.edu/abs/2016A%26A...593A.126B}
}

@ARTICLE{hsc2019,
       author = {{Aihara}, Hiroaki and {AlSayyad}, Yusra and {Ando}, Makoto and
         {Armstrong}, Robert and {Bosch}, James and {Egami}, Eiichi and
         {Furusawa}, Hisanori and {Furusawa}, Junko and {Goulding}, Andy and
         {Harikane}, Yuichi and {Hikage}, Chiaki and {Ho}, Paul T.~P. and
         {Hsieh}, Bau-Ching and {Huang}, Song and {Ikeda}, Hiroyuki and
         {Imanishi}, Masatoshi and {Ito}, Kei and {Iwata}, Ikuru and
         {Jaelani}, Anton T. and {Kakuma}, Ryota and {Kawana}, Kojiro and
         {Kikuta}, Satoshi and {Kobayashi}, Umi and {Koike}, Michitaro and
         {Komiyama}, Yutaka and {Li}, Xiangchong and {Liang}, Yongming and
         {Lin}, Yen-Ting and {Luo}, Wentao and {Lupton}, Robert and
         {Lust}, Nate B. and {MacArthur}, Lauren A. and {Matsuoka}, Yoshiki and
         {Mineo}, Sogo and {Miyatake}, Hironao and {Miyazaki}, Satoshi and
         {More}, Surhud and {Murata}, Ryoma and {Namiki}, Shigeru V. and
         {Nishizawa}, Atsushi J. and {Oguri}, Masamune and {Okabe}, Nobuhiro and
         {Okamoto}, Sakurako and {Okura}, Yuki and {Ono}, Yoshiaki and
         {Onodera}, Masato and {Onoue}, Masafusa and {Osato}, Ken and
         {Ouchi}, Masami and {Shibuya}, Takatoshi and {Strauss}, Michael A. and
         {Sugiyama}, Naoshi and {Suto}, Yasushi and {Takada}, Masahiro and
         {Takagi}, Yuhei and {Takata}, Tadafumi and {Takita}, Satoshi and
         {Tanaka}, Masayuki and {Terai}, Tsuyoshi and {Toba}, Yoshiki and
         {Uchiyama}, Hisakazu and {Utsumi}, Yousuke and {Wang}, Shiang-Yu and
         {Wang}, Wenting and {Yamada}, Yoshihiko},
        title = "{Second data release of the Hyper Suprime-Cam Subaru Strategic Program}",
      journal = {\pasj},
         year = 2019,
        month = dec,
       volume = {71},
       number = {6},
          eid = {114},
        pages = {114},
          doi = {10.1093/pasj/psz103},
archivePrefix = {arXiv},
       eprint = {1905.12221},
 primaryClass = {astro-ph.IM},
       adsurl = {https://ui.adsabs.harvard.edu/abs/2019PASJ...71..114A}
}

@INPROCEEDINGS{Putman2017ASSL..430....1P,
       author = {{Putman}, Mary E.},
        title = "{An Introduction to Gas Accretion onto Galaxies}",
    booktitle = {Gas Accretion onto Galaxies},
         year = 2017,
       editor = {{Fox}, Andrew and {Dav{\'e}}, Romeel},
       series = {Astrophysics and Space Science Library},
       volume = {430},
        month = jan,
        pages = {1},
          doi = {10.1007/978-3-319-52512-9_1},
archivePrefix = {arXiv},
       eprint = {1612.00461},
 primaryClass = {astro-ph.GA},
       adsurl = {https://ui.adsabs.harvard.edu/abs/2017ASSL..430....1P}
}

@ARTICLE{Bevacqua2022MNRAS.511..139B,
       author = {{Bevacqua}, Davide and {Cappellari}, Michele and {Pellegrini}, Silvia},
        title = "{SDSS-IV MaNGA: integral-field kinematics and stellar population of a sample of galaxies with counter-rotating stellar discs selected from about 4000 galaxies}",
      journal = {\mnras},
         year = 2022,
        month = mar,
       volume = {511},
       number = {1},
        pages = {139-157},
          doi = {10.1093/mnras/stab3732},
archivePrefix = {arXiv},
       eprint = {2107.09528},
 primaryClass = {astro-ph.GA},
       adsurl = {https://ui.adsabs.harvard.edu/abs/2022MNRAS.511..139B}
}

@ARTICLE{Bao2022ApJ...926L..13B,
       author = {{Bao}, Min and {Chen}, Yanmei and {Zhu}, Pengpei and {Shi}, Yong and {Bizyaev}, Dmitry and {Zhu}, Ling and {Yang}, Meng and {Beom}, Minje and {Brownstein}, Joel R. and {Lane}, Richard R.},
        title = "{Different Formation Scenarios for Counterrotating Stellar Disks in Nearby Galaxies}",
      journal = {\apjl},
         year = 2022,
        month = feb,
       volume = {926},
       number = {2},
          eid = {L13},
        pages = {L13},
          doi = {10.3847/2041-8213/ac52ad},
       adsurl = {https://ui.adsabs.harvard.edu/abs/2022ApJ...926L..13B}
}

@ARTICLE{Algorry2014MNRAS.437.3596A,
       author = {{Algorry}, David G. and {Navarro}, Julio F. and {Abadi}, Mario G. and
         {Sales}, Laura V. and {Steinmetz}, Matthias and {Piontek}, Franziska},
        title = "{Counterrotating stars in simulated galaxy discs}",
      journal = {\mnras},
         year = 2014,
        month = feb,
       volume = {437},
       number = {4},
        pages = {3596-3602},
          doi = {10.1093/mnras/stt2154},
archivePrefix = {arXiv},
       eprint = {1311.1215},
 primaryClass = {astro-ph.CO},
       adsurl = {https://ui.adsabs.harvard.edu/abs/2014MNRAS.437.3596A}
}

@ARTICLE{Thakar1996ApJ...461...55T,
       author = {{Thakar}, Aniruddha R. and {Ryden}, Barbara S.},
        title = "{Formation of Massive Counterrotating Disks in Spiral Galaxies}",
      journal = {\apj},
         year = 1996,
        month = apr,
       volume = {461},
        pages = {55},
          doi = {10.1086/177037},
archivePrefix = {arXiv},
       eprint = {astro-ph/9510053},
 primaryClass = {astro-ph},
       adsurl = {https://ui.adsabs.harvard.edu/abs/1996ApJ...461...55T}
}

@ARTICLE{Thakar1998ApJ...506...93T,
       author = {{Thakar}, Aniruddha R. and {Ryden}, Barbara S.},
        title = "{Smoothed Particle Hydrodynamics Simulations of Counterrotating Disk Formation in Spiral Galaxies}",
      journal = {\apj},
         year = 1998,
        month = oct,
       volume = {506},
       number = {1},
        pages = {93-115},
          doi = {10.1086/306223},
archivePrefix = {arXiv},
       eprint = {astro-ph/9805082},
 primaryClass = {astro-ph},
       adsurl = {https://ui.adsabs.harvard.edu/abs/1998ApJ...506...93T}
}

@ARTICLE{Mitzkus2017MNRAS.464.4789M,
       author = {{Mitzkus}, Martin and {Cappellari}, Michele and {Walcher}, C. Jakob},
        title = "{Dominant dark matter and a counter-rotating disc: MUSE view of the low-luminosity S0 galaxy NGC 5102}",
      journal = {\mnras},
         year = 2017,
        month = feb,
       volume = {464},
       number = {4},
        pages = {4789-4806},
          doi = {10.1093/mnras/stw2677},
archivePrefix = {arXiv},
       eprint = {1610.04516},
 primaryClass = {astro-ph.GA},
       adsurl = {https://ui.adsabs.harvard.edu/abs/2017MNRAS.464.4789M}
}

@ARTICLE{Johnston2013MNRAS.428.1296J,
       author = {{Johnston}, Evelyn J. and {Merrifield}, Michael R. and
         {Arag{\'o}n-Salamanca}, Alfonso and {Cappellari}, Michele},
        title = "{Disentangling the stellar populations in the counter-rotating disc galaxy NGC 4550}",
      journal = {\mnras},
         year = 2013,
        month = jan,
       volume = {428},
       number = {2},
        pages = {1296-1302},
          doi = {10.1093/mnras/sts121},
archivePrefix = {arXiv},
       eprint = {1210.0535},
 primaryClass = {astro-ph.CO},
       adsurl = {https://ui.adsabs.harvard.edu/abs/2013MNRAS.428.1296J}
}

@ARTICLE{Khrapov2024OAst...3320231K,
       author = {{Khrapov}, Sergey S. and {Khoperskov}, Alexander V.},
        title = "{Retrograde infall of the intergalactic gas onto S-galaxy and activity of galactic nuclei}",
      journal = {Open Astronomy},
         year = 2024,
        month = mar,
       volume = {33},
       number = {1},
          eid = {20220231},
        pages = {20220231},
          doi = {10.1515/astro-2022-0231},
       adsurl = {https://ui.adsabs.harvard.edu/abs/2024OAst...3320231K}
}

@ARTICLE{Egorov2019MNRAS.486.4186E,
       author = {{Egorov}, Oleg V. and {Moiseev}, Alexei V.},
        title = "{Metallicity and ionization state of the gas in polar-ring galaxies}",
      journal = {\mnras},
         year = 2019,
        month = jul,
       volume = {486},
       number = {3},
        pages = {4186-4197},
          doi = {10.1093/mnras/stz1112},
archivePrefix = {arXiv},
       eprint = {1904.02513},
 primaryClass = {astro-ph.GA},
       adsurl = {https://ui.adsabs.harvard.edu/abs/2019MNRAS.486.4186E}
}

@ARTICLE{kewley01,
       author = {{Kewley}, L.~J. and {Dopita}, M.~A. and {Sutherland}, R.~S. and {Heisler}, C.~A. and {Trevena}, J.},
        title = "{Theoretical Modeling of Starburst Galaxies}",
      journal = {\apj},
         year = 2001,
        month = jul,
       volume = {556},
       number = {1},
        pages = {121-140},
          doi = {10.1086/321545},
archivePrefix = {arXiv},
       eprint = {astro-ph/0106324},
 primaryClass = {astro-ph},
       adsurl = {https://ui.adsabs.harvard.edu/abs/2001ApJ...556..121K}
}

@ARTICLE{kauffmann03,
       author = {{Kauffmann}, Guinevere and {Heckman}, Timothy M. and {Tremonti}, Christy and {Brinchmann}, Jarle and {Charlot}, St{\'e}phane and {White}, Simon D.~M. and {Ridgway}, Susan E. and {Brinkmann}, Jon and {Fukugita}, Masataka and {Hall}, Patrick B. and {Ivezi{\'c}}, {\v{Z}}eljko and {Richards}, Gordon T. and {Schneider}, Donald P.},
        title = "{The host galaxies of active galactic nuclei}",
      journal = {\mnras},
         year = 2003,
        month = dec,
       volume = {346},
       number = {4},
        pages = {1055-1077},
          doi = {10.1111/j.1365-2966.2003.07154.x},
archivePrefix = {arXiv},
       eprint = {astro-ph/0304239},
 primaryClass = {astro-ph},
       adsurl = {https://ui.adsabs.harvard.edu/abs/2003MNRAS.346.1055K}
}

@ARTICLE{Chilingarian2017ApJS..228...14C,
       author = {{Chilingarian}, Igor V. and {Zolotukhin}, Ivan Yu. and {Katkov}, Ivan Yu. and {Melchior}, Anne-Laure and {Rubtsov}, Evgeniy V. and {Grishin}, Kirill A.},
        title = "{RCSED{\textemdash}A Value-added Reference Catalog of Spectral Energy Distributions of 800,299 Galaxies in 11 Ultraviolet, Optical, and Near-infrared Bands: Morphologies, Colors, Ionized Gas, and Stellar Population Properties}",
      journal = {\apjs},
         year = 2017,
        month = feb,
       volume = {228},
       number = {2},
          eid = {14},
        pages = {14},
          doi = {10.3847/1538-4365/228/2/14},
archivePrefix = {arXiv},
       eprint = {1612.02047},
 primaryClass = {astro-ph.GA},
       adsurl = {https://ui.adsabs.harvard.edu/abs/2017ApJS..228...14C}
}

@ARTICLE{Rubino2021A&A...654A..30R,
       author = {{Rubino}, M. and {Pizzella}, A. and {Morelli}, L. and {Coccato}, L. and {Portaluri}, E. and {Debattista}, V.~P. and {Corsini}, E.~M. and {Dalla Bont{\`a}}, E.},
        title = "{Detectability of large-scale counter-rotating stellar disks in galaxies with integral-field spectroscopy}",
      journal = {\aap},
         year = 2021,
        month = oct,
       volume = {654},
          eid = {A30},
        pages = {A30},
          doi = {10.1051/0004-6361/202140702},
archivePrefix = {arXiv},
       eprint = {2107.02226},
 primaryClass = {astro-ph.GA},
       adsurl = {https://ui.adsabs.harvard.edu/abs/2021A&A...654A..30R}
}

@ARTICLE{Dey2019_legacysurveys,
       author = {{Dey}, Arjun and {Schlegel}, David J. and {Lang}, Dustin and
         {Blum}, Robert and {Burleigh}, Kaylan and {Fan}, Xiaohui and
         {Findlay}, Joseph R. and {Finkbeiner}, Doug and {Herrera}, David and
         {Juneau}, St{\'e}phanie and {Landriau}, Martin and {Levi}, Michael and
         {McGreer}, Ian and {Meisner}, Aaron and {Myers}, Adam D. and
         {Moustakas}, John and {Nugent}, Peter and {Patej}, Anna and
         {Schlafly}, Edward F. and {Walker}, Alistair R. and
         {Valdes}, Francisco and {Weaver}, Benjamin A. and
         {Y{\`e}che}, Christophe and {Zou}, Hu and {Zhou}, Xu and
         {Abareshi}, Behzad and {Abbott}, T.~M.~C. and {Abolfathi}, Bela and
         {Aguilera}, C. and {Alam}, Shadab and {Allen}, Lori and {Alvarez}, A. and
         {Annis}, James and {Ansarinejad}, Behzad and {Aubert}, Marie and
         {Beechert}, Jacqueline and {Bell}, Eric F. and {BenZvi}, Segev Y. and
         {Beutler}, Florian and {Bielby}, Richard M. and {Bolton}, Adam S. and
         {Brice{\~n}o}, C{\'e}sar and {Buckley-Geer}, Elizabeth J. and
         {Butler}, Karen and {Calamida}, Annalisa and {Carlberg}, Raymond G. and
         {Carter}, Paul and {Casas}, Ricard and {Castander}, Francisco J. and
         {Choi}, Yumi and {Comparat}, Johan and {Cukanovaite}, Elena and
         {Delubac}, Timoth{\'e}e and {DeVries}, Kaitlin and {Dey}, Sharmila and
         {Dhungana}, Govinda and {Dickinson}, Mark and {Ding}, Zhejie and
         {Donaldson}, John B. and {Duan}, Yutong and
         {Duckworth}, Christopher J. and {Eftekharzadeh}, Sarah and
         {Eisenstein}, Daniel J. and {Etourneau}, Thomas and
         {Fagrelius}, Parker A. and {Farihi}, Jay and {Fitzpatrick}, Mike and
         {Font-Ribera}, Andreu and {Fulmer}, Leah and {G{\"a}nsicke}, Boris T. and
         {Gaztanaga}, Enrique and {George}, Koshy and {Gerdes}, David W. and
         {Gontcho}, Satya Gontcho A. and {Gorgoni}, Claudio and
         {Green}, Gregory and {Guy}, Julien and {Harmer}, Diane and {Hernand
        ez}, M. and {Honscheid}, Klaus and {Huang}, Lijuan Wendy and
         {James}, David J. and {Jannuzi}, Buell T. and {Jiang}, Linhua and
         {Joyce}, Richard and {Karcher}, Armin and {Karkar}, Sonia and
         {Kehoe}, Robert and {Kneib}, Jean-Paul and {Kueter-Young}, Andrea and
         {Lan}, Ting-Wen and {Lauer}, Tod R. and {Le Guillou}, Laurent and
         {Le Van Suu}, Auguste and {Lee}, Jae Hyeon and {Lesser}, Michael and
         {Perreault Levasseur}, Laurence and {Li}, Ting S. and
         {Mann}, Justin L. and {Marshall}, Robert and
         {Mart{\'\i}nez-V{\'a}zquez}, C.~E. and {Martini}, Paul and
         {du Mas des Bourboux}, H{\'e}lion and {McManus}, Sean and
         {Meier}, Tobias Gabriel and {M{\'e}nard}, Brice and {Metcalfe}, Nigel and
         {Mu{\~n}oz-Guti{\'e}rrez}, Andrea and {Najita}, Joan and
         {Napier}, Kevin and {Narayan}, Gautham and {Newman}, Jeffrey A. and
         {Nie}, Jundan and {Nord}, Brian and {Norman}, Dara J. and
         {Olsen}, Knut A.~G. and {Paat}, Anthony and
         {Palanque-Delabrouille}, Nathalie and {Peng}, Xiyan and
         {Poppett}, Claire L. and {Poremba}, Megan R. and {Prakash}, Abhishek and
         {Rabinowitz}, David and {Raichoor}, Anand and {Rezaie}, Mehdi and
         {Robertson}, A.~N. and {Roe}, Natalie A. and {Ross}, Ashley J. and
         {Ross}, Nicholas P. and {Rudnick}, Gregory and {Safonova}, Sasha and
         {Saha}, Abhijit and {S{\'a}nchez}, F. Javier and {Savary}, Elodie and
         {Schweiker}, Heidi and {Scott}, Adam and {Seo}, Hee-Jong and
         {Shan}, Huanyuan and {Silva}, David R. and {Slepian}, Zachary and
         {Soto}, Christian and {Sprayberry}, David and {Staten}, Ryan and
         {Stillman}, Coley M. and {Stupak}, Robert J. and {Summers}, David L. and
         {Sien Tie}, Suk and {Tirado}, H. and {Vargas-Maga{\~n}a}, Mariana and
         {Vivas}, A. Katherina and {Wechsler}, Risa H. and {Williams}, Doug and
         {Yang}, Jinyi and {Yang}, Qian and {Yapici}, Tolga and
         {Zaritsky}, Dennis and {Zenteno}, A. and {Zhang}, Kai and
         {Zhang}, Tianmeng and {Zhou}, Rongpu and {Zhou}, Zhimin},
        title = "{Overview of the DESI Legacy Imaging Surveys}",
      journal = {\aj},
         year = "2019",
        month = "May",
       volume = {157},
       number = {5},
          eid = {168},
        pages = {168},
          doi = {10.3847/1538-3881/ab089d},
archivePrefix = {arXiv},
       eprint = {1804.08657},
 primaryClass = {astro-ph.IM},
       adsurl = {https://ui.adsabs.harvard.edu/abs/2019AJ....157..168D}
}

@article{astropy:2013,
Adsurl = {http://adsabs.harvard.edu/abs/2013A%26A...558A..33A},
Archiveprefix = {arXiv},
Author = {{Astropy Collaboration} and {Robitaille}, T.~P. and {Tollerud}, E.~J. and {Greenfield}, P. and {Droettboom}, M. and {Bray}, E. and {Aldcroft}, T. and {Davis}, M. and {Ginsburg}, A. and {Price-Whelan}, A.~M. and {Kerzendorf}, W.~E. and {Conley}, A. and {Crighton}, N. and {Barbary}, K. and {Muna}, D. and {Ferguson}, H. and {Grollier}, F. and {Parikh}, M.~M. and {Nair}, P.~H. and {Unther}, H.~M. and {Deil}, C. and {Woillez}, J. and {Conseil}, S. and {Kramer}, R. and {Turner}, J.~E.~H. and {Singer}, L. and {Fox}, R. and {Weaver}, B.~A. and {Zabalza}, V. and {Edwards}, Z.~I. and {Azalee Bostroem}, K. and {Burke}, D.~J. and {Casey}, A.~R. and {Crawford}, S.~M. and {Dencheva}, N. and {Ely}, J. and {Jenness}, T. and {Labrie}, K. and {Lim}, P.~L. and {Pierfederici}, F. and {Pontzen}, A. and {Ptak}, A. and {Refsdal}, B. and {Servillat}, M. and {Streicher}, O.},
Doi = {10.1051/0004-6361/201322068},
Eid = {A33},
Eprint = {1307.6212},
Journal = {\aap},
Month = oct,
Pages = {A33},
Primaryclass = {astro-ph.IM},
Title = {{Astropy: A community Python package for astronomy}},
Volume = 558,
Year = 2013}

@ARTICLE{astropy:2018,
       author = {{Astropy Collaboration} and {Price-Whelan}, A.~M. and
         {Sip{\H{o}}cz}, B.~M. and {G{\"u}nther}, H.~M. and {Lim}, P.~L. and
         {Crawford}, S.~M. and {Conseil}, S. and {Shupe}, D.~L. and
         {Craig}, M.~W. and {Dencheva}, N. and {Ginsburg}, A. and {Vand
        erPlas}, J.~T. and {Bradley}, L.~D. and {P{\'e}rez-Su{\'a}rez}, D. and
         {de Val-Borro}, M. and {Aldcroft}, T.~L. and {Cruz}, K.~L. and
         {Robitaille}, T.~P. and {Tollerud}, E.~J. and {Ardelean}, C. and
         {Babej}, T. and {Bach}, Y.~P. and {Bachetti}, M. and {Bakanov}, A.~V. and
         {Bamford}, S.~P. and {Barentsen}, G. and {Barmby}, P. and
         {Baumbach}, A. and {Berry}, K.~L. and {Biscani}, F. and {Boquien}, M. and
         {Bostroem}, K.~A. and {Bouma}, L.~G. and {Brammer}, G.~B. and
         {Bray}, E.~M. and {Breytenbach}, H. and {Buddelmeijer}, H. and
         {Burke}, D.~J. and {Calderone}, G. and {Cano Rodr{\'\i}guez}, J.~L. and
         {Cara}, M. and {Cardoso}, J.~V.~M. and {Cheedella}, S. and {Copin}, Y. and
         {Corrales}, L. and {Crichton}, D. and {D'Avella}, D. and {Deil}, C. and
         {Depagne}, {\'E}. and {Dietrich}, J.~P. and {Donath}, A. and
         {Droettboom}, M. and {Earl}, N. and {Erben}, T. and {Fabbro}, S. and
         {Ferreira}, L.~A. and {Finethy}, T. and {Fox}, R.~T. and
         {Garrison}, L.~H. and {Gibbons}, S.~L.~J. and {Goldstein}, D.~A. and
         {Gommers}, R. and {Greco}, J.~P. and {Greenfield}, P. and
         {Groener}, A.~M. and {Grollier}, F. and {Hagen}, A. and {Hirst}, P. and
         {Homeier}, D. and {Horton}, A.~J. and {Hosseinzadeh}, G. and {Hu}, L. and
         {Hunkeler}, J.~S. and {Ivezi{\'c}}, {\v{Z}}. and {Jain}, A. and
         {Jenness}, T. and {Kanarek}, G. and {Kendrew}, S. and {Kern}, N.~S. and
         {Kerzendorf}, W.~E. and {Khvalko}, A. and {King}, J. and {Kirkby}, D. and
         {Kulkarni}, A.~M. and {Kumar}, A. and {Lee}, A. and {Lenz}, D. and
         {Littlefair}, S.~P. and {Ma}, Z. and {Macleod}, D.~M. and
         {Mastropietro}, M. and {McCully}, C. and {Montagnac}, S. and
         {Morris}, B.~M. and {Mueller}, M. and {Mumford}, S.~J. and {Muna}, D. and
         {Murphy}, N.~A. and {Nelson}, S. and {Nguyen}, G.~H. and
         {Ninan}, J.~P. and {N{\"o}the}, M. and {Ogaz}, S. and {Oh}, S. and
         {Parejko}, J.~K. and {Parley}, N. and {Pascual}, S. and {Patil}, R. and
         {Patil}, A.~A. and {Plunkett}, A.~L. and {Prochaska}, J.~X. and
         {Rastogi}, T. and {Reddy Janga}, V. and {Sabater}, J. and
         {Sakurikar}, P. and {Seifert}, M. and {Sherbert}, L.~E. and
         {Sherwood-Taylor}, H. and {Shih}, A.~Y. and {Sick}, J. and
         {Silbiger}, M.~T. and {Singanamalla}, S. and {Singer}, L.~P. and
         {Sladen}, P.~H. and {Sooley}, K.~A. and {Sornarajah}, S. and
         {Streicher}, O. and {Teuben}, P. and {Thomas}, S.~W. and
         {Tremblay}, G.~R. and {Turner}, J.~E.~H. and {Terr{\'o}n}, V. and
         {van Kerkwijk}, M.~H. and {de la Vega}, A. and {Watkins}, L.~L. and
         {Weaver}, B.~A. and {Whitmore}, J.~B. and {Woillez}, J. and
         {Zabalza}, V. and {Astropy Contributors}},
        title = "{The Astropy Project: Building an Open-science Project and Status of the v2.0 Core Package}",
      journal = {\aj},
         year = 2018,
        month = sep,
       volume = {156},
       number = {3},
          eid = {123},
        pages = {123},
          doi = {10.3847/1538-3881/aabc4f},
archivePrefix = {arXiv},
       eprint = {1801.02634},
 primaryClass = {astro-ph.IM},
       adsurl = {https://ui.adsabs.harvard.edu/abs/2018AJ....156..123A}
}

@ARTICLE{astropy:2022,
       author = {{Astropy Collaboration} and {Price-Whelan}, Adrian M. and {Lim}, Pey Lian and {Earl}, Nicholas and {Starkman}, Nathaniel and {Bradley}, Larry and {Shupe}, David L. and {Patil}, Aarya A. and {Corrales}, Lia and {Brasseur}, C.~E. and {N{"o}the}, Maximilian and {Donath}, Axel and {Tollerud}, Erik and {Morris}, Brett M. and {Ginsburg}, Adam and {Vaher}, Eero and {Weaver}, Benjamin A. and {Tocknell}, James and {Jamieson}, William and {van Kerkwijk}, Marten H. and {Robitaille}, Thomas P. and {Merry}, Bruce and {Bachetti}, Matteo and {G{"u}nther}, H. Moritz and {Aldcroft}, Thomas L. and {Alvarado-Montes}, Jaime A. and {Archibald}, Anne M. and {B{'o}di}, Attila and {Bapat}, Shreyas and {Barentsen}, Geert and {Baz{'a}n}, Juanjo and {Biswas}, Manish and {Boquien}, M{'e}d{'e}ric and {Burke}, D.~J. and {Cara}, Daria and {Cara}, Mihai and {Conroy}, Kyle E. and {Conseil}, Simon and {Craig}, Matthew W. and {Cross}, Robert M. and {Cruz}, Kelle L. and {D'Eugenio}, Francesco and {Dencheva}, Nadia and {Devillepoix}, Hadrien A.~R. and {Dietrich}, J{"o}rg P. and {Eigenbrot}, Arthur Davis and {Erben}, Thomas and {Ferreira}, Leonardo and {Foreman-Mackey}, Daniel and {Fox}, Ryan and {Freij}, Nabil and {Garg}, Suyog and {Geda}, Robel and {Glattly}, Lauren and {Gondhalekar}, Yash and {Gordon}, Karl D. and {Grant}, David and {Greenfield}, Perry and {Groener}, Austen M. and {Guest}, Steve and {Gurovich}, Sebastian and {Handberg}, Rasmus and {Hart}, Akeem and {Hatfield-Dodds}, Zac and {Homeier}, Derek and {Hosseinzadeh}, Griffin and {Jenness}, Tim and {Jones}, Craig K. and {Joseph}, Prajwel and {Kalmbach}, J. Bryce and {Karamehmetoglu}, Emir and {Ka{l}uszy{'n}ski}, Miko{l}aj and {Kelley}, Michael S.~P. and {Kern}, Nicholas and {Kerzendorf}, Wolfgang E. and {Koch}, Eric W. and {Kulumani}, Shankar and {Lee}, Antony and {Ly}, Chun and {Ma}, Zhiyuan and {MacBride}, Conor and {Maljaars}, Jakob M. and {Muna}, Demitri and {Murphy}, N.~A. and {Norman}, Henrik and {O'Steen}, Richard and {Oman}, Kyle A. and {Pacifici}, Camilla and {Pascual}, Sergio and {Pascual-Granado}, J. and {Patil}, Rohit R. and {Perren}, Gabriel I. and {Pickering}, Timothy E. and {Rastogi}, Tanuj and {Roulston}, Benjamin R. and {Ryan}, Daniel F. and {Rykoff}, Eli S. and {Sabater}, Jose and {Sakurikar}, Parikshit and {Salgado}, Jes{'u}s and {Sanghi}, Aniket and {Saunders}, Nicholas and {Savchenko}, Volodymyr and {Schwardt}, Ludwig and {Seifert-Eckert}, Michael and {Shih}, Albert Y. and {Jain}, Anany Shrey and {Shukla}, Gyanendra and {Sick}, Jonathan and {Simpson}, Chris and {Singanamalla}, Sudheesh and {Singer}, Leo P. and {Singhal}, Jaladh and {Sinha}, Manodeep and {Sip{H{o}}cz}, Brigitta M. and {Spitler}, Lee R. and {Stansby}, David and {Streicher}, Ole and {{{S}}umak}, Jani and {Swinbank}, John D. and {Taranu}, Dan S. and {Tewary}, Nikita and {Tremblay}, Grant R. and {Val-Borro}, Miguel de and {Van Kooten}, Samuel J. and {Vasovi{'c}}, Zlatan and {Verma}, Shresth and {de Miranda Cardoso}, Jos{'e} Vin{'i}cius and {Williams}, Peter K.~G. and {Wilson}, Tom J. and {Winkel}, Benjamin and {Wood-Vasey}, W.~M. and {Xue}, Rui and {Yoachim}, Peter and {Zhang}, Chen and {Zonca}, Andrea and {Astropy Project Contributors}},
        title = "{The Astropy Project: Sustaining and Growing a Community-oriented Open-source Project and the Latest Major Release (v5.0) of the Core Package}",
      journal = {apj},
         year = 2022,
        month = aug,
       volume = {935},
       number = {2},
          eid = {167},
        pages = {167},
          doi = {10.3847/1538-4357/ac7c74},
archivePrefix = {arXiv},
       eprint = {2206.14220},
 primaryClass = {astro-ph.IM},
       adsurl = {https://ui.adsabs.harvard.edu/abs/2022ApJ...935..167A}
}

@ARTICLE{Saburova2024,
       author = {{Saburova}, Anna S. and {Gasymov}, Damir and {Rubtsov}, Evgenii V. and {Chilingarian}, Igor V. and {Borisov}, Sviatoslav and {Gerasimov}, Ivan and {Kolganov}, Fedor and {Kasparova}, Anastasia V. and {Uklein}, Roman I. and {B{\'\i}lek}, Michal and {Grishin}, Kirill A. and {Zasov}, Anatoly and {Demianenko}, Mariia and {Katkov}, Ivan Yu. and {Lalovi{\'c}}, Ana and {Samurovi{\'c}}, Srdjan},
        title = "{A Closer Look at the Extended Edge-on Low-surface Brightness Galaxies}",
      journal = {\apj},
         year = 2024,
        month = oct,
       volume = {973},
       number = {2},
          eid = {167},
        pages = {167},
          doi = {10.3847/1538-4357/ad67e0},
archivePrefix = {arXiv},
       eprint = {2407.17548},
 primaryClass = {astro-ph.GA},
       adsurl = {https://ui.adsabs.harvard.edu/abs/2024ApJ...973..167S}
}

@ARTICLE{ManceraPina2021,
       author = {{Mancera Pi{\~n}a}, Pavel E. and {Posti}, Lorenzo and {Pezzulli}, Gabriele and {Fraternali}, Filippo and {Fall}, S. Michael and {Oosterloo}, Tom and {Adams}, Elizabeth A.~K.},
        title = "{A tight angular-momentum plane for disc galaxies}",
      journal = {\aap},
         year = 2021,
        month = jul,
       volume = {651},
          eid = {L15},
        pages = {L15},
          doi = {10.1051/0004-6361/202141574},
archivePrefix = {arXiv},
       eprint = {2107.02809},
 primaryClass = {astro-ph.GA},
       adsurl = {https://ui.adsabs.harvard.edu/abs/2021A&A...651L..15M}
}

@ARTICLE{khoperskov_illustris,
       author = {{Khoperskov}, Sergey and {Zinchenko}, Igor and {Avramov}, Branislav and {Khrapov}, Sergey and {Berczik}, Peter and {Saburova}, Anna and {Ishchenko}, Marina and {Khoperskov}, Alexander and {Pulsoni}, Claudia and {Venichenko}, Yulia and {Bizyaev}, Dmitry and {Moiseev}, Alexei},
        title = "{Extreme kinematic misalignment in IllustrisTNG galaxies: the origin, structure, and internal dynamics of galaxies with a large-scale counterrotation}",
      journal = {\mnras},
         year = 2021,
        month = jan,
       volume = {500},
       number = {3},
        pages = {3870-3888},
          doi = {10.1093/mnras/staa3330},
archivePrefix = {arXiv},
       eprint = {2010.11581},
 primaryClass = {astro-ph.GA},
       adsurl = {https://ui.adsabs.harvard.edu/abs/2021MNRAS.500.3870K}
}

@ARTICLE{Khim2021ApJS..254...27K,
       author = {{Khim}, Donghyeon J. and {Yi}, Sukyoung K. and {Pichon}, Christophe and {Dubois}, Yohan and {Devriendt}, Julien and {Choi}, Hoseung and {Bryant}, Julia J. and {Croom}, Scott M.},
        title = "{Star-Gas Misalignment in Galaxies. II. Origins Found from the Horizon-AGN Simulation}",
      journal = {\apjs},
         year = 2021,
        month = jun,
       volume = {254},
       number = {2},
          eid = {27},
        pages = {27},
          doi = {10.3847/1538-4365/abf043},
archivePrefix = {arXiv},
       eprint = {2012.04659},
 primaryClass = {astro-ph.GA},
       adsurl = {https://ui.adsabs.harvard.edu/abs/2021ApJS..254...27K}
}

@INPROCEEDINGS{nburst_a,
   author = {{Chilingarian}, I. and {Prugniel}, P. and {Sil'Chenko}, O. and
  {Koleva}, M.},
    title = "{NBursts: Simultaneous Extraction of Internal Kinematics and Parametrized SFH from Integrated Light Spectra}",
booktitle = {Stellar Populations as Building Blocks of Galaxies},
     year = 2007,
   series = {IAU Symposium},
   volume = 241,
archivePrefix = "arXiv",
   eprint = {0709.3047},
   editor = {{Vazdekis}, A. and {Peletier}, R.},
    month = aug,
    pages = {175-176},
      doi = {10.1017/S1743921307007752},
   adsurl = {http://adsabs.harvard.edu/abs/2007IAUS..241..175C}
}

@ARTICLE{nburst_b,
   author = {{Chilingarian}, I.~V. and {Prugniel}, P. and {Sil'Chenko}, O.~K. and
  {Afanasiev}, V.~L.},
    title = "{Kinematics and stellar populations of the dwarf elliptical galaxy IC 3653}",
  journal = {\mnras},
   eprint = {astro-ph/0701842},
     year = 2007,
    month = apr,
   volume = 376,
    pages = {1033-1046},
      doi = {10.1111/j.1365-2966.2007.11549.x},
   adsurl = {http://adsabs.harvard.edu/abs/2007MNRAS.376.1033C}
}

@ARTICLE{bpt,
       author = {{Baldwin}, J.~A. and {Phillips}, M.~M. and {Terlevich}, R.},
        title = "{Classification parameters for the emission-line spectra of extragalactic objects.}",
      journal = {\pasp},
         year = 1981,
        month = feb,
       volume = {93},
        pages = {5-19},
          doi = {10.1086/130766},
       adsurl = {https://ui.adsabs.harvard.edu/abs/1981PASP...93....5B}
}

@ARTICLE{Kasparova2020MNRAS.493.5464K,
       author = {{Kasparova}, Anastasia V. and {Katkov}, Ivan Yu and {Chilingarian}, Igor V.},
        title = "{An excessively massive thick disc of the enormous edge-on lenticular galaxy NGC 7572}",
      journal = {\mnras},
         year = 2020,
        month = apr,
       volume = {493},
       number = {4},
        pages = {5464-5478},
          doi = {10.1093/mnras/staa611},
archivePrefix = {arXiv},
       eprint = {1912.04887},
 primaryClass = {astro-ph.GA},
       adsurl = {https://ui.adsabs.harvard.edu/abs/2020MNRAS.493.5464K}
}

@ARTICLE{Katkov2016MNRAS.461.2068K,
       author = {{Katkov}, Ivan Yu. and {Sil'chenko}, Olga K. and {Chilingarian}, Igor V. and {Uklein}, Roman I. and {Egorov}, Oleg V.},
        title = "{Stellar counter-rotation in lenticular galaxy NGC 448}",
      journal = {\mnras},
         year = 2016,
        month = sep,
       volume = {461},
       number = {2},
        pages = {2068-2076},
          doi = {10.1093/mnras/stw1452},
archivePrefix = {arXiv},
       eprint = {1606.04862},
 primaryClass = {astro-ph.GA},
       adsurl = {https://ui.adsabs.harvard.edu/abs/2016MNRAS.461.2068K}
}

@ARTICLE{Katkov2013ApJ...769..105K,
       author = {{Katkov}, Ivan Yu. and {Sil'chenko}, Olga K. and {Afanasiev}, Victor L.},
        title = "{Lenticular Galaxy IC 719: Current Building of the Counterrotating Large-scale Stellar Disk}",
      journal = {\apj},
         year = 2013,
        month = jun,
       volume = {769},
       number = {2},
          eid = {105},
        pages = {105},
          doi = {10.1088/0004-637X/769/2/105},
archivePrefix = {arXiv},
       eprint = {1304.3339},
 primaryClass = {astro-ph.CO},
       adsurl = {https://ui.adsabs.harvard.edu/abs/2013ApJ...769..105K}
}

@ARTICLE{Fioc1997A&A...326..950F,
       author = {{Fioc}, M. and {Rocca-Volmerange}, B.},
        title = "{PEGASE: a UV to NIR spectral evolution model of galaxies. Application to the calibration of bright galaxy counts.}",
      journal = {\aap},
         year = 1997,
        month = oct,
       volume = {500},
        pages = {507-519},
archivePrefix = {arXiv},
       eprint = {astro-ph/9707017},
 primaryClass = {astro-ph},
       adsurl = {https://ui.adsabs.harvard.edu/abs/1997A&A...326..950F}
}

@ARTICLE{Lejeune1997A&AS..125..229L,
       author = {{Lejeune}, Th. and {Cuisinier}, F. and {Buser}, R.},
        title = "{Standard stellar library for evolutionary synthesis. I. Calibration of theoretical spectra}",
      journal = {\aaps},
         year = 1997,
        month = oct,
       volume = {125},
        pages = {229-246},
          doi = {10.1051/aas:1997373},
archivePrefix = {arXiv},
       eprint = {astro-ph/9701019},
 primaryClass = {astro-ph},
       adsurl = {https://ui.adsabs.harvard.edu/abs/1997A&AS..125..229L}
}

@ARTICLE{Fitzpatrick1999PASP..111...63F,
       author = {{Fitzpatrick}, Edward L.},
        title = "{Correcting for the Effects of Interstellar Extinction}",
      journal = {\pasp},
         year = 1999,
        month = jan,
       volume = {111},
       number = {755},
        pages = {63-75},
          doi = {10.1086/316293},
archivePrefix = {arXiv},
       eprint = {astro-ph/9809387},
 primaryClass = {astro-ph},
       adsurl = {https://ui.adsabs.harvard.edu/abs/1999PASP..111...63F}
}

@MISC{Newville2014zndo.....11813N,
       author = {{Newville}, Matthew and {Stensitzki}, Till and {Allen}, Daniel B. and {Ingargiola}, Antonino},
        title = "{LMFIT: Non-Linear Least-Square Minimization and Curve-Fitting for Python}",
 howpublished = {Zenodo},
         year = 2014,
        month = sep,
          eid = {10.5281/zenodo.11813},
          doi = {10.5281/zenodo.11813},
      version = {0.8.0},
    publisher = {Zenodo},
       adsurl = {https://ui.adsabs.harvard.edu/abs/2014zndo.....11813N}
}

@ARTICLE{Krajnovic2011MNRAS.414.2923K,
       author = {{Krajnovi{\'c}}, Davor and {Emsellem}, Eric and {Cappellari}, Michele and {Alatalo}, Katherine and {Blitz}, Leo and {Bois}, Maxime and {Bournaud}, Fr{\'e}d{\'e}ric and {Bureau}, Martin and {Davies}, Roger L. and {Davis}, Timothy A. and {de Zeeuw}, P.~T. and {Khochfar}, Sadegh and {Kuntschner}, Harald and {Lablanche}, Pierre-Yves and {McDermid}, Richard M. and {Morganti}, Raffaella and {Naab}, Thorsten and {Oosterloo}, Tom and {Sarzi}, Marc and {Scott}, Nicholas and {Serra}, Paolo and {Weijmans}, Anne-Marie and {Young}, Lisa M.},
        title = "{The ATLAS$^{3D}$ project - II. Morphologies, kinemetric features and alignment between photometric and kinematic axes of early-type galaxies}",
      journal = {\mnras},
         year = 2011,
        month = jul,
       volume = {414},
       number = {4},
        pages = {2923-2949},
          doi = {10.1111/j.1365-2966.2011.18560.x},
archivePrefix = {arXiv},
       eprint = {1102.3801},
 primaryClass = {astro-ph.CO},
       adsurl = {https://ui.adsabs.harvard.edu/abs/2011MNRAS.414.2923K}
}

@ARTICLE{Drory2015AJ....149...77D,
       author = {{Drory}, N. and {MacDonald}, N. and {Bershady}, M.~A. and {Bundy}, K. and
         {Gunn}, J. and {Law}, D.~R. and {Smith}, M. and {Stoll}, R. and
         {Tremonti}, C.~A. and {Wake}, D.~A. and {Yan}, R. and
         {Weijmans}, A.~M. and {Byler}, N. and {Cherinka}, B. and {Cope}, F. and
         {Eigenbrot}, A. and {Harding}, P. and {Holder}, D. and
         {Huehnerhoff}, J. and {Jaehnig}, K. and {Jansen}, T.~C. and
         {Klaene}, M. and {Paat}, A.~M. and {Percival}, J. and {Sayres}, C.},
        title = "{The MaNGA Integral Field Unit Fiber Feed System for the Sloan 2.5 m Telescope}",
      journal = {\aj},
         year = 2015,
        month = feb,
       volume = {149},
       number = {2},
          eid = {77},
        pages = {77},
          doi = {10.1088/0004-6256/149/2/77},
archivePrefix = {arXiv},
       eprint = {1412.1535},
 primaryClass = {astro-ph.IM},
       adsurl = {https://ui.adsabs.harvard.edu/abs/2015AJ....149...77D}
}

@ARTICLE{Law2015AJ....150...19L,
       author = {{Law}, David R. and {Yan}, Renbin and {Bershady}, Matthew A. and
         {Bundy}, Kevin and {Cherinka}, Brian and {Drory}, Niv and
         {MacDonald}, Nicholas and {S{\'a}nchez-Gallego}, Jos{\'e} R. and
         {Wake}, David A. and {Weijmans}, Anne-Marie and {Blanton}, Michael R. and
         {Klaene}, Mark A. and {Moran}, Sean M. and {Sanchez}, Sebastian F. and
         {Zhang}, Kai},
        title = "{Observing Strategy for the SDSS-IV/MaNGA IFU Galaxy Survey}",
      journal = {\aj},
         year = 2015,
        month = jul,
       volume = {150},
       number = {1},
          eid = {19},
        pages = {19},
          doi = {10.1088/0004-6256/150/1/19},
archivePrefix = {arXiv},
       eprint = {1505.04285},
 primaryClass = {astro-ph.IM},
       adsurl = {https://ui.adsabs.harvard.edu/abs/2015AJ....150...19L}
}

@ARTICLE{Yan2016AJ....152..197Y,
       author = {{Yan}, Renbin and {Bundy}, Kevin and {Law}, David R. and
         {Bershady}, Matthew A. and {Andrews}, Brett and {Cherinka}, Brian and
         {Diamond-Stanic}, Aleksandar M. and {Drory}, Niv and
         {MacDonald}, Nicholas and {S{\'a}nchez-Gallego}, Jos{\'e} R. and
         {Thomas}, Daniel and {Wake}, David A. and {Weijmans}, Anne-Marie and
         {Westfall}, Kyle B. and {Zhang}, Kai and
         {Arag{\'o}n-Salamanca}, Alfonso and {Belfiore}, Francesco and
         {Bizyaev}, Dmitry and {Blanc}, Guillermo A. and {Blanton}, Michael R. and
         {Brownstein}, Joel and {Cappellari}, Michele and {D'Souza}, Richard and
         {Emsellem}, Eric and {Fu}, Hai and {Gaulme}, Patrick and
         {Graham}, Mark T. and {Goddard}, Daniel and {Gunn}, James E. and
         {Harding}, Paul and {Jones}, Amy and {Kinemuchi}, Karen and
         {Li}, Cheng and {Li}, Hongyu and {Maiolino}, Roberto and {Mao}, Shude and
         {Maraston}, Claudia and {Masters}, Karen and {Merrifield}, Michael R. and
         {Oravetz}, Daniel and {Pan}, Kaike and {Parejko}, John K. and
         {Sanchez}, Sebastian F. and {Schlegel}, David and {Simmons}, Audrey and
         {Thanjavur}, Karun and {Tinker}, Jeremy and {Tremonti}, Christy and
         {van den Bosch}, Remco and {Zheng}, Zheng},
        title = "{SDSS-IV MaNGA IFS Galaxy Survey{\textemdash}Survey Design, Execution, and Initial Data Quality}",
      journal = {\aj},
         year = 2016,
        month = dec,
       volume = {152},
       number = {6},
          eid = {197},
        pages = {197},
          doi = {10.3847/0004-6256/152/6/197},
archivePrefix = {arXiv},
       eprint = {1607.08613},
 primaryClass = {astro-ph.GA},
       adsurl = {https://ui.adsabs.harvard.edu/abs/2016AJ....152..197Y}
}

@ARTICLE{Law2016AJ....152...83L,
       author = {{Law}, David R. and {Cherinka}, Brian and {Yan}, Renbin and
         {Andrews}, Brett H. and {Bershady}, Matthew A. and {Bizyaev}, Dmitry and
         {Blanc}, Guillermo A. and {Blanton}, Michael R. and {Bolton}, Adam S. and
         {Brownstein}, Joel R. and {Bundy}, Kevin and {Chen}, Yanmei and
         {Drory}, Niv and {D'Souza}, Richard and {Fu}, Hai and {Jones}, Amy and
         {Kauffmann}, Guinevere and {MacDonald}, Nicholas and
         {Masters}, Karen L. and {Newman}, Jeffrey A. and {Parejko}, John K. and
         {S{\'a}nchez-Gallego}, Jos{\'e} R. and {S{\'a}nchez}, Sebastian F. and
         {Schlegel}, David J. and {Thomas}, Daniel and {Wake}, David A. and
         {Weijmans}, Anne-Marie and {Westfall}, Kyle B. and {Zhang}, Kai},
        title = "{The Data Reduction Pipeline for the SDSS-IV MaNGA IFU Galaxy Survey}",
      journal = {\aj},
         year = 2016,
        month = oct,
       volume = {152},
       number = {4},
          eid = {83},
        pages = {83},
          doi = {10.3847/0004-6256/152/4/83},
archivePrefix = {arXiv},
       eprint = {1607.08619},
 primaryClass = {astro-ph.IM},
       adsurl = {https://ui.adsabs.harvard.edu/abs/2016AJ....152...83L}
}

@ARTICLE{Westfall2019AJ....158..231W,
       author = {{Westfall}, Kyle B. and {Cappellari}, Michele and
         {Bershady}, Matthew A. and {Bundy}, Kevin and {Belfiore}, Francesco and
         {Ji}, Xihan and {Law}, David R. and {Schaefer}, Adam and
         {Shetty}, Shravan and {Tremonti}, Christy A. and {Yan}, Renbin and
         {Andrews}, Brett H. and {Brownstein}, Joel R. and {Cherinka}, Brian and
         {Coccato}, Lodovico and {Drory}, Niv and {Maraston}, Claudia and
         {Parikh}, Taniya and {S{\'a}nchez-Gallego}, Jos{\'e} R. and
         {Thomas}, Daniel and {Weijmans}, Anne-Marie and
         {Barrera-Ballesteros}, Jorge and {Du}, Cheng and {Goddard}, Daniel and
         {Li}, Niu and {Masters}, Karen and {Ibarra Medel}, H{\'e}ctor Javier and
         {S{\'a}nchez}, Sebasti{\'a}n F. and {Yang}, Meng and {Zheng}, Zheng and
         {Zhou}, Shuang},
        title = "{The Data Analysis Pipeline for the SDSS-IV MaNGA IFU Galaxy Survey: Overview}",
      journal = {\aj},
         year = 2019,
        month = dec,
       volume = {158},
       number = {6},
          eid = {231},
        pages = {231},
          doi = {10.3847/1538-3881/ab44a2},
archivePrefix = {arXiv},
       eprint = {1901.00856},
 primaryClass = {astro-ph.GA},
       adsurl = {https://ui.adsabs.harvard.edu/abs/2019AJ....158..231W}
}

@ARTICLE{Belfiore2019AJ....158..160B,
       author = {{Belfiore}, Francesco and {Westfall}, Kyle B. and {Schaefer}, Adam and
         {Cappellari}, Michele and {Ji}, Xihan and {Bershady}, Matthew A. and
         {Tremonti}, Christy and {Law}, David R. and {Yan}, Renbin and
         {Bundy}, Kevin and {Shetty}, Shravan and {Drory}, Niv and
         {Thomas}, Daniel and {Emsellem}, Eric and
         {S{\'a}nchez}, Sebasti{\'a}n F.},
        title = "{The Data Analysis Pipeline for the SDSS-IV MaNGA IFU Galaxy Survey: Emission-line Modeling}",
      journal = {\aj},
         year = 2019,
        month = oct,
       volume = {158},
       number = {4},
          eid = {160},
        pages = {160},
          doi = {10.3847/1538-3881/ab3e4e},
archivePrefix = {arXiv},
       eprint = {1901.00866},
 primaryClass = {astro-ph.GA},
       adsurl = {https://ui.adsabs.harvard.edu/abs/2019AJ....158..160B}
}

@ARTICLE{Bundy2015ApJ...798....7B,
       author = {{Bundy}, Kevin and {Bershady}, Matthew A. and {Law}, David R. and
         {Yan}, Renbin and {Drory}, Niv and {MacDonald}, Nicholas and
         {Wake}, David A. and {Cherinka}, Brian and
         {S{\'a}nchez-Gallego}, Jos{\'e} R. and {Weijmans}, Anne-Marie and
         {Thomas}, Daniel and {Tremonti}, Christy and {Masters}, Karen and
         {Coccato}, Lodovico and {Diamond-Stanic}, Aleksandar M. and
         {Arag{\'o}n-Salamanca}, Alfonso and {Avila-Reese}, Vladimir and
         {Badenes}, Carles and {Falc{\'o}n-Barroso}, J{\'e}sus and
         {Belfiore}, Francesco and {Bizyaev}, Dmitry and {Blanc}, Guillermo A. and
         {Bland-Hawthorn}, Joss and {Blanton}, Michael R. and
         {Brownstein}, Joel R. and {Byler}, Nell and {Cappellari}, Michele and
         {Conroy}, Charlie and {Dutton}, Aaron A. and {Emsellem}, Eric and
         {Etherington}, James and {Frinchaboy}, Peter M. and {Fu}, Hai and
         {Gunn}, James E. and {Harding}, Paul and {Johnston}, Evelyn J. and
         {Kauffmann}, Guinevere and {Kinemuchi}, Karen and {Klaene}, Mark A. and
         {Knapen}, Johan H. and {Leauthaud}, Alexie and {Li}, Cheng and
         {Lin}, Lihwai and {Maiolino}, Roberto and {Malanushenko}, Viktor and
         {Malanushenko}, Elena and {Mao}, Shude and {Maraston}, Claudia and
         {McDermid}, Richard M. and {Merrifield}, Michael R. and
         {Nichol}, Robert C. and {Oravetz}, Daniel and {Pan}, Kaike and
         {Parejko}, John K. and {Sanchez}, Sebastian F. and {Schlegel}, David and
         {Simmons}, Audrey and {Steele}, Oliver and {Steinmetz}, Matthias and
         {Thanjavur}, Karun and {Thompson}, Benjamin A. and {Tinker}, Jeremy L. and
         {van den Bosch}, Remco C.~E. and {Westfall}, Kyle B. and
         {Wilkinson}, David and {Wright}, Shelley and {Xiao}, Ting and
         {Zhang}, Kai},
        title = "{Overview of the SDSS-IV MaNGA Survey: Mapping nearby Galaxies at Apache Point Observatory}",
      journal = {\apj},
         year = 2015,
        month = jan,
       volume = {798},
       number = {1},
          eid = {7},
        pages = {7},
          doi = {10.1088/0004-637X/798/1/7},
archivePrefix = {arXiv},
       eprint = {1412.1482},
 primaryClass = {astro-ph.GA},
       adsurl = {https://ui.adsabs.harvard.edu/abs/2015ApJ...798....7B}
}

@INPROCEEDINGS{Gasymov2022muto.confE..17G,
       author = {{Gasymov}, D.~F.~O. and {Katkov}, I.},
        title = "{Detailed study of galaxies with the stellar counter-rotation phenomenon}",
    booktitle = {The Multifaceted Universe: Theory and Observations - 2000},
         year = 2022,
        month = dec,
          eid = {17},
        pages = {17},
          doi = {10.48550/arXiv.2209.11240},
archivePrefix = {arXiv},
       eprint = {2209.11240},
 primaryClass = {astro-ph.GA},
       adsurl = {https://ui.adsabs.harvard.edu/abs/2022muto.confE..17G}
}

@ARTICLE{Katkov2022A&A...658A.154K_n254,
       author = {{Katkov}, Ivan Yu. and {Kniazev}, Alexei Yu. and {Sil'chenko}, Olga K. and {Gasymov}, Damir},
        title = "{Star formation in outer rings of S0 galaxies. IV. NGC 254: A double-ringed S0 with gas counter-rotation}",
      journal = {\aap},
         year = 2022,
        month = feb,
       volume = {658},
          eid = {A154},
        pages = {A154},
          doi = {10.1051/0004-6361/202141934},
archivePrefix = {arXiv},
       eprint = {2112.03289},
 primaryClass = {astro-ph.GA},
       adsurl = {https://ui.adsabs.harvard.edu/abs/2022A&A...658A.154K}
}

@ARTICLE{Falcon-Barroso2021A&A...646A..31F,
       author = {{Falc{\'o}n-Barroso}, J. and {Martig}, M.},
        title = "{BAYES-LOSVD: A Bayesian framework for non-parametric extraction of the line-of-sight velocity distribution of galaxies}",
      journal = {\aap},
         year = 2021,
        month = feb,
       volume = {646},
          eid = {A31},
        pages = {A31},
          doi = {10.1051/0004-6361/202039624},
archivePrefix = {arXiv},
       eprint = {2011.12023},
 primaryClass = {astro-ph.GA},
       adsurl = {https://ui.adsabs.harvard.edu/abs/2021A&A...646A..31F}
}

@ARTICLE{Coccato2011MNRAS.412L.113C,
       author = {{Coccato}, L. and {Morelli}, L. and {Corsini}, E.~M. and {Buson}, L. and {Pizzella}, A. and {Vergani}, D. and {Bertola}, F.},
        title = "{Dating the formation of the counter-rotating stellar disc in the spiral galaxy NGC 5719 by disentangling its stellar populations}",
      journal = {\mnras},
         year = 2011,
        month = mar,
       volume = {412},
       number = {1},
        pages = {L113-L117},
          doi = {10.1111/j.1745-3933.2011.01016.x},
archivePrefix = {arXiv},
       eprint = {1101.3092},
 primaryClass = {astro-ph.GA},
       adsurl = {https://ui.adsabs.harvard.edu/abs/2011MNRAS.412L.113C}
}

@INPROCEEDINGS{Combes2014ASPC..480..211C,
       author = {{Combes}, F.},
        title = "{Gas Accretion in Disk Galaxies}",
    booktitle = {Structure and Dynamics of Disk Galaxies},
         year = 2014,
       editor = {{Seigar}, M.~S. and {Treuthardt}, P.},
       series = {Astronomical Society of the Pacific Conference Series},
       volume = {480},
        month = mar,
        pages = {211},
          doi = {10.48550/arXiv.1309.1603},
archivePrefix = {arXiv},
       eprint = {1309.1603},
 primaryClass = {astro-ph.CO},
       adsurl = {https://ui.adsabs.harvard.edu/abs/2014ASPC..480..211C}
}

@ARTICLE{Kroupa2001MNRAS.322..231K,
       author = {{Kroupa}, Pavel},
        title = "{On the variation of the initial mass function}",
      journal = {\mnras},
         year = 2001,
        month = apr,
       volume = {322},
       number = {2},
        pages = {231-246},
          doi = {10.1046/j.1365-8711.2001.04022.x},
archivePrefix = {arXiv},
       eprint = {astro-ph/0009005},
 primaryClass = {astro-ph},
       adsurl = {https://ui.adsabs.harvard.edu/abs/2001MNRAS.322..231K}
}

@ARTICLE{vandeVoort2012MNRAS.423.2991V,
       author = {{van de Voort}, Freeke and {Schaye}, Joop},
        title = "{Properties of gas in and around galaxy haloes}",
      journal = {\mnras},
         year = 2012,
        month = jul,
       volume = {423},
       number = {4},
        pages = {2991-3010},
          doi = {10.1111/j.1365-2966.2012.20949.x},
archivePrefix = {arXiv},
       eprint = {1111.5039},
 primaryClass = {astro-ph.CO},
       adsurl = {https://ui.adsabs.harvard.edu/abs/2012MNRAS.423.2991V}
}

@ARTICLE{Hafen2017MNRAS.469.2292H,
       author = {{Hafen}, Zachary and {Faucher-Gigu{\`e}re}, Claude-Andr{\'e} and {Angl{\'e}s-Alc{\'a}zar}, Daniel and {Kere{\v{s}}}, Du{\v{s}}an and {Feldmann}, Robert and {Chan}, T.~K. and {Quataert}, Eliot and {Murray}, Norman and {Hopkins}, Philip F.},
        title = "{Low-redshift Lyman limit systems as diagnostics of cosmological inflows and outflows}",
      journal = {\mnras},
         year = 2017,
        month = aug,
       volume = {469},
       number = {2},
        pages = {2292-2304},
          doi = {10.1093/mnras/stx952},
archivePrefix = {arXiv},
       eprint = {1608.05712},
 primaryClass = {astro-ph.GA},
       adsurl = {https://ui.adsabs.harvard.edu/abs/2017MNRAS.469.2292H}
}

@ARTICLE{Smee2013AJ....146...32S,
       author = {{Smee}, Stephen A. and {Gunn}, James E. and {Uomoto}, Alan and {Roe}, Natalie and {Schlegel}, David and {Rockosi}, Constance M. and {Carr}, Michael A. and {Leger}, French and {Dawson}, Kyle S. and {Olmstead}, Matthew D. and {Brinkmann}, Jon and {Owen}, Russell and {Barkhouser}, Robert H. and {Honscheid}, Klaus and {Harding}, Paul and {Long}, Dan and {Lupton}, Robert H. and {Loomis}, Craig and {Anderson}, Lauren and {Annis}, James and {Bernardi}, Mariangela and {Bhardwaj}, Vaishali and {Bizyaev}, Dmitry and {Bolton}, Adam S. and {Brewington}, Howard and {Briggs}, John W. and {Burles}, Scott and {Burns}, James G. and {Castander}, Francisco Javier and {Connolly}, Andrew and {Davenport}, James R.~A. and {Ebelke}, Garrett and {Epps}, Harland and {Feldman}, Paul D. and {Friedman}, Scott D. and {Frieman}, Joshua and {Heckman}, Timothy and {Hull}, Charles L. and {Knapp}, Gillian R. and {Lawrence}, David M. and {Loveday}, Jon and {Mannery}, Edward J. and {Malanushenko}, Elena and {Malanushenko}, Viktor and {Merrelli}, Aronne James and {Muna}, Demitri and {Newman}, Peter R. and {Nichol}, Robert C. and {Oravetz}, Daniel and {Pan}, Kaike and {Pope}, Adrian C. and {Ricketts}, Paul G. and {Shelden}, Alaina and {Sandford}, Dale and {Siegmund}, Walter and {Simmons}, Audrey and {Smith}, D. Shane and {Snedden}, Stephanie and {Schneider}, Donald P. and {SubbaRao}, Mark and {Tremonti}, Christy and {Waddell}, Patrick and {York}, Donald G.},
        title = "{The Multi-object, Fiber-fed Spectrographs for the Sloan Digital Sky Survey and the Baryon Oscillation Spectroscopic Survey}",
      journal = {\aj},
         year = 2013,
        month = aug,
       volume = {146},
       number = {2},
          eid = {32},
        pages = {32},
          doi = {10.1088/0004-6256/146/2/32},
archivePrefix = {arXiv},
       eprint = {1208.2233},
 primaryClass = {astro-ph.IM},
       adsurl = {https://ui.adsabs.harvard.edu/abs/2013AJ....146...32S}
}

@ARTICLE{Gunn2006AJ....131.2332G,
       author = {{Gunn}, James E. and {Siegmund}, Walter A. and {Mannery}, Edward J. and {Owen}, Russell E. and {Hull}, Charles L. and {Leger}, R. French and {Carey}, Larry N. and {Knapp}, Gillian R. and {York}, Donald G. and {Boroski}, William N. and {Kent}, Stephen M. and {Lupton}, Robert H. and {Rockosi}, Constance M. and {Evans}, Michael L. and {Waddell}, Patrick and {Anderson}, John E. and {Annis}, James and {Barentine}, John C. and {Bartoszek}, Larry M. and {Bastian}, Steven and {Bracker}, Stephen B. and {Brewington}, Howard J. and {Briegel}, Charles I. and {Brinkmann}, Jon and {Brown}, Yorke J. and {Carr}, Michael A. and {Czarapata}, Paul C. and {Drennan}, Craig C. and {Dombeck}, Thomas and {Federwitz}, Glenn R. and {Gillespie}, Bruce A. and {Gonzales}, Carlos and {Hansen}, Sten U. and {Harvanek}, Michael and {Hayes}, Jeffrey and {Jordan}, Wendell and {Kinney}, Ellyne and {Klaene}, Mark and {Kleinman}, S.~J. and {Kron}, Richard G. and {Kresinski}, Jurek and {Lee}, Glenn and {Limmongkol}, Siriluk and {Lindenmeyer}, Carl W. and {Long}, Daniel C. and {Loomis}, Craig L. and {McGehee}, Peregrine M. and {Mantsch}, Paul M. and {Neilsen}, Eric H., Jr. and {Neswold}, Richard M. and {Newman}, Peter R. and {Nitta}, Atsuko and {Peoples}, John, Jr. and {Pier}, Jeffrey R. and {Prieto}, Peter S. and {Prosapio}, Angela and {Rivetta}, Claudio and {Schneider}, Donald P. and {Snedden}, Stephanie and {Wang}, Shu-i.},
        title = "{The 2.5 m Telescope of the Sloan Digital Sky Survey}",
      journal = {\aj},
         year = 2006,
        month = apr,
       volume = {131},
       number = {4},
        pages = {2332-2359},
          doi = {10.1086/500975},
archivePrefix = {arXiv},
       eprint = {astro-ph/0602326},
 primaryClass = {astro-ph},
       adsurl = {https://ui.adsabs.harvard.edu/abs/2006AJ....131.2332G}
}

@ARTICLE{Lu2021MNRAS.503..726L,
       author = {{Lu}, Shengdong and {Xu}, Dandan and {Wang}, Yunchong and {Chen}, Yanmei and {Zhu}, Ling and {Mao}, Shude and {Springel}, Volker and {Wang}, Jing and {Vogelsberger}, Mark and {Hernquist}, Lars},
        title = "{Hot and counter-rotating star-forming disc galaxies in IllustrisTNG and their real-world counterparts}",
      journal = {\mnras},
         year = 2021,
        month = may,
       volume = {503},
       number = {1},
        pages = {726-742},
          doi = {10.1093/mnras/stab497},
archivePrefix = {arXiv},
       eprint = {2011.01949},
 primaryClass = {astro-ph.GA},
       adsurl = {https://ui.adsabs.harvard.edu/abs/2021MNRAS.503..726L}
}

@ARTICLE{Stark2021MNRAS.503.1345S,
       author = {{Stark}, David V. and {Masters}, Karen L. and {Avila-Reese}, Vladimir and {Riffel}, Rogemar and {Riffel}, Rogerio and {Boardman}, Nicholas Fraser and {Zheng}, Zheng and {Weijmans}, Anne-Marie and {Dillon}, Sean and {Fielder}, Catherine and {Finnegan}, Daniel and {Fofie}, Patricia and {Goddy}, Julian and {Harrington}, Emily and {Pace}, Zachary and {Rujopakarn}, Wiphu and {Samanso}, Nattida and {Shamsi}, Shoaib and {Sharma}, Anubhav and {Warrick}, Elizabeth and {Witherspoon}, Catherine and {Wolthuis}, Nathan},
        title = "{H I-MaNGA: tracing the physics of the neutral and ionized ISM with the second data release}",
      journal = {\mnras},
         year = 2021,
        month = may,
       volume = {503},
       number = {1},
        pages = {1345-1366},
          doi = {10.1093/mnras/stab566},
archivePrefix = {arXiv},
       eprint = {2101.12680},
 primaryClass = {astro-ph.GA},
       adsurl = {https://ui.adsabs.harvard.edu/abs/2021MNRAS.503.1345S}
}

@ARTICLE{Blanton2011AJ....142...31B,
       author = {{Blanton}, Michael R. and {Kazin}, Eyal and {Muna}, Demitri and {Weaver}, Benjamin A. and {Price-Whelan}, Adrian},
        title = "{Improved Background Subtraction for the Sloan Digital Sky Survey Images}",
      journal = {\aj},
         year = 2011,
        month = jul,
       volume = {142},
       number = {1},
          eid = {31},
        pages = {31},
          doi = {10.1088/0004-6256/142/1/31},
archivePrefix = {arXiv},
       eprint = {1105.1960},
 primaryClass = {astro-ph.IM},
       adsurl = {https://ui.adsabs.harvard.edu/abs/2011AJ....142...31B}
}

@ARTICLE{Wake2017AJ....154...86W,
       author = {{Wake}, David A. and {Bundy}, Kevin and {Diamond-Stanic}, Aleksandar M. and {Yan}, Renbin and {Blanton}, Michael R. and {Bershady}, Matthew A. and {S{\'a}nchez-Gallego}, Jos{\'e} R. and {Drory}, Niv and {Jones}, Amy and {Kauffmann}, Guinevere and {Law}, David R. and {Li}, Cheng and {MacDonald}, Nicholas and {Masters}, Karen and {Thomas}, Daniel and {Tinker}, Jeremy and {Weijmans}, Anne-Marie and {Brownstein}, Joel R.},
        title = "{The SDSS-IV MaNGA Sample: Design, Optimization, and Usage Considerations}",
      journal = {\aj},
         year = 2017,
        month = sep,
       volume = {154},
       number = {3},
          eid = {86},
        pages = {86},
          doi = {10.3847/1538-3881/aa7ecc},
archivePrefix = {arXiv},
       eprint = {1707.02989},
 primaryClass = {astro-ph.GA},
       adsurl = {https://ui.adsabs.harvard.edu/abs/2017AJ....154...86W}
}

@ARTICLE{Haynes2018ApJ...861...49H,
       author = {{Haynes}, Martha P. and {Giovanelli}, Riccardo and {Kent}, Brian R. and {Adams}, Elizabeth A.~K. and {Balonek}, Thomas J. and {Craig}, David W. and {Fertig}, Derek and {Finn}, Rose and {Giovanardi}, Carlo and {Hallenbeck}, Gregory and {Hess}, Kelley M. and {Hoffman}, G. Lyle and {Huang}, Shan and {Jones}, Michael G. and {Koopmann}, Rebecca A. and {Kornreich}, David A. and {Leisman}, Lukas and {Miller}, Jeffrey and {Moorman}, Crystal and {O'Connor}, Jessica and {O'Donoghue}, Aileen and {Papastergis}, Emmanouil and {Troischt}, Parker and {Stark}, David and {Xiao}, Li},
        title = "{The Arecibo Legacy Fast ALFA Survey: The ALFALFA Extragalactic H I Source Catalog}",
      journal = {\apj},
         year = 2018,
        month = jul,
       volume = {861},
       number = {1},
          eid = {49},
        pages = {49},
          doi = {10.3847/1538-4357/aac956},
archivePrefix = {arXiv},
       eprint = {1805.11499},
 primaryClass = {astro-ph.GA},
       adsurl = {https://ui.adsabs.harvard.edu/abs/2018ApJ...861...49H}
}

@ARTICLE{Argudo-Fernandez2015A&A...578A.110A,
       author = {{Argudo-Fern{\'a}ndez}, M. and {Verley}, S. and {Bergond}, G. and {Duarte Puertas}, S. and {Ramos Carmona}, E. and {Sabater}, J. and {Fern{\'a}ndez Lorenzo}, M. and {Espada}, D. and {Sulentic}, J. and {Ruiz}, J.~E. and {Leon}, S.},
        title = "{Catalogues of isolated galaxies, isolated pairs, and isolated triplets in the local Universe}",
      journal = {\aap},
         year = 2015,
        month = jun,
       volume = {578},
          eid = {A110},
        pages = {A110},
          doi = {10.1051/0004-6361/201526016},
archivePrefix = {arXiv},
       eprint = {1504.00117},
 primaryClass = {astro-ph.GA},
       adsurl = {https://ui.adsabs.harvard.edu/abs/2015A&A...578A.110A}
}

@ARTICLE{Etherington2015MNRAS.451..660E,
       author = {{Etherington}, James and {Thomas}, Daniel},
        title = "{Measuring galaxy environments in large-scale photometric surveys}",
      journal = {\mnras},
         year = 2015,
        month = jul,
       volume = {451},
       number = {1},
        pages = {660-679},
          doi = {10.1093/mnras/stv999},
archivePrefix = {arXiv},
       eprint = {1505.01171},
 primaryClass = {astro-ph.GA},
       adsurl = {https://ui.adsabs.harvard.edu/abs/2015MNRAS.451..660E}
}

@ARTICLE{Wang2016ApJ...831..164W,
       author = {{Wang}, Huiyuan and {Mo}, H.~J. and {Yang}, Xiaohu and {Zhang}, Youcai and {Shi}, JingJing and {Jing}, Y.~P. and {Liu}, Chengze and {Li}, Shijie and {Kang}, Xi and {Gao}, Yang},
        title = "{ELUCID - Exploring the Local Universe with ReConstructed Initial Density Field III: Constrained Simulation in the SDSS Volume}",
      journal = {\apj},
         year = 2016,
        month = nov,
       volume = {831},
       number = {2},
          eid = {164},
        pages = {164},
          doi = {10.3847/0004-637X/831/2/164},
archivePrefix = {arXiv},
       eprint = {1608.01763},
 primaryClass = {astro-ph.CO},
       adsurl = {https://ui.adsabs.harvard.edu/abs/2016ApJ...831..164W}
}

@ARTICLE{Salim2016ApJS..227....2S,
       author = {{Salim}, Samir and {Lee}, Janice C. and {Janowiecki}, Steven and {da Cunha}, Elisabete and {Dickinson}, Mark and {Boquien}, M{\'e}d{\'e}ric and {Burgarella}, Denis and {Salzer}, John J. and {Charlot}, St{\'e}phane},
        title = "{GALEX-SDSS-WISE Legacy Catalog (GSWLC): Star Formation Rates, Stellar Masses, and Dust Attenuations of 700,000 Low-redshift Galaxies}",
      journal = {\apjs},
         year = 2016,
        month = nov,
       volume = {227},
       number = {1},
          eid = {2},
        pages = {2},
          doi = {10.3847/0067-0049/227/1/2},
archivePrefix = {arXiv},
       eprint = {1610.00712},
 primaryClass = {astro-ph.GA},
       adsurl = {https://ui.adsabs.harvard.edu/abs/2016ApJS..227....2S}
}

@ARTICLE{Law2021AJ....161...52L,
       author = {{Law}, David R. and {Westfall}, Kyle B. and {Bershady}, Matthew A. and {Cappellari}, Michele and {Yan}, Renbin and {Belfiore}, Francesco and {Bizyaev}, Dmitry and {Brownstein}, Joel R. and {Chen}, Yanping and {Cherinka}, Brian and {Drory}, Niv and {Lazarz}, Daniel and {Shetty}, Shravan},
        title = "{SDSS-IV MaNGA: Modeling the Spectral Line-spread Function to Subpercent Accuracy}",
      journal = {\aj},
         year = 2021,
        month = feb,
       volume = {161},
       number = {2},
          eid = {52},
        pages = {52},
          doi = {10.3847/1538-3881/abcaa2},
archivePrefix = {arXiv},
       eprint = {2011.04675},
 primaryClass = {astro-ph.IM},
       adsurl = {https://ui.adsabs.harvard.edu/abs/2021AJ....161...52L}
}

@ARTICLE{Katkov2024ApJ...962...27K,
       author = {{Katkov}, Ivan Yu. and {Gasymov}, Damir and {Kniazev}, Alexei Yu. and {Gelfand}, Joseph D. and {Rubtsov}, Evgenii V. and {Chilingarian}, Igor V. and {Sil'chenko}, Olga K.},
        title = "{Probing the History of the Galaxy Assembly of the Counterrotating Disk Galaxy PGC 66551}",
      journal = {\apj},
         year = 2024,
        month = feb,
       volume = {962},
       number = {1},
          eid = {27},
        pages = {27},
          doi = {10.3847/1538-4357/ad1331},
archivePrefix = {arXiv},
       eprint = {2305.01719},
 primaryClass = {astro-ph.GA},
       adsurl = {https://ui.adsabs.harvard.edu/abs/2024ApJ...962...27K}
}

@ARTICLE{Zinchenko2023A&A...674L...7Z,
       author = {{Zinchenko}, I.~A.},
        title = "{Gas and stellar kinematic misalignment in MaNGA galaxies: What is the origin of counter-rotating gas?}",
      journal = {\aap},
         year = 2023,
        month = jun,
       volume = {674},
          eid = {L7},
        pages = {L7},
          doi = {10.1051/0004-6361/202346846},
archivePrefix = {arXiv},
       eprint = {2305.13387},
 primaryClass = {astro-ph.GA},
       adsurl = {https://ui.adsabs.harvard.edu/abs/2023A&A...674L...7Z}
}

@ARTICLE{Pilyugin2016MNRAS.457.3678P,
       author = {{Pilyugin}, L.~S. and {Grebel}, E.~K.},
        title = "{New calibrations for abundance determinations in H II regions}",
      journal = {\mnras},
         year = 2016,
        month = apr,
       volume = {457},
       number = {4},
        pages = {3678-3692},
          doi = {10.1093/mnras/stw238},
archivePrefix = {arXiv},
       eprint = {1601.08217},
 primaryClass = {astro-ph.GA},
       adsurl = {https://ui.adsabs.harvard.edu/abs/2016MNRAS.457.3678P}
}

@ARTICLE{DuartePuertas2022,
       author = {{Duarte Puertas}, S. and {Vilchez}, J.~M. and {Iglesias-P{\'a}ramo}, J. and {Moll{\'a}}, M. and {P{\'e}rez-Montero}, E. and {Kehrig}, C. and {Pilyugin}, L.~S. and {Zinchenko}, I.~A.},
        title = "{Mass-metallicity and star formation rate in galaxies: A complex relation tuned to stellar age}",
      journal = {\aap},
         year = 2022,
        month = oct,
       volume = {666},
          eid = {A186},
        pages = {A186},
          doi = {10.1051/0004-6361/202141571},
archivePrefix = {arXiv},
       eprint = {2205.01203},
 primaryClass = {astro-ph.GA},
       adsurl = {https://ui.adsabs.harvard.edu/abs/2022A&A...666A.186D}
}

@ARTICLE{Bao2024MNRAS.528.2643B,
       author = {{Bao}, Min and {Chen}, Yanmei and {Yang}, Meng and {Zhu}, Ling and {Shi}, Yong and {Gu}, Qiusheng},
        title = "{Uncovering the formation of the counter-rotating stellar discs in SDSS J074834.64+444117.8}",
      journal = {\mnras},
         year = 2024,
        month = feb,
       volume = {528},
       number = {2},
        pages = {2643-2652},
          doi = {10.1093/mnras/stae243},
archivePrefix = {arXiv},
       eprint = {2401.11179},
 primaryClass = {astro-ph.GA},
       adsurl = {https://ui.adsabs.harvard.edu/abs/2024MNRAS.528.2643B}
}

@ARTICLE{Chang2015ApJS..219....8C,
       author = {{Chang}, Yu-Yen and {van der Wel}, Arjen and {da Cunha}, Elisabete and {Rix}, Hans-Walter},
        title = "{Stellar Masses and Star Formation Rates for 1M Galaxies from SDSS+WISE}",
      journal = {\apjs},
         year = 2015,
        month = jul,
       volume = {219},
       number = {1},
          eid = {8},
        pages = {8},
          doi = {10.1088/0067-0049/219/1/8},
archivePrefix = {arXiv},
       eprint = {1506.00648},
 primaryClass = {astro-ph.GA},
       adsurl = {https://ui.adsabs.harvard.edu/abs/2015ApJS..219....8C}
}

@ARTICLE{Efstathiou1982MNRAS.201..975E,
       author = {{Efstathiou}, G. and {Ellis}, R.~S. and {Carter}, D.},
        title = "{Further observations of the elliptical galaxy NGC 5813.}",
      journal = {\mnras},
         year = 1982,
        month = dec,
       volume = {201},
        pages = {975-990},
          doi = {10.1093/mnras/201.4.975},
       adsurl = {https://ui.adsabs.harvard.edu/abs/1982MNRAS.201..975E}
}

@ARTICLE{Vazquez-Mata2022MNRAS.512.2222V,
       author = {{V{\'a}zquez-Mata}, J.~A. and {Hern{\'a}ndez-Toledo}, H.~M. and {Avila-Reese}, V. and {Herrera-Endoqui}, M. and {Rodr{\'\i}guez-Puebla}, A. and {Cano-D{\'\i}az}, M. and {Lacerna}, I. and {Mart{\'\i}nez-V{\'a}zquez}, L.~A. and {Lane}, R.},
        title = "{SDSS IV MaNGA: visual morphological and statistical characterization of the DR15 sample}",
      journal = {\mnras},
         year = 2022,
        month = may,
       volume = {512},
       number = {2},
        pages = {2222-2244},
          doi = {10.1093/mnras/stac635},
archivePrefix = {arXiv},
       eprint = {2203.02565},
 primaryClass = {astro-ph.GA},
       adsurl = {https://ui.adsabs.harvard.edu/abs/2022MNRAS.512.2222V}
}

@ARTICLE{Asplund2021A&A...653A.141A,
       author = {{Asplund}, M. and {Amarsi}, A.~M. and {Grevesse}, N.},
        title = "{The chemical make-up of the Sun: A 2020 vision}",
      journal = {\aap},
         year = 2021,
        month = sep,
       volume = {653},
          eid = {A141},
        pages = {A141},
          doi = {10.1051/0004-6361/202140445},
archivePrefix = {arXiv},
       eprint = {2105.01661},
 primaryClass = {astro-ph.SR},
       adsurl = {https://ui.adsabs.harvard.edu/abs/2021A&A...653A.141A}
}

@ARTICLE{Katkov2014MNRAS.438.2798K,
       author = {{Katkov}, Ivan Yu. and {Sil'chenko}, Olga K. and {Afanasiev}, Victor L.},
        title = "{Decoupled gas kinematics in isolated S0 galaxies}",
      journal = {\mnras},
         year = 2014,
        month = mar,
       volume = {438},
       number = {4},
        pages = {2798-2803},
          doi = {10.1093/mnras/stt2365},
archivePrefix = {arXiv},
       eprint = {1312.6701},
 primaryClass = {astro-ph.GA},
       adsurl = {https://ui.adsabs.harvard.edu/abs/2014MNRAS.438.2798K}
}

@software{Ginsburg2016ascl.soft08010G,
       author = {{Ginsburg}, Adam and {Robitaille}, Thomas and {Beaumont}, Chris},
        title = "{pvextractor: Position-Velocity Diagram Extractor}",
 howpublished = {Astrophysics Source Code Library, record ascl:1608.010},
         year = 2016,
        month = aug,
          eid = {ascl:1608.010},
       adsurl = {https://ui.adsabs.harvard.edu/abs/2016ascl.soft08010G}
}

@ARTICLE{Kuijken1996MNRAS.283..543K,
       author = {{Kuijken}, K. and {Fisher}, D. and {Merrifield}, M.~R.},
        title = "{A search for counter-rotating stars in S0 galaxies.}",
      journal = {\mnras},
         year = 1996,
        month = dec,
       volume = {283},
       number = {2},
        pages = {543-550},
          doi = {10.1093/mnras/283.2.543},
archivePrefix = {arXiv},
       eprint = {astro-ph/9606099},
 primaryClass = {astro-ph},
       adsurl = {https://ui.adsabs.harvard.edu/abs/1996MNRAS.283..543K}
}

@ARTICLE{Pizzella2004A&A...424..447P,
       author = {{Pizzella}, A. and {Corsini}, E.~M. and {Vega Beltr{\'a}n}, J.~C. and {Bertola}, F.},
        title = "{Ionized gas and stellar kinematics of seventeen nearby spiral galaxies}",
      journal = {\aap},
         year = 2004,
        month = sep,
       volume = {424},
        pages = {447-454},
          doi = {10.1051/0004-6361:20047183},
archivePrefix = {arXiv},
       eprint = {astro-ph/0404558},
 primaryClass = {astro-ph},
       adsurl = {https://ui.adsabs.harvard.edu/abs/2004A&A...424..447P}
}

@INPROCEEDINGS{Katkov2024ASPC..535..239K,
       author = {{Katkov}, I. and {Gasymov}, D. and {Gelfand}, J.~D. and {Toptun}, V. and {Grishin}, K. and {Chilingarian}, I. and {Kasparova}, A. and {Klochkov}, V. and {Rubtsov}, E. and {Goradzhanov}, V.},
        title = "{Fast interactive web-based data visualizer of panoramic spectroscopic surveys}",
    booktitle = {Astromical Data Analysis Software and Systems XXXI},
         year = 2024,
       editor = {{Hugo}, B.~V. and {Van Rooyen}, R. and {Smirnov}, O.~M.},
       series = {Astronomical Society of the Pacific Conference Series},
       volume = {535},
        month = may,
        pages = {239},
          doi = {10.48550/arXiv.2112.03291},
archivePrefix = {arXiv},
       eprint = {2112.03291},
 primaryClass = {astro-ph.IM},
       adsurl = {https://ui.adsabs.harvard.edu/abs/2024ASPC..535..239K}
}

@ARTICLE{Vazdekis2016MNRAS.463.3409V,
       author = {{Vazdekis}, A. and {Koleva}, M. and {Ricciardelli}, E. and {R{\"o}ck}, B. and {Falc{\'o}n-Barroso}, J.},
        title = "{UV-extended E-MILES stellar population models: young components in massive early-type galaxies}",
      journal = {\mnras},
         year = 2016,
        month = dec,
       volume = {463},
       number = {4},
        pages = {3409-3436},
          doi = {10.1093/mnras/stw2231},
archivePrefix = {arXiv},
       eprint = {1612.01187},
 primaryClass = {astro-ph.GA},
       adsurl = {https://ui.adsabs.harvard.edu/abs/2016MNRAS.463.3409V}
}

@INPROCEEDINGS{Gasymov2024ASPC..535..279G,
       author = {{Gasymov}, D. and {Katkov}, I.},
        title = "{Non-parametric stellar LOSVD analysis}",
    booktitle = {Astromical Data Analysis Software and Systems XXXI},
         year = 2024,
       editor = {{Hugo}, B.~V. and {Van Rooyen}, R. and {Smirnov}, O.~M.},
       series = {Astronomical Society of the Pacific Conference Series},
       volume = {535},
        month = may,
        pages = {279},
       adsurl = {https://ui.adsabs.harvard.edu/abs/2024ASPC..535..279G}
}

@ARTICLE{Hinshaw2013ApJS..208...19H,
       author = {{Hinshaw}, G. and {Larson}, D. and {Komatsu}, E. and {Spergel}, D.~N. and {Bennett}, C.~L. and {Dunkley}, J. and {Nolta}, M.~R. and {Halpern}, M. and {Hill}, R.~S. and {Odegard}, N. and {Page}, L. and {Smith}, K.~M. and {Weiland}, J.~L. and {Gold}, B. and {Jarosik}, N. and {Kogut}, A. and {Limon}, M. and {Meyer}, S.~S. and {Tucker}, G.~S. and {Wollack}, E. and {Wright}, E.~L.},
        title = "{Nine-year Wilkinson Microwave Anisotropy Probe (WMAP) Observations: Cosmological Parameter Results}",
      journal = {\apjs},
         year = 2013,
        month = oct,
       volume = {208},
       number = {2},
          eid = {19},
        pages = {19},
          doi = {10.1088/0067-0049/208/2/19},
archivePrefix = {arXiv},
       eprint = {1212.5226},
 primaryClass = {astro-ph.CO},
       adsurl = {https://ui.adsabs.harvard.edu/abs/2013ApJS..208...19H}
}

@ARTICLE{Rubin1992ApJ...394L...9R,
       author = {{Rubin}, Vera C. and {Graham}, J.~A. and {Kenney}, Jeffrey D.~P.},
        title = "{Cospatial Counterrotating Stellar Disks in the Virgo E7/S0 Galaxy NGC 4550}",
      journal = {\apjl},
         year = 1992,
        month = jul,
       volume = {394},
        pages = {L9},
          doi = {10.1086/186460},
       adsurl = {https://ui.adsabs.harvard.edu/abs/1992ApJ...394L...9R}
}

@ARTICLE{Franx1988ApJ...327L..55F,
       author = {{Franx}, Marijn and {Illingworth}, Garth D.},
        title = "{A Counterrotating Core in IC 1459}",
      journal = {\apjl},
         year = 1988,
        month = apr,
       volume = {327},
        pages = {L55},
          doi = {10.1086/185139},
       adsurl = {https://ui.adsabs.harvard.edu/abs/1988ApJ...327L..55F}
}

@ARTICLE{1996ApJ...461...55T,
       author = {{Thakar}, Aniruddha R. and {Ryden}, Barbara S.},
        title = "{Formation of Massive Counterrotating Disks in Spiral Galaxies}",
      journal = {\apj},
         year = 1996,
        month = apr,
       volume = {461},
        pages = {55},
          doi = {10.1086/177037},
archivePrefix = {arXiv},
       eprint = {astro-ph/9510053},
 primaryClass = {astro-ph},
       adsurl = {https://ui.adsabs.harvard.edu/abs/1996ApJ...461...55T}
}

@ARTICLE{2014MNRAS.437.3596A,
       author = {{Algorry}, David G. and {Navarro}, Julio F. and {Abadi}, Mario G. and {Sales}, Laura V. and {Steinmetz}, Matthias and {Piontek}, Franziska},
        title = "{Counterrotating stars in simulated galaxy discs}",
      journal = {\mnras},
         year = 2014,
        month = feb,
       volume = {437},
       number = {4},
        pages = {3596-3602},
          doi = {10.1093/mnras/stt2154},
archivePrefix = {arXiv},
       eprint = {1311.1215},
 primaryClass = {astro-ph.CO},
       adsurl = {https://ui.adsabs.harvard.edu/abs/2014MNRAS.437.3596A}
}

@ARTICLE{Smirnov2020AstL...46..501S,
       author = {{Smirnov}, D.~V. and {Reshetnikov}, V.~P.},
        title = "{Active Galactic Nuclei among Polar-Ring Galaxies}",
      journal = {Astronomy Letters},
         year = 2020,
        month = aug,
       volume = {46},
       number = {8},
        pages = {501-508},
          doi = {10.1134/S1063773720080046},
archivePrefix = {arXiv},
       eprint = {2010.03349},
 primaryClass = {astro-ph.GA},
       adsurl = {https://ui.adsabs.harvard.edu/abs/2020AstL...46..501S}
}

@ARTICLE{2001Ap&SS.276..909P,
       author = {{Puerari}, I. and {Pfenniger}, D.},
        title = "{Formation of Massive Counter-Rotating Discs: An Alternative Scenario}",
      journal = {\apss},
         year = 2001,
        month = mar,
       volume = {276},
        pages = {909-914},
          doi = {10.1023/A:1017581325673},
archivePrefix = {arXiv},
       eprint = {astro-ph/9903096},
 primaryClass = {astro-ph},
       adsurl = {https://ui.adsabs.harvard.edu/abs/2001Ap&SS.276..909P}
}

@ARTICLE{2009MNRAS.393.1255C,
       author = {{Crocker}, Alison F. and {Jeong}, Hyunjin and {Komugi}, Shinya and {Combes}, Francoise and {Bureau}, Martin and {Young}, Lisa M. and {Yi}, Sukyoung},
        title = "{Molecular gas and star formation in the red-sequence counter-rotating disc galaxy NGC 4550}",
      journal = {\mnras},
         year = 2009,
        month = mar,
       volume = {393},
       number = {4},
        pages = {1255-1264},
          doi = {10.1111/j.1365-2966.2008.14295.x},
archivePrefix = {arXiv},
       eprint = {0812.0178},
 primaryClass = {astro-ph},
       adsurl = {https://ui.adsabs.harvard.edu/abs/2009MNRAS.393.1255C}
}

@ARTICLE{Evans1994ApJ...420L..67E,
       author = {{Evans}, N.~W. and {Collett}, J.~L.},
        title = "{Separatrix Crossing and the Enigma of NGC 4550}",
      journal = {\apjl},
         year = 1994,
        month = jan,
       volume = {420},
        pages = {L67},
          doi = {10.1086/187164},
       adsurl = {https://ui.adsabs.harvard.edu/abs/1994ApJ...420L..67E}
}

@ARTICLE{Graham2018MNRAS.477.4711G,
       author = {{Graham}, Mark T. and {Cappellari}, Michele and {Li}, Hongyu and {Mao}, Shude and {Bershady}, Matthew A. and {Bizyaev}, Dmitry and {Brinkmann}, Jonathan and {Brownstein}, Joel R. and {Bundy}, Kevin and {Drory}, Niv and {Law}, David R. and {Pan}, Kaike and {Thomas}, Daniel and {Wake}, David A. and {Weijmans}, Anne-Marie and {Westfall}, Kyle B. and {Yan}, Renbin},
        title = "{SDSS-IV MaNGA: stellar angular momentum of about 2300 galaxies: unveiling the bimodality of massive galaxy properties}",
      journal = {\mnras},
         year = 2018,
        month = jul,
       volume = {477},
       number = {4},
        pages = {4711-4737},
          doi = {10.1093/mnras/sty504},
archivePrefix = {arXiv},
       eprint = {1802.08213},
 primaryClass = {astro-ph.GA},
       adsurl = {https://ui.adsabs.harvard.edu/abs/2018MNRAS.477.4711G}
}

@ARTICLE{Beom2024AJ....168..197B,
       author = {{Beom}, Minje and {Walterbos}, Ren{\'e} A.~M. and {Bizyaev}, Dmitry},
        title = "{SDSS. IV. MaNGA: The Impact of the Acquisition of Gas with Opposite Angular Momentum on the Evolution of Galaxies}",
      journal = {\aj},
         year = 2024,
        month = nov,
       volume = {168},
       number = {5},
          eid = {197},
        pages = {197},
          doi = {10.3847/1538-3881/ad6f0b},
archivePrefix = {arXiv},
       eprint = {2410.06256},
 primaryClass = {astro-ph.GA},
       adsurl = {https://ui.adsabs.harvard.edu/abs/2024AJ....168..197B}
}

@ARTICLE{2022MNRAS.514.1006I,
       author = {{Izquierdo-Villalba}, David and {Bonoli}, Silvia and {Rosas-Guevara}, Yetli and {Springel}, Volker and {White}, Simon D.~M. and {Zana}, Tommaso and {Dotti}, Massimo and {Spinoso}, Daniele and {Bonetti}, Matteo and {Lupi}, Alessandro},
        title = "{Disc instability and bar formation: view from the IllustrisTNG simulations}",
      journal = {\mnras},
         year = 2022,
        month = jul,
       volume = {514},
       number = {1},
        pages = {1006-1020},
          doi = {10.1093/mnras/stac1413},
archivePrefix = {arXiv},
       eprint = {2203.07734},
 primaryClass = {astro-ph.GA},
       adsurl = {https://ui.adsabs.harvard.edu/abs/2022MNRAS.514.1006I}
}

@ARTICLE{Toomre1972ApJ...178..623T,
       author = {{Toomre}, Alar and {Toomre}, Juri},
        title = "{Galactic Bridges and Tails}",
      journal = {\apj},
         year = 1972,
        month = dec,
       volume = {178},
        pages = {623-666},
          doi = {10.1086/151823},
       adsurl = {https://ui.adsabs.harvard.edu/abs/1972ApJ...178..623T}
}

@ARTICLE{Barnes1988ApJ...331..699B,
       author = {{Barnes}, Joshua E.},
        title = "{Encounters of Disk/Halo Galaxies}",
      journal = {\apj},
         year = 1988,
        month = aug,
       volume = {331},
        pages = {699},
          doi = {10.1086/166593},
       adsurl = {https://ui.adsabs.harvard.edu/abs/1988ApJ...331..699B}
}

@ARTICLE{Sancisi2008A&ARv..15..189S,
       author = {{Sancisi}, Renzo and {Fraternali}, Filippo and {Oosterloo}, Tom and {van der Hulst}, Thijs},
        title = "{Cold gas accretion in galaxies}",
      journal = {\aapr},
         year = 2008,
        month = jun,
       volume = {15},
       number = {3},
        pages = {189-223},
          doi = {10.1007/s00159-008-0010-0},
archivePrefix = {arXiv},
       eprint = {0803.0109},
 primaryClass = {astro-ph},
       adsurl = {https://ui.adsabs.harvard.edu/abs/2008A&ARv..15..189S}
}

@ARTICLE{Rubin1994AJ....108..456R,
       author = {{Rubin}, Vera C.},
        title = "{Multi-Spin Galaxies}",
      journal = {\aj},
         year = 1994,
        month = aug,
       volume = {108},
        pages = {456},
          doi = {10.1086/117083},
       adsurl = {https://ui.adsabs.harvard.edu/abs/1994AJ....108..456R}
}

@ARTICLE{Rix1992ApJ...400L...5R,
       author = {{Rix}, Hans-Walter and {Franx}, Marijn and {Fisher}, David and {Illingworth}, Garth},
        title = "{NGC 4550: A Laboratory for Testing Galaxy Formation}",
      journal = {\apjl},
         year = 1992,
        month = nov,
       volume = {400},
        pages = {L5},
          doi = {10.1086/186635},
       adsurl = {https://ui.adsabs.harvard.edu/abs/1992ApJ...400L...5R}
}

@ARTICLE{Bender1988A&A...202L...5B,
       author = {{Bender}, R.},
        title = "{Rotating and counter-rotating cores in elliptical galaxies.}",
      journal = {\aap},
         year = 1988,
        month = aug,
       volume = {202},
        pages = {L5-L8},
       adsurl = {https://ui.adsabs.harvard.edu/abs/1988A&A...202L...5B}
}

@ARTICLE{Jedrzejewski1988ApJ...330L..87J,
       author = {{Jedrzejewski}, Robert and {Schechter}, Paul L.},
        title = "{Evidence for Dynamical Subsystems in Elliptical Galaxies}",
      journal = {\apjl},
         year = 1988,
        month = jul,
       volume = {330},
        pages = {L87},
          doi = {10.1086/185211},
       adsurl = {https://ui.adsabs.harvard.edu/abs/1988ApJ...330L..87J}
}

@ARTICLE{Franx1989ApJ...344..613F,
       author = {{Franx}, Marijn and {Illingworth}, Garth and {Heckman}, Timothy},
        title = "{Major and Minor Axis Kinematics of 22 Ellipticals}",
      journal = {\apj},
         year = 1989,
        month = sep,
       volume = {344},
        pages = {613},
          doi = {10.1086/167830},
       adsurl = {https://ui.adsabs.harvard.edu/abs/1989ApJ...344..613F}
}

@ARTICLE{Bertola1992ApJ...401L..79B,
       author = {{Bertola}, F. and {Buson}, L.~M. and {Zeilinger}, W.~W.},
        title = "{The External Origin of the Gas in S0 Galaxies}",
      journal = {\apjl},
         year = 1992,
        month = dec,
       volume = {401},
        pages = {L79},
          doi = {10.1086/186675},
       adsurl = {https://ui.adsabs.harvard.edu/abs/1992ApJ...401L..79B}
}

@ARTICLE{Merrifield1994ApJ...432..575M,
       author = {{Merrifield}, Michael R. and {Kuijken}, Konrad},
        title = "{Counterrotating Stars in the Disk of the SAB Galaxy NGC 7217}",
      journal = {\apj},
         year = 1994,
        month = sep,
       volume = {432},
        pages = {575},
          doi = {10.1086/174596},
       adsurl = {https://ui.adsabs.harvard.edu/abs/1994ApJ...432..575M}
}

@ARTICLE{Bertola1996ApJ...458L..67B,
       author = {{Bertola}, Francesco and {Cinzano}, Pierantonio and {Corsini}, Enrico Maria and {Pizzella}, Alessandro and {Persic}, Massimo and {Salucci}, Paolo},
        title = "{Counterrotating Stellar Disks in Early-Type Spirals: NGC 3593}",
      journal = {\apjl},
         year = 1996,
        month = feb,
       volume = {458},
        pages = {L67},
          doi = {10.1086/309924},
       adsurl = {https://ui.adsabs.harvard.edu/abs/1996ApJ...458L..67B}
}

@ARTICLE{Jore1996AJ....112..438J,
       author = {{Jore}, Katherine P. and {Broeils}, Adrick H. and {Haynes}, Martha P.},
        title = "{A Counter-Rotating Disk in the Normal SA Galaxy NGC 4138}",
      journal = {\aj},
         year = 1996,
        month = aug,
       volume = {112},
        pages = {438},
          doi = {10.1086/118027},
       adsurl = {https://ui.adsabs.harvard.edu/abs/1996AJ....112..438J}
}

@ARTICLE{MazzilliCiraulo2021A&A...653A..47M,
       author = {{Mazzilli Ciraulo}, Barbara and {Melchior}, Anne-Laure and {Maschmann}, Daniel and {Katkov}, Ivan Yu. and {Halle}, Ana{\"e}lle and {Combes}, Fran{\c{c}}oise and {Gelfand}, Joseph D. and {Al Yazeedi}, Aisha},
        title = "{Two interacting galaxies hiding as one, revealed by MaNGA}",
      journal = {\aap},
         year = 2021,
        month = sep,
       volume = {653},
          eid = {A47},
        pages = {A47},
          doi = {10.1051/0004-6361/202141319},
archivePrefix = {arXiv},
       eprint = {2106.07060},
 primaryClass = {astro-ph.GA},
       adsurl = {https://ui.adsabs.harvard.edu/abs/2021A&A...653A..47M}
}

@ARTICLE{Katkov2015AJ....150...24K,
       author = {{Katkov}, Ivan Yu. and {Kniazev}, Alexei Yu. and {Sil'chenko}, Olga K.},
        title = "{Kinematics and Stellar Populations in Isolated Lenticular Galaxies}",
      journal = {\aj},
         year = 2015,
        month = jul,
       volume = {150},
       number = {1},
          eid = {24},
        pages = {24},
          doi = {10.1088/0004-6256/150/1/24},
archivePrefix = {arXiv},
       eprint = {1505.01386},
 primaryClass = {astro-ph.GA},
       adsurl = {https://ui.adsabs.harvard.edu/abs/2015AJ....150...24K}
}

@ARTICLE{Freeman1970ApJ...160..811F,
       author = {{Freeman}, K.~C.},
        title = "{On the Disks of Spiral and S0 Galaxies}",
      journal = {\apj},
         year = 1970,
        month = jun,
       volume = {160},
        pages = {811},
          doi = {10.1086/150474},
       adsurl = {https://ui.adsabs.harvard.edu/abs/1970ApJ...160..811F}
}

@ARTICLE{Peirani2025arXiv250217902P,
       author = {{Peirani}, S. and {Suto}, Y. and {Han}, S. and {Yi}, S.~K. and {Dubois}, Y. and {Kraljic}, K. and {Park}, M. and {Pichon}, C.},
        title = "{Dissecting the formation of gas-versus-star counter-rotating galaxies from the NewHorizon simulation}",
      journal = {arXiv e-prints},
         year = 2025,
        month = feb,
          eid = {arXiv:2502.17902},
        pages = {arXiv:2502.17902},
          doi = {10.48550/arXiv.2502.17902},
archivePrefix = {arXiv},
       eprint = {2502.17902},
 primaryClass = {astro-ph.GA},
       adsurl = {https://ui.adsabs.harvard.edu/abs/2025arXiv250217902P}
}

@ARTICLE{Chilingarian2011MNRAS.412.1627C,
       author = {{Chilingarian}, Igor V. and {Mieske}, Steffen and {Hilker}, Michael and {Infante}, Leopoldo},
        title = "{Dynamical versus stellar masses of ultracompact dwarf galaxies in the Fornax cluster}",
      journal = {\mnras},
         year = 2011,
        month = apr,
       volume = {412},
       number = {3},
        pages = {1627-1638},
          doi = {10.1111/j.1365-2966.2010.18000.x},
archivePrefix = {arXiv},
       eprint = {1011.1852},
 primaryClass = {astro-ph.CO},
       adsurl = {https://ui.adsabs.harvard.edu/abs/2011MNRAS.412.1627C}
}

@ARTICLE{Cenci2024ApJ...961L..40C,
       author = {{Cenci}, Elia and {Feldmann}, Robert and {Gensior}, Jindra and {Bullock}, James S. and {Moreno}, Jorge and {Bassini}, Luigi and {Bernardini}, Mauro},
        title = "{Starburst-induced Gas{\textendash}Star Kinematic Misalignment}",
      journal = {\apjl},
         year = 2024,
        month = feb,
       volume = {961},
       number = {2},
          eid = {L40},
        pages = {L40},
          doi = {10.3847/2041-8213/ad1ffb},
archivePrefix = {arXiv},
       eprint = {2312.07334},
 primaryClass = {astro-ph.GA},
       adsurl = {https://ui.adsabs.harvard.edu/abs/2024ApJ...961L..40C}
}

@ARTICLE{Masters2019MNRAS.488.3396M,
       author = {{Masters}, Karen L. and {Stark}, David V. and {Pace}, Zachary J. and {Phipps}, Frederika and {Rujopakarn}, Wiphu and {Samanso}, Nattida and {Harrington}, Emily and {S{\'a}nchez-Gallego}, Jos{\'e} R. and {Avila-Reese}, Vladimir and {Bershady}, Matthew and {Cherinka}, Brian and {Fielder}, Catherine E. and {Finnegan}, Daniel and {Riffel}, Rogemar A. and {Rowlands}, Kate and {Shamsi}, Shoaib and {Newnham}, Lucy and {Weijmans}, Anne-Marie and {Witherspoon}, Catherine A.},
        title = "{H I-MaNGA: H I follow-up for the MaNGA survey}",
      journal = {\mnras},
         year = 2019,
        month = sep,
       volume = {488},
       number = {3},
        pages = {3396-3405},
          doi = {10.1093/mnras/stz1889},
archivePrefix = {arXiv},
       eprint = {1901.05579},
 primaryClass = {astro-ph.GA},
       adsurl = {https://ui.adsabs.harvard.edu/abs/2019MNRAS.488.3396M}
}

@book{tikhonov1977solutions,
  address = {Washington, D.C.: John Wiley \& Sons, New York},
  author = {Tikhonov, Andrey N. and Arsenin, Vasiliy Y.},
  description = {MR: Publications results for "MR Number=(455365)"},
  mrclass = {65J05 (65R05 65NXX)},
  mrnumber = {0455365 (56 \#13604)},
  mrreviewer = {M. Z. Nashed},
  note = {Translated from the Russian, Preface by translation editor Fritz John, Scripta Series in Mathematics},
  pages = {xiii+258},
  publisher = {V. H. Winston \& Sons},
  title = {Solutions of ill-posed problems},
  year = 1977
}

@ARTICLE{Tully1998AJ....115.2264T,
       author = {{Tully}, R. Brent and {Pierce}, Michael J. and {Huang}, Jia-Sheng and {Saunders}, Will and {Verheijen}, Marc A.~W. and {Witchalls}, Peter L.},
        title = "{Global Extinction in Spiral Galaxies}",
      journal = {\aj},
         year = 1998,
        month = jun,
       volume = {115},
       number = {6},
        pages = {2264-2272},
          doi = {10.1086/300379},
archivePrefix = {arXiv},
       eprint = {astro-ph/9802247},
 primaryClass = {astro-ph},
       adsurl = {https://ui.adsabs.harvard.edu/abs/1998AJ....115.2264T}
}

@ARTICLE{Chung2021ApJS..257...66C,
       author = {{Chung}, Haeun and {Park}, Changbom and {Park}, Yong-Sun},
        title = "{Point-spread Function Deconvolution of the IFU Data and Restoration of Galaxy Stellar Kinematics}",
      journal = {\apjs},
         year = 2021,
        month = dec,
       volume = {257},
       number = {2},
          eid = {66},
        pages = {66},
          doi = {10.3847/1538-4365/ac2828},
archivePrefix = {arXiv},
       eprint = {2008.04313},
 primaryClass = {astro-ph.GA},
       adsurl = {https://ui.adsabs.harvard.edu/abs/2021ApJS..257...66C}
}
\bibliographystyle{aasjournal}

\appendix

\section{Comparison with other searches}
\label{sec:App_A}

In this section, we  compare our sample with previous studies \cite{Graham2018MNRAS.477.4711G}, \cite{Bevacqua2022MNRAS.511..139B}, \cite{Bao2022ApJ...926L..13B, Bao2024MNRAS.528.2643B}, and \cite{Beom2024AJ....168..197B}. 

\subsection{\cite{Graham2018MNRAS.477.4711G}}

The first extensive analysis of angular momentum in the MaNGA SDSS-IV sample ($\sim$2300 targets) was conducted by \cite{Graham2018MNRAS.477.4711G}. 
A total of 22 2$\sigma$ galaxies were identified. 
Of these, 12 galaxies are included in our sample, while galaxy 1-113520 was classified as a gas-polar system.
The remaining galaxies do not exhibit clear CR features.

\subsection{\citet{Bevacqua2022MNRAS.511..139B}}

The next sample of galaxies with the stellar counter-rotation phenomenon was selected from MaNGA DR16 ($\sim$4600 targets) by \citet{Bevacqua2022MNRAS.511..139B}.
Limiting our sample to MaNGA DR16 leaves 63 galaxies (64 in the \cite{Bevacqua2022MNRAS.511..139B}).
Of these, 35 galaxies match between the samples, while 29 are found only in the \cite{Bevacqua2022MNRAS.511..139B} sample.
Among these 29, four exhibit kinematic misalignment between gas and stars (Sec.~\ref{sec:app_full_sample}).
The remaining 25 galaxies were rechecked, revealing no signs of $2\sigma$, S-rotation, or CR-GS features, except for one (1-549076), which exhibits a strong $2\sigma$ feature.
We analyzed this galaxy using our pipeline, but the recovered non-parametric LOSVD did not show any noticeable X-shape.

We expect counter-rotating disks to be younger than the main component and to co-rotate with the gaseous component, as found in all our galaxies with two-component spectral decomposition.
\cite{Bevacqua2022MNRAS.511..139B} identified two galaxies (1-38543 and 1-248410) where the gas co-rotates with an older disk.
We fitted the first galaxy (1-38543) using a two-component \nb\ decomposition, confirming co-rotation between the gas and a younger component (Fig.~\ref{fig:full_analysis_example}).
The next galaxy (1-248410) is a part of our probable sample (Tab.~\ref{Tab:Prob_samp}). 
Due to low SNR, our pipeline retrieves only the luminosity-weighted SSP equivalent age, which does not exhibit a reliable gradient from the center to the outer regions.
This does not allow us to determine whether the gas disk is associated with a younger or older component.

\subsection{\cite{Bao2022ApJ...926L..13B, Bao2024MNRAS.528.2643B}}

The sample from \cite{Bao2022ApJ...926L..13B}, based on MaNGA DR17 ($\sim$10,000 galaxies), includes 101 galaxies exhibiting stellar counter-rotation with a regularly rotating gas disk.
Our sample is slightly larger (120 galaxies), but after filtering out those without ionized gas, 101 counter-rotating galaxies remain.

The \cite{Bao2022ApJ...926L..13B} sample was categorized into four types (Types 1 and 2, each with subtypes a and b). Type ``1'' in their classification corresponds to S-rotation in our notation, while Type ``2'' lacks this feature but exhibits either 2$\sigma$ or $\sigma$-elongated structures.
Subtypes ``1a'' and ``2b'' indicate cases where the ionized gas rotates in the opposite direction to the outer regions (CR-GS feature in our notation), whereas ``1b'' and ``2a'' denote gas co-rotation with the outer parts. 
In our classification scheme, the former represents inner counter-rotation, while the latter corresponds to outer counter-rotation.

A key aspect highlighted by \cite{Bao2022ApJ...926L..13B} is the distinction between ``a'' and ``b'' types based on their specific star formation rate (sSFR) distributions. 
These galaxies were divided into three categories: the blue cloud (BC, $\log$~sSFR $> -11$ yr$^{-1}$), the red sequence (RS, $\log$~sSFR $< -14$ yr$^{-1}$), and the green valley (GV), where sSFR lies between these limits.
The fraction of ``b'' type galaxies increases from the BC to RS populations, whereas the ``a'' type decreases.
However, in our sample, where sSFR data is taken from GSWLC \citep{Salim2016ApJS..227....2S}, this correlation is absent. Notably, we lack RS galaxies, as the minimum sSFR in our sample is $\log$ sSFR $\approx -13.3$ yr$^{-1}$.
To assess consistency, we cross-matched the sSFR estimates from Salim \citep{Salim2016ApJS..227....2S} and Chang \citep{Chang2015ApJS..219....8C}, finding general agreement within the margin of error. 
However, Chang’s estimates for $\log$~sSFR $< -11$ yr$^{-1}$ exhibit significant uncertainties, with large 1$\sigma$ errors of about 1-1.5 dex.
We used demarcation values of $\log$~sSFR$=-10.8$ yr$^{-1}$ and $\log$~sSFR$=-11.8$ yr$^{-1}$ (Fig.~\ref{fig:cmd_M_SFR}) and confirmed sSFR differences between the ``a'' and ``b'' types, consistent with findings in \citet{Bao2022ApJ...926L..13B} (Fig.~2):
\begin{deluxetable}{c|cc|cc}[H]
\tablecaption{Population difference on SFR--M$_\star$ diagram (Fig.~\ref{fig:cmd_M_SFR}) in ``a'' and ``b'' types of CR.}
\tablehead{
   & \colhead{1a} & \colhead{1b} & \colhead{2a} & \colhead{2b}
}
\startdata
BC &  100\%  &  0\%   &  62\%  &  38\%  \\
GV &  62\%   &  38\%  &  25\%  &  75\%  \\
RS &  43\%   &  57\%  &  29\%  &  71\%  \\
\enddata
\end{deluxetable}

The galaxy 1-339061, which hosts an inner CR disk, was studied in detail by \citet{Bao2024MNRAS.528.2643B}.
The flux ratio between the primary and secondary disks is about 1.8, which could be converted to a CR disk luminosity weight of $1/(1+1.8) \approx 36\%$, close to our estimate of $\textrm{W}_\textrm{L, CR} \approx 33\%$.
The parameters of stellar populations of both disks agree within error bounds (Tab.~\ref{tab:np_losvd_2comp} and Fig.~9 in \citet{Bao2024MNRAS.528.2643B}). 
Mass-weight estimates differ (16\% compared to our $\approx$8\%), but this discrepancy arises from variations in the definition of mass-to-light ratios.
However, the CR disk masses in this galaxy converge because we applied a $(100 / H_0)^{2}$ correction to NSA masses, which increased the stellar mass of the galaxy by a factor of $\sim 2$.

\subsection{\citet{Beom2024AJ....168..197B}}

\cite{Beom2024AJ....168..197B} focused on galaxies exhibiting gaseous CR, though they also marked galaxies with stellar CR (type ``CS'').
This sample contains 29 CR galaxies, of which 25 have inner CR (ICS, in their notation) and 4 have outer (OCS, in their notation).
20 galaxies match our sample, but one galaxy (1-26197) does not conform to our CR disk configuration.
In the \cite{Beom2024AJ....168..197B} sample, this galaxy is marked as ICS, but the recovered LOSVD clearly shows a large-scale CR disk (third row from the bottom in the central column of Fig.~\ref{fig:np_losvd}).
The remaining nine galaxies are included in the full sample of kinematically decoupled galaxies (Tab.~\ref{Tab:Mis_sample}): 1-195979 (CR-GS), 1-295542 (CR-GS, NRR), 1-92702 (CR-GS), 1-246298 (CR-GS, NRR), 1-261386 (Gas-mis), 1-546345 (CR-GS), 1-294016 ($2\sigma$?), 1-35832 (CR-GS), 1-634718 (Srot, CR-GS), but they do not show clear CR features.
We have rechecked their kinematic maps and recovered LOSVDs, confirming our classification and the absence of CR.

\section{Sample of galaxies with kinematical misalignment}
\label{sec:app_full_sample}

\startlongtable
\begin{deluxetable*}{clrrcccccc}
\label{Tab:Mis_sample}
\tablecaption{Table with kinematically decoupled galaxies (excluding stellar CR). Description is similar to Tab.~\ref{Tab:Rel_samp}.}
\tablewidth{700pt}
\tabletypesize{\tiny}
\tablehead{
\colhead{No.} & 
\colhead{ID} & 
\colhead{R.A.} & 
\colhead{Dec.} & 
\colhead{Type} & 
\colhead{$\log$ (M$_\star$/M$_\odot$)} & 
\colhead{z} & 
\colhead{Features} & 
\colhead{Environment} & 
\colhead{Morphology}} 
\decimalcolnumbers
\startdata
1 & \href{https://manga.voxastro.org/visualiser/7962-6104}{1-178443} & 17:23:01.1 & 27:34:10.10 & SABbc &10.22 & 0.048 & 2$\sigma$ & \textbf{Cl}, \textbf{G} & \ldots \\
2 & \href{https://manga.voxastro.org/visualiser/8486-12704}{1-209407} & 15:53:02.8 & 46:46:04.79 & SABbc &9.40 & 0.020 & 2$\sigma$ & \textbf{Cl}, \textbf{G} & \textbf{E} \\
3 & \href{https://manga.voxastro.org/visualiser/11944-6103}{1-248565} & 16:10:50.2 & 38:42:29.67 & SABa &9.83 & 0.038 & 2$\sigma$ & \textbf{Cl}, \textbf{G} & \ldots \\
4 & \href{https://manga.voxastro.org/visualiser/11823-6102}{1-269886} & 16:33:59.1 & 26:53:52.71 & Edc &10.95 & 0.071 & 2$\sigma$ & \textbf{Cl}, \textbf{G} & \ldots \\
5 & \href{https://manga.voxastro.org/visualiser/8262-6101}{1-282020} & 12:18:39.6 & 43:32:19.40 & SBd &9.61 & 0.024 & 2$\sigma$ & \textbf{F}, \textbf{G} & \textbf{B} \\
6 & \href{https://manga.voxastro.org/visualiser/9194-3701}{1-603965} & 03:07:32.3 & -00:57:52.40 & Edc &11.40 & 0.039 & 2$\sigma$ & \textbf{V}, \textbf{G} & \textbf{Sh} \\
7 & \href{https://manga.voxastro.org/visualiser/11753-6104}{1-78294} & 09:48:16.2 & 03:54:21.08 & Sbc &10.08 & 0.024 & 2$\sigma$ & \textbf{F}, \textbf{G} & \textbf{E} \\
8 & \href{https://manga.voxastro.org/visualiser/12679-3701}{1-97176} & 21:13:09.8 & -07:37:54.36 & Sa &10.44 & 0.028 & 2$\sigma$ & \textbf{V}, \textbf{G} & \textbf{Sh} \\
\enddata
\end{deluxetable*}

\section{Analysis of recovered LOSVD and two-component full-spectrum fitting}
\label{sec:app_C}
Figures like Fig.~\ref{fig:full_analysis_example} have been prepared for all galaxies in the sample and are available at \url{https://zenodo.org/records/15132551}.

\section{Table with parameters from LOSVD analysis}

\begin{deluxetable*}{lcccccccccccccc}
\centering
\scriptsize
\tablecaption{Parameters of the LOSVD fit for the sample (Sec.~\ref{sec:np_losvd}). Parameters marked with $^{\ast}$ have high relative uncertainties ($>50\%$), while italicized values indicate fixed parameters used to stabilize fits for noisy LOSVDs.}
\tablehead{
\colhead{ID} & \colhead{I$_0$} & \colhead{h} & \colhead{V$_{01}$} & \colhead{V$_{02}$} & \colhead{$\sigma_{01}$} & \colhead{$\sigma_{02}$} & \colhead{$\sigma_{11}$} & \colhead{$\sigma_{12}$} & \colhead{R$_{01}$} & \colhead{R$_{02}$} & \colhead{k$_1$} & \colhead{k$_2$} & \colhead{$\log{\textrm{J}_m}$} & \colhead{$\log{\textrm{J}_{\textrm{CR}}}$} \\
}
\startdata
1-42660 & $0.8 \pm 0.1$ & $9.1 \pm 2.0$ & $95 \pm 1$ & $94 \pm 6$ & $\mathit{40}$ & $\mathit{40}$ & $\mathit{75}$ & $\mathit{75}$ & $6.1 \pm 0.5$ & $10.5 \pm 0.0$ & $\mathit{-15}$ & $\mathit{-4}$ & 12.6 & 12.6 \\
1-42247 & $0.6 \pm 0.1$ & $-4.0 \pm 0.8$ & $74 \pm 5$ & $74 \pm 3$ & $27 \pm 0$ & $34 \pm 0$ & $67 \pm 0$ & $75 \pm 0$ & $4.9 \pm 1.1$ & $5.0 \pm 1.2$ & $-7.0 \pm 0.0$ & $-11.0 \pm 0.1$ & \ldots & \ldots \\
1-37062 & $0.1 \pm 0.0$ & $33.9^{\ast}$ & $150 \pm 0$ & $195 \pm 5$ & $25 \pm 0$ & $25 \pm 0$ & $82 \pm 3$ & $55 \pm 22$ & $24.9 \pm 0.4$ & $9.8 \pm 1.0$ & $-6.8 \pm 0.5$ & $-8.0^{\ast}$ & 12.6 & 11.2 \\
1-37478 & $0.4 \pm 0.0$ & $-1978.9^{\ast}$ & $116 \pm 2$ & $131 \pm 6$ & $49 \pm 3$ & $60 \pm 0$ & $145 \pm 9$ & $143 \pm 18$ & $4.5 \pm 0.0$ & $2.5^{\ast}$ & $-18.2 \pm 2.0$ & $-29.4 \pm 8.8$ & \ldots & \ldots \\
1-38543 & $1.2 \pm 0.1$ & $8.6 \pm 1.4$ & $130 \pm 1$ & $158 \pm 4$ & $35 \pm 1$ & $36 \pm 2$ & $105 \pm 6$ & $98 \pm 2$ & $4.4 \pm 0.7$ & $7.1 \pm 0.8$ & $-14.2 \pm 1.4$ & $-15.0 \pm 0.0$ & 13.0 & 12.3 \\
1-339061 & $0.8 \pm 0.1$ & $14.4 \pm 4.1$ & $192 \pm 4$ & $147 \pm 3$ & $66 \pm 2$ & $50 \pm 4$ & $101 \pm 6$ & $123 \pm 10$ & $8.4 \pm 0.2$ & $6.8 \pm 1.0$ & $-5.6 \pm 1.2$ & $-9.7 \pm 1.7$ & 13.2 & 12.0 \\
1-44047 & $1.0 \pm 0.0$ & $-19.9 \pm 1.4$ & $134 \pm 2$ & $140 \pm 2$ & $40 \pm 2$ & $43 \pm 0$ & $113 \pm 6$ & $137 \pm 0$ & $4.0 \pm 0.5$ & $6.4 \pm 0.7$ & $-14.3 \pm 1.6$ & $-19.7 \pm 0.0$ & 12.6 & 12.3 \\
1-44483 & $0.5 \pm 0.0$ & $-5.1 \pm 0.4$ & $129 \pm 1$ & $155 \pm 1$ & $27 \pm 2$ & $44 \pm 0$ & $121 \pm 4$ & $121 \pm 0$ & $6.9 \pm 0.5$ & $12.8 \pm 0.0$ & $-14.7 \pm 0.9$ & $-12.2 \pm 0.1$ & 12.8 & 12.9 \\
\enddata
\end{deluxetable*}



\end{document}